\documentclass[aps,prl,twocolumn,showpacs,superscriptaddress,amsmath,amssymb]{revtex4-2}
\usepackage{graphicx}
\usepackage{latexsym}
\usepackage{bm} 
\usepackage{color}
\usepackage{epsfig}
\usepackage{multirow}
\usepackage{xcolor}
\usepackage{colortbl}
\usepackage{hhline}
\usepackage{simplewick}
\usepackage{soul}

\usepackage{graphicx}
\usepackage[colorlinks=true,linkcolor=blue,citecolor=blue,urlcolor=blue]{hyperref}
\usepackage{bbold}
\usepackage{gensymb}

\AtBeginDocument{%
    \newwrite\bibnotes
    \def\bibnotesext{Notes.bib}
    \immediate\openout\bibnotes=\jobname\bibnotesext
    \immediate\write\bibnotes{@CONTROL{REVTEX42Control}}
    \immediate\write\bibnotes{@CONTROL{%
    apsrev42Control,author="08",editor="1",pages="0",title="0",year="1"}}
     \if@filesw
     \immediate\write\@auxout{\string\citation{apsrev42Control}}%
    \fi
}%

\def \mbf {\mathbf}

\newcommand{\epvl}[1]{\langle#1\rangle}

\def \Tr {\mathrm{Tr}}
\newcommand{\comment}[1]{}

\definecolor{mygreen}{rgb}{0, 0.7, 0}

\begin{document}

\title{Complex magnetic and spatial symmetry breaking from correlations in kagome flat bands}

\author{Yu-Ping Lin}
\affiliation{Department of Physics, University of California, Berkeley, California 94720, USA}
\author{Chunxiao Liu}
\affiliation{Department of Physics, University of California, Berkeley, California 94720, USA}
\author{Joel E. Moore}
\affiliation{Department of Physics, University of California, Berkeley, California 94720, USA}
\affiliation{Materials Sciences Division, Lawrence Berkeley National Laboratory, Berkeley, California 94720, USA}

\date{\today}

\begin{abstract}
We present the mean-field phase diagram of electrons in a kagome flat band with repulsive interactions. In addition to flat-band ferromagnetism, the Hartree-Fock analysis yields cascades of unconventional magnetic orders driven by onsite repulsion as filling changes. These include a series of antiferromagnetic (AFM) spin-charge stripe orders, as well as an evolution from $120^\circ$AFM to intriguing noncoplanar spin orders with tetrahedral structures. We also map out the phase diagram under extended repulsion at half and empty fillings of the flat band. To examine the possibilities beyond the mean-field level, we conduct a projective symmetry group analysis and identify the feasible $\mathbb Z_2$ spin liquids and the magnetic orders derivable from them. The theoretical phase diagrams are compared with recent experiments on FeSn and FeGe, enabling a determination of the most likely magnetic instabilities in these and similar flat-band kagome materials.
\end{abstract}

\maketitle

\textit{Introduction.---}The study of flat bands has become a major focus of condensed matter research in the past decade. Among various flat-band systems, there is a family in which the flat bands are inherent to geometric frustration of the lattices \cite{mielke92jpa,bergman08prb,leykam18apx,maimaiti21prb,calugaru22np,regnault22n,graf21prb,hwang21prb,neves24npjcm}. Under the destructive interference of hoppings, compact localized states (CLSs) appear as dispersionless eigenstates and form the flat bands, in combination with noncontractible loop states. The search among frustrated materials has identified many candidates. In particular, recent experiments have observed (approximately) flat bands in the quasi-two-dimensional (2D) kagome metals FeSn \cite{kang20nm} and CoSn \cite{kang20nc} and their relevance to FeGe \cite{teng22n,yin22prl,teng23np}.

The kagome lattice is known as a fruitful basis for correlated phases. In particular, the massive density of states carried by the flat band can enhance correlation effects, giving rise to a rich phase diagram. The famous Stoner criterion for ferromagnetism (FM) \cite{stoner38prsa} can be easily satisfied, supposedly accounting for the intralayer FM below 300--400 K in FeSn \cite{kang20nm,xie21cp} and FeGe \cite{teng22n,yin22prl,teng23np}. Interestingly, FeGe further shows a charge density wave (CDW) below 100 K \cite{teng22n,yin22prl,teng23np}, similar to the nonmagnetic kagome metals $A$V$_3$Sb$_5$ with $A=$ K, Rb, and Cs \cite{jiang21nm,zhao21n}. FM is proposed to be the ground state under onsite repulsion in the kagome flat band, especially at and above half filling \cite{mielke92jpa,mielke93cmp,hanisch95annp,pollmann08prl}. Meanwhile, the large number of kagome materials suggests an opportunity for many other correlated phases to be explored.
Various unconventional correlated phases, such as quantum anomalous Hall FM (QAHFM) and combinations of spin and charge orders, have been proposed at spinful half filling \cite{ren21prl}, spinless \cite{nishimoto10prl,obrien10prb,wen10prb,zhu16prl,ren18prb} or spinful \cite{liu10prb,wen10prb,pollmann14prb} empty filling, and fractional filling under topology \cite{okamoto22cp,setty22ax}.

\begin{figure*}
\centering
\includegraphics[scale=1]{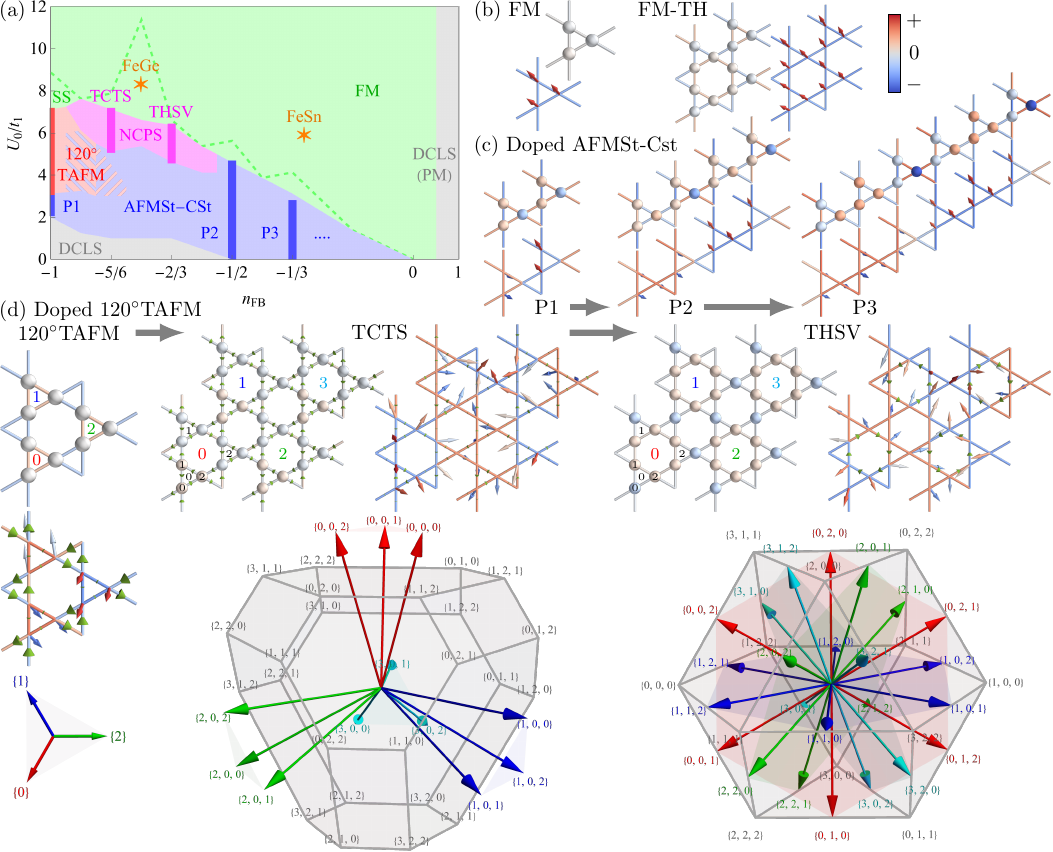}
\caption{\label{fig:pd} Mean-field phase diagram in kagome flat band. (a) The phase diagram of pure Hubbard model. The colored bars indicate commensurate-filling orders, and the hatches indicate phase separation. (b)-(d) S-C patterns of the ground states. For the charge patterns, the warm (cool) colors indicate higher (lower) densities on the sites and bonds from their averages (see the color bar). The green bond arrowheads indicate the bond currents. For the spin patterns, the site arrows represent in-plane components, while the colors indicate out-of-plane components on both sites and bonds. The bond arrowheads again indicate the bond currents. In the noncollinear spin orders (d), the spin structures are further presented. The arrow colors match the labels of chosen cells. The labels for 120$^\circ$TAFM apply to the 1-triangle cells. For the TCTS and THSV orders, the labels \{3-triangle cell, 1-triangle cell, site\} are defined. The colored arrows and gray-polyhedron corners represent the spins on high- and low-density sites, respectively.}
\end{figure*}

Given the rich range of correlated phases in both experiment and theory, it is natural to seek the interaction-driven phase diagram in the kagome flat band. In this \textit{Letter}, we conduct various theoretical analyses to achieve this goal. Our central result is a mean-field phase diagram obtained through Hartree-Fock analysis. This phase diagram, as shown in Fig.~\ref{fig:pd}, is controlled by onsite repulsion and electron filling. We highlight that our Hartree-Fock analysis, compared with previous works \cite{hanisch95annp,liu10prb,wen10prb}, does not impose any symmetry constraints or ans\"atze. This fully unrestricted approach allows us to uncover many unconventional ordered states existing in the phase diagram (Fig.~\ref{fig:pd}). Interestingly, we find doping cascades of spin-charge (S-C) stripe and $120^\circ$ antiferromagnetic (AFM)-to-noncoplanar spin (NCPS) orders, which fill the weak-to-moderate-coupling regime below half filling \cite{hanisch95annp}. Adding extended repulsions further expands the phase diagrams at half \cite{ren21prl} and empty fillings \cite{liu10prb,wen10prb}. Going beyond the mean-field level, we conduct a projective symmetry group (PSG) analysis, thereby identifying a $\mathbb Z_2$ spin liquid whose derived magnetic orders match the Hartree-Fock results. Finally, by comparing our theory with existing experiments on FeSn and FeGe, we guide the search for unconventional ordering in future studies of flat-band kagome metals.

\begin{figure}[t]
\centering
\includegraphics[scale=1]{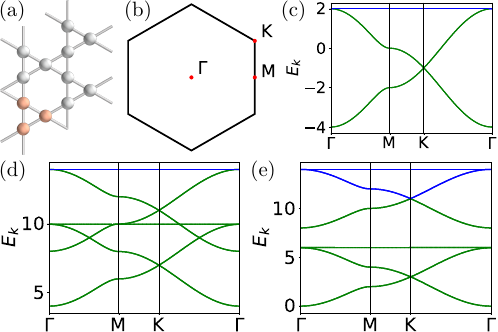}
\caption{\label{fig:kagome} Kagome lattice. (a) The lattice with a unit cell marked (orange). The triangular-Bravais-lattice vectors are $\mbf a_1=(0,2)$ and $\mbf a_2=(\sqrt3,1)$ with intersite distance $1$. (b) The Brillouin zone and high-symmetry points. (c) The band structure with filled (green) and empty (blue) band segments. (d)(e) The FM band structures with $U_0/t=12$ at $n_\text{FB}=0$ and $-1$.}
\end{figure}

\textit{Kagome lattice and flat band.---}We study the repulsive fermionic Hubbard model \cite{arovas22arcmp,qin22arcmp} on the kagome lattice [Fig.~\ref{fig:kagome}(a)]
\begin{equation}
\label{eq:hubbard}
\begin{aligned}
H&=-\sum_{ii'\tau\tau'}\sum_\sigma t_{ii'\tau\tau'}c_{i\tau\sigma}^\dagger c_{i'\tau'\sigma}\\&\quad+\frac{1}{2}\sum_{ii'\tau\tau'}\sum_{\sigma\sigma'}U_{ii'\tau\tau'}c_{i\tau\sigma}^\dagger c_{i'\tau'\sigma'}^\dagger c_{i'\tau'\sigma'}c_{i\tau\sigma},
\end{aligned}
\end{equation}
where $c_{i\tau\sigma}^{(\dagger)}$ annihilates (creates) a fermion at the Bravais-lattice site $i$ and sublattice $\tau=0,1,2$ with spin $\sigma=\uparrow\downarrow$. The tight-binding hoppings $t_{ii'\tau\tau'}$ and density-density repulsions $U_{ii'\tau\tau'}$ are defined according to their ranges of action as onsite $t_0$ and $U_0$, nearest-neighbor $t_1$ and $U_1$, second-neighbor $t_2$ and $U_2$, etc.

The tight-binding Hamiltonian hosts three bands in the Brillouin zone [Figs.~\ref{fig:kagome}(b) and \ref{fig:kagome}(c)]. The topmost band is completely flat under nearest-neighbor hopping $t_1>0$ \cite{mielke92jpa,bergman08prb}. Focusing on this flat band, we define the flat-band (FB) filling $n_\text{FB}\in[-1,1]$. Half filling is $n_\text{FB}=0$, and $n_\text{FB}=\pm1$ correspond to full and empty fillings, respectively.

\textit{Mean-field phase diagram.---}We employ the Hartree-Fock analysis to obtain the repulsion-driven ground states at the mean-field level. Our spatially unrestricted formalism \cite{supp} is effective for unbiased search of symmetry-breaking ground states with enlarged unit cells. We first focus on the pure Hubbard model with nearest-neighbor hopping $t_1=1$ and onsite repulsion $U_0>0$. Sweeping across all flat-band fillings, we map out the $n_\text{FB}$-$U_0$ phase diagram [Fig.~\ref{fig:pd}(a)]. One immediate observation is the wide FM phase [Fig.~\ref{fig:pd}(b)] at strong coupling \cite{hanisch95annp}, which extends to the weak-coupling limit $U_0=0$ at and above half filling $n_\text{FB}\geq0$. Meanwhile, the other competing orders arise at weak-to-moderate coupling below half filling $n_\text{FB}<0$.

\textit{Ferromagnetism.---}When FM develops, the bands split into majority- and minority-spin branches. The density imbalance $\delta n_\text{S}=n_\text{major}-n_\text{minor}$ 
determines the energy of this Stoner splitting
\begin{equation}
\label{eq:stoner}
\Delta E_\text{S}=U_0\delta n_\text{S}.
\end{equation}
To saturate the maximal magnetization $m=\delta n_\text{S}/2$, the Stoner splitting should fully fill the majority-spin bands by pushing them below the minority-spin Fermi level. For half filling $n_\text{FB}=0$, the saturated minority-spin Fermi level lies at the quadratic band-crossing point (QBCP) $\boldsymbol\Gamma$ [Fig.~\ref{fig:kagome}(d)]. The saturation occurs under arbitrarily small repulsion, indicating a strong weak-coupling instability towards FM. This scenario also applies above half filling $n_\text{FB}>0$. On the other hand, the saturation splitting increases below half filling $n_\text{FB}<0$, with the maximum at empty filling $n_\text{FB}=-1$ [Fig.~\ref{fig:kagome}(e)]. The FM is destabilized gradually, giving way to other competing orders at weak-to-moderate coupling. Note that the FM in the phase diagram [Fig.~\ref{fig:pd}(a)] is always saturated and is a half (semi)metal.

\textit{Competing orders.---}Next we explore the plethora of phases below half filling $n_\text{FB}<0$. We begin with the moderate-coupling regime at empty filling $n_\text{FB}=-1$. An $\mbf M$-point S-C stripe order arises at $U_0=2$. Here AFM stripes develop on charge nematic high-density stripes and form an AFM-stripe-charge-stripe-period-1 (AFMSt-CStP1) pattern [Fig.~\ref{fig:pd}(c)]. For $U_0\geq3$, the ground state is taken over by a $\mbf K$-point order. This order shows a $\sqrt3\times\sqrt3$ $120^\circ$ triangle AFM ($120^\circ$TAFM) [Fig.~\ref{fig:pd}(d)], where coplanar $120^\circ$ order \cite{huse88prl} forms among three nearest-neighbor triangles.

\textit{Cascade of S-C stripe orders.---} Moving away from empty filling $n_\text{FB}=-1$, many orders appear.  In the S-C-stripe regime, the AFMSt-CStP1 order deforms into the FM as $n_\text{FB}$ goes from $-1$ to $0$. Intuitively, the FM is an AFMSt-CSt order with infinite period (P$\infty$). There is an elegant cascade of P$n$ orders at the reciprocal integer fillings $n_\text{FB}=-1/n$ [Fig.~\ref{fig:pd}(c)], where one out of $n$ low-density stripes is unfilled at $n_\text{FB}=-1/n$. Note that the cascade is also visible away from the reciprocal integer fillings $n_\text{FB}=-1/n$ \cite{supp}, despite some intertwinement with the $120^\circ$TAFM.

\textit{Cascade of noncoplanar spin orders.---} More intriguing magnetic orders appear when the $120^\circ$TAFM is doped. Remarkably, the energetically favorable path breaks coplanarity and travels through noncoplanar spin orders. At commensurate fillings $n_\text{FB}=-5/6$ and $-2/3$, which are $1/3$ and $2/3$ between $n_\text{FB}=-1$ and $-1/2$, we observe enriched tetrahedral spin orders [Fig.~\ref{fig:pd}(d)]. The S-C unit cell is $2\sqrt{3}\times2\sqrt{3}$ at both fillings. The charge unit cell is $\sqrt{3}\times\sqrt{3}$, containing both high- and low-density sites. The tetrahedral spin orders further enlarge the period by $2\times2$. At $n_\text{FB}=-5/6$, each high-density triangle hosts $\text{C}_3$-symmetric canted spins about a principal order. The principal orders are tetrahedral, making the ground state a tetrahedral canted triangle-spin (TCTS) order. Note that the low-density sites also host TCTS structures, which couple to the high-density TCTS order and induce the canting. At $n_\text{FB}=-2/3$, we observe a coplanar $2\pi$ spin winding on each high-density hexagon, which we term the hexagon spin vortex (HSV). Amazingly, the HSVs are again tetrahedral, and the ground state is a tetrahedral HSV (THSV) order. This order is actually a canted tripling of the cuboc order in the classical $J_1$-$J_2$ model \cite{domenge05prb,domenge08prb,messio11prb}. The low-density kagome superlattice hosts the $12$-site cuboc order. Meanwhile, the high-density sites show a $24$-site canted doubling, where each spin is the middle of its two low-density neighbors.

Due to nontrivial spin configurations, noncollinear spin orders may act as sources of time-reversal-symmetry-breaking fluxes, thereby inducing nontrivial band topology \cite{ohgushi00prb,taguchi01sc,martin08prl,barros14prb}. To examine the possibility of noncollinear-spin Chern insulators, we conduct a thorough band-structure computation \cite{supp}. Remarkably, we find nonzero quantized Chern number $C=\pm1$ for the TCTS order, consistent with nonzero scalar spin chirality from noncoplanar spin textures \cite{ohgushi00prb,taguchi01sc,martin08prl,barros14prb}.

\begin{figure}[t]
\centering
\includegraphics[scale=1]{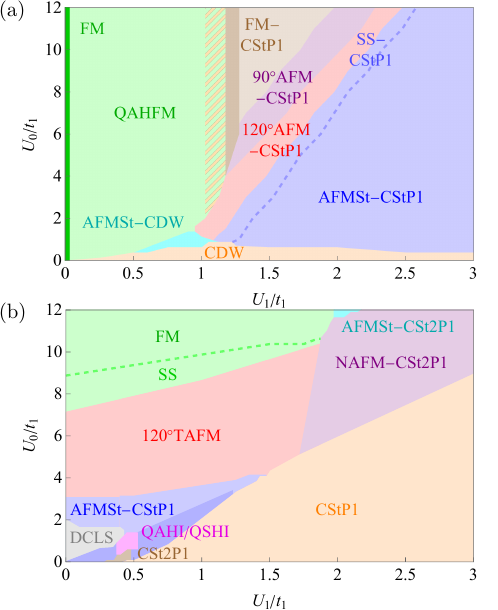}
\caption{\label{fig:pdnfb0n1} Extended-repulsion phase diagrams at (a) half filling $n_\text{FB}=0$ and (b) empty filling $n_\text{FB}=-1$.}
\end{figure}

\textit{Completion of phase diagram.---}The phase diagram [Fig.~\ref{fig:pd}(a)] exhibits a few more features apart from the uniform orders, including phase separation \cite{emery90prl,hanisch95annp,qin22jpcm}, coplanar spiral spin (SS) orders \cite{hanisch95annp,qin22jpcm}, and dilute CLS (DCLS). Furthermore, we go beyond the pure Hubbard model by adding extended repulsions $U_1=2U_2>0$ (Fig.~\ref{fig:pdnfb0n1}). While onsite repulsion favors magnetic orders, extended repulsion enhances charge orders and alters the ground states. See Supplemental Material \cite{supp} for the detailed discussions.

\textit{Possible spin liquids.---}The rich structure in the phase diagrams reveals an intricate relation among the competing magnetic orders. An interesting scenario may happen when they arise as symmetry-breaking instabilities of a \emph{parent} quantum spin liquid \cite{explanation}. Here, from a symmetry perspective, we explore this possibility by selecting the most likely $\mathbb{Z}_2$ kagome spin liquids \cite{PhysRevB.65.224412,yan2011spin,nishimoto2013controlling,jiang2012identifying} and analyzing the various magnetic orders derived from them \cite{messio12prl,supp}.

Our starting point is the two $\mathbb{Z}_2$-spin-liquid Hamiltonians obtained from the PSG classification \cite{PhysRevB.74.174423} for bosonic spinons. This conveniently allows us to consider the leading instability upon condensing the spinons. The resulting orders are given in Table \ref{Tab:mag_order_psg}. Starting from the $(1,0)$ class of $\mathbb Z_2$ parent spin liquid, we reproduce the FM, $120^\circ$TAFM, and SS orders in Hartree-Fock analysis \cite{note1}. Crucially, the SS orders are much weaker, suggesting the possible discovery of $\mathbb Z_2$ parent spin liquid in its mean-field phase region [Fig.~\ref{fig:pd}(a)]. On the other hand, we find an interesting sublattice-momentum-locked magnetic order from the $(0,1)$ class. This magnetic order may share similar structure to the tetrahedral chiral spin density wave \cite{martin08prl}.

\begin{table}[t]
\centering
\caption{Summary of magnetic orders derived from the PSG classes $(0,1)$ and $(1,0)$. Ordering structure in the enlarged magnetic unit cell: $\dag$: non-uniform, umbrella order. $\ddag$: Coplanar $120^\circ$. The cases 1, 2, and 3--5 of class (1,0) correspond to the FM, $120^\circ$TAFM, and SS orders in the Hartree-Fock analysis, respectively.}\label{Tab:mag_order_psg}
\begin{tabular}{cccccc}
\hline
Class&No.&Spinon $k_c$&  Mag. unit cell & Sublattice structure \\
\hline
\multirow{5}{*}{(0,1)}&1&$\Gamma$&Not enlarged&Collinear\\
&2&$\Gamma$&Not enlarged&Collinear\\
&3&$\pm\mathrm{K}$&$\sqrt{3}\times \sqrt{3}^\dag$&Collinear\\
&4&$\pm\mathrm{K}$&$\sqrt{3}\times \sqrt{3}^\dag$&Coplanar $120^\circ$\\
&5&$\mathrm{M}_{1,2,3}$&$2\times 2^\dag$& $\tau$--$\mathrm{M}_\tau$-locked stripe\\
\hline
\multirow{4}{*}{(1,0)}&1&$\Gamma$&Not enlarged&Collinear\\
&2&$\pm\mathrm{K}$&$\sqrt{3}\times \sqrt{3}^\ddag$&Collinear\\
&\multirow{2}{*}{3--5}&$\Gamma\leftrightarrow\pm\mathrm{K}$&\multirow{2}{*}{Incommensurate}&\multirow{2}{*}{$-$}\\
&&$\leftrightarrow \mathrm{M}_{1,2,3}$&&\\
\hline
\end{tabular}
\end{table}

\textit{Experimental realization.---}We examine the flat-band kagome metals FeSn \cite{kang20nm} and FeGe \cite{teng22n,yin22prl,teng23np} in light of our phase diagrams. Motivated by the Fe-$d_{xz/yz}$-orbital band structures of FeSn \cite{xie21cp} and FeGe [assuming the Van Hove singularity (VHS) at the Fermi level as was experimentally observed] \cite{teng23np} in first-principles computations, we consider the single-orbital pure Hubbard model on the kagome lattice. The flat-band filling $n_\text{FB}$ and the onsite repulsion $U_0$ are estimated from the minority-spin filling and Stoner splitting energy (\ref{eq:stoner}), respectively. This estimation locates FeSn and FeGe in our phase diagram, as shown in Fig.~\ref{fig:pd}(a). FeSn falls in the FM phase, consistent with the experiments. Although FeGe sits in the SS-order phase, the interlayer and multi-orbital effects may stabilize the FM. We further include the nearest-neighbor repulsion $U_1/t_1\approx1.07$ for FeGe by fitting the CDW splitting \cite{teng23np}. This induces the secondary tri-hexagonal bond order (FM-TH) [Fig.~\ref{fig:pd}(b)], consistent with the experimental $2\times2$ CDW \cite{teng22n,yin22prl,teng23np} and the theoretical CDW at VHS \cite{tan21prl,lin21prb,park21prb,feng21prb,christensen21prb}.

It is worth noting the possible implications of the results in our work. The majority of existing compounds (such as FeSn and FeGe) are located in the large-$U_0$ FM phase. Meanwhile, our analysis finds intriguing magnetic orders at smaller repulsion $U_0$. We thus propose the reduction of electronic repulsion as a feasible route toward unconventional magnetism in a kagome flat band. In addition to the binary compounds $AB$, another feasible family is the ternary series $AB_3C_5$. These materials host possibly weaker correlations than FeSn and FeGe, and the Fermi level can be tuned flexibly by element substitutions \cite{jiang22cpl}. On the other hand, with the ultracold-atom simulation of a kagome lattice \cite{jo12prl}, the flat-band correlated phases can be induced by adding the Hubbard interaction \cite{mazurenko17n}.

\textit{Discussion.---}Our analysis includes a detailed study of kagome flat bands at the mean-field level, but there remains considerable scope for further investigation. First, the Hartree-Fock analysis can overestimate the symmetry-breaking orders and miss essential intertwinement between different orders. While the former may be less serious in the flat band, the latter may be enhanced due to strong correlations. Second, there are important correlated phases that are not captured. These include superconductivity, which does not appear at the mean-field level under electronic repulsion. It will be interesting to search for superconductivity by doping the unconventional magnetic orders \cite{watanabe05jpsj,song21prb,peng21njp,zhu22prb,huang23prl,zhu23prb} in our phase diagram. Meanwhile, spin-triplet superconductivity may occur in the FM half metal \cite{cheng10prb,gneist22epjb,he23prr}. On the other hand, the ground states away from the fermion-bilinear condensates, such as spin liquids \cite{savary16rpp} and fractional Chern insulators \cite{liu23bk}, are also invisible. While we suggest candidate $\mathbb Z_2$ spin liquids with a PSG analysis, a future confirmation with strong-correlation numerical methods will be informative.

\textit{Note added.---}After the original version of this work, the synthesis of CsCr$_3$Sb$_5$ in the $AB_3C_5$ kagome family was reported \cite{liu23ax}. A non-FM S-C-density-wave order develops close to the flat band, consistent with expectations from our Hartree-Fock analysis.

\begin{acknowledgments}
The authors thank Shubhayu Chatterjee, Yi-Ping Huang, Haining Pan, and Linda Ye for fruitful discussions. This work was primarily supported by the Air Force Office of Scientific Research under Grant No. FA9550-22-1-0270. Y.P.L. and C.L. acknowledge fellowship support from the Gordon and Betty Moore Foundation through the Emergent Phenomena in Quantum Systems (EPiQS) program.  J.E.M. acknowledges a Simons Investigatorship. Parts of the numerical computations were performed on the Lawrencium cluster at Lawrence Berkeley National Laboratory.
\end{acknowledgments}




\bibliography{reference}

\begin{thebibliography}{79}%
\makeatletter
\providecommand \@ifxundefined [1]{%
 \@ifx{#1\undefined}
}%
\providecommand \@ifnum [1]{%
 \ifnum #1\expandafter \@firstoftwo
 \else \expandafter \@secondoftwo
 \fi
}%
\providecommand \@ifx [1]{%
 \ifx #1\expandafter \@firstoftwo
 \else \expandafter \@secondoftwo
 \fi
}%
\providecommand \natexlab [1]{#1}%
\providecommand \enquote  [1]{``#1''}%
\providecommand \bibnamefont  [1]{#1}%
\providecommand \bibfnamefont [1]{#1}%
\providecommand \citenamefont [1]{#1}%
\providecommand \href@noop [0]{\@secondoftwo}%
\providecommand \href [0]{\begingroup \@sanitize@url \@href}%
\providecommand \@href[1]{\@@startlink{#1}\@@href}%
\providecommand \@@href[1]{\endgroup#1\@@endlink}%
\providecommand \@sanitize@url [0]{\catcode `\\12\catcode `\$12\catcode
  `\&12\catcode `\#12\catcode `\^12\catcode `\_12\catcode `\%12\relax}%
\providecommand \@@startlink[1]{}%
\providecommand \@@endlink[0]{}%
\providecommand \url  [0]{\begingroup\@sanitize@url \@url }%
\providecommand \@url [1]{\endgroup\@href {#1}{\urlprefix }}%
\providecommand \urlprefix  [0]{URL }%
\providecommand \Eprint [0]{\href }%
\providecommand \doibase [0]{https://doi.org/}%
\providecommand \selectlanguage [0]{\@gobble}%
\providecommand \bibinfo  [0]{\@secondoftwo}%
\providecommand \bibfield  [0]{\@secondoftwo}%
\providecommand \translation [1]{[#1]}%
\providecommand \BibitemOpen [0]{}%
\providecommand \bibitemStop [0]{}%
\providecommand \bibitemNoStop [0]{.\EOS\space}%
\providecommand \EOS [0]{\spacefactor3000\relax}%
\providecommand \BibitemShut  [1]{\csname bibitem#1\endcsname}%
\let\auto@bib@innerbib\@empty
\bibitem [{\citenamefont {Mielke}(1992)}]{mielke92jpa}%
  \BibitemOpen
  \bibfield  {author} {\bibinfo {author} {\bibfnamefont {A.}~\bibnamefont
  {Mielke}},\ }\bibfield  {title} {\bibinfo {title} {Exact ground states for
  the hubbard model on the kagome lattice},\ }\href
  {https://doi.org/10.1088/0305-4470/25/16/011} {\bibfield  {journal} {\bibinfo
   {journal} {J. Phys. A: Math. Gen.}\ }\textbf {\bibinfo {volume} {25}},\
  \bibinfo {pages} {4335} (\bibinfo {year} {1992})}\BibitemShut {NoStop}%
\bibitem [{\citenamefont {Bergman}\ \emph {et~al.}(2008)\citenamefont
  {Bergman}, \citenamefont {Wu},\ and\ \citenamefont {Balents}}]{bergman08prb}%
  \BibitemOpen
  \bibfield  {author} {\bibinfo {author} {\bibfnamefont {D.~L.}\ \bibnamefont
  {Bergman}}, \bibinfo {author} {\bibfnamefont {C.}~\bibnamefont {Wu}},\ and\
  \bibinfo {author} {\bibfnamefont {L.}~\bibnamefont {Balents}},\ }\bibfield
  {title} {\bibinfo {title} {Band touching from real-space topology in
  frustrated hopping models},\ }\href
  {https://doi.org/10.1103/PhysRevB.78.125104} {\bibfield  {journal} {\bibinfo
  {journal} {Phys. Rev. B}\ }\textbf {\bibinfo {volume} {78}},\ \bibinfo
  {pages} {125104} (\bibinfo {year} {2008})}\BibitemShut {NoStop}%
\bibitem [{\citenamefont {Leykam}\ \emph {et~al.}(2018)\citenamefont {Leykam},
  \citenamefont {Andreanov},\ and\ \citenamefont {Flach}}]{leykam18apx}%
  \BibitemOpen
  \bibfield  {author} {\bibinfo {author} {\bibfnamefont {D.}~\bibnamefont
  {Leykam}}, \bibinfo {author} {\bibfnamefont {A.}~\bibnamefont {Andreanov}},\
  and\ \bibinfo {author} {\bibfnamefont {S.}~\bibnamefont {Flach}},\ }\bibfield
   {title} {\bibinfo {title} {Artificial flat band systems: from lattice models
  to experiments},\ }\href {https://doi.org/10.1080/23746149.2018.1473052}
  {\bibfield  {journal} {\bibinfo  {journal} {Adv. Phys. X}\ }\textbf {\bibinfo
  {volume} {3}},\ \bibinfo {pages} {1473052} (\bibinfo {year}
  {2018})}\BibitemShut {NoStop}%
\bibitem [{\citenamefont {Maimaiti}\ \emph {et~al.}(2021)\citenamefont
  {Maimaiti}, \citenamefont {Andreanov},\ and\ \citenamefont
  {Flach}}]{maimaiti21prb}%
  \BibitemOpen
  \bibfield  {author} {\bibinfo {author} {\bibfnamefont {W.}~\bibnamefont
  {Maimaiti}}, \bibinfo {author} {\bibfnamefont {A.}~\bibnamefont
  {Andreanov}},\ and\ \bibinfo {author} {\bibfnamefont {S.}~\bibnamefont
  {Flach}},\ }\bibfield  {title} {\bibinfo {title} {Flat-band generator in two
  dimensions},\ }\href {https://doi.org/10.1103/PhysRevB.103.165116} {\bibfield
   {journal} {\bibinfo  {journal} {Phys. Rev. B}\ }\textbf {\bibinfo {volume}
  {103}},\ \bibinfo {pages} {165116} (\bibinfo {year} {2021})}\BibitemShut
  {NoStop}%
\bibitem [{\citenamefont {C\ifmmode \u{a}\else \u{a}\fi{}lug\ifmmode~\u{a}\else
  \u{a}\fi{}ru}\ \emph {et~al.}(2022)\citenamefont {C\ifmmode \u{a}\else
  \u{a}\fi{}lug\ifmmode~\u{a}\else \u{a}\fi{}ru}, \citenamefont {Chew},
  \citenamefont {Elcoro}, \citenamefont {Xu}, \citenamefont {Regnault},
  \citenamefont {Song},\ and\ \citenamefont {Bernevig}}]{calugaru22np}%
  \BibitemOpen
  \bibfield  {author} {\bibinfo {author} {\bibfnamefont {D.}~\bibnamefont
  {C\ifmmode \u{a}\else \u{a}\fi{}lug\ifmmode~\u{a}\else \u{a}\fi{}ru}},
  \bibinfo {author} {\bibfnamefont {A.}~\bibnamefont {Chew}}, \bibinfo {author}
  {\bibfnamefont {L.}~\bibnamefont {Elcoro}}, \bibinfo {author} {\bibfnamefont
  {Y.}~\bibnamefont {Xu}}, \bibinfo {author} {\bibfnamefont {N.}~\bibnamefont
  {Regnault}}, \bibinfo {author} {\bibfnamefont {Z.-D.}\ \bibnamefont {Song}},\
  and\ \bibinfo {author} {\bibfnamefont {B.~A.}\ \bibnamefont {Bernevig}},\
  }\bibfield  {title} {\bibinfo {title} {General construction and topological
  classification of crystalline flat bands},\ }\href
  {https://doi.org/10.1038/s41567-021-01445-3} {\bibfield  {journal} {\bibinfo
  {journal} {Nat. Phys.}\ }\textbf {\bibinfo {volume} {18}},\ \bibinfo {pages}
  {185} (\bibinfo {year} {2022})}\BibitemShut {NoStop}%
\bibitem [{\citenamefont {Regnault}\ \emph {et~al.}(2022)\citenamefont
  {Regnault}, \citenamefont {Xu}, \citenamefont {Li}, \citenamefont {Ma},
  \citenamefont {Jovanovic}, \citenamefont {Yazdani}, \citenamefont {Parkin},
  \citenamefont {Felser}, \citenamefont {Schoop}, \citenamefont {Ong},
  \citenamefont {Cava}, \citenamefont {Elcoro}, \citenamefont {Song},\ and\
  \citenamefont {Bernevig}}]{regnault22n}%
  \BibitemOpen
  \bibfield  {author} {\bibinfo {author} {\bibfnamefont {N.}~\bibnamefont
  {Regnault}}, \bibinfo {author} {\bibfnamefont {Y.}~\bibnamefont {Xu}},
  \bibinfo {author} {\bibfnamefont {M.-R.}\ \bibnamefont {Li}}, \bibinfo
  {author} {\bibfnamefont {D.-S.}\ \bibnamefont {Ma}}, \bibinfo {author}
  {\bibfnamefont {M.}~\bibnamefont {Jovanovic}}, \bibinfo {author}
  {\bibfnamefont {A.}~\bibnamefont {Yazdani}}, \bibinfo {author} {\bibfnamefont
  {S.~S.~P.}\ \bibnamefont {Parkin}}, \bibinfo {author} {\bibfnamefont
  {C.}~\bibnamefont {Felser}}, \bibinfo {author} {\bibfnamefont {L.~M.}\
  \bibnamefont {Schoop}}, \bibinfo {author} {\bibfnamefont {N.~P.}\
  \bibnamefont {Ong}}, \bibinfo {author} {\bibfnamefont {R.~J.}\ \bibnamefont
  {Cava}}, \bibinfo {author} {\bibfnamefont {L.}~\bibnamefont {Elcoro}},
  \bibinfo {author} {\bibfnamefont {Z.-D.}\ \bibnamefont {Song}},\ and\
  \bibinfo {author} {\bibfnamefont {B.~A.}\ \bibnamefont {Bernevig}},\
  }\bibfield  {title} {\bibinfo {title} {Catalogue of flat-band stoichiometric
  materials},\ }\href {https://doi.org/10.1038/s41586-022-04519-1} {\bibfield
  {journal} {\bibinfo  {journal} {Nature}\ }\textbf {\bibinfo {volume} {603}},\
  \bibinfo {pages} {824} (\bibinfo {year} {2022})}\BibitemShut {NoStop}%
\bibitem [{\citenamefont {Graf}\ and\ \citenamefont
  {Pi\'echon}(2021)}]{graf21prb}%
  \BibitemOpen
  \bibfield  {author} {\bibinfo {author} {\bibfnamefont {A.}~\bibnamefont
  {Graf}}\ and\ \bibinfo {author} {\bibfnamefont {F.}~\bibnamefont
  {Pi\'echon}},\ }\bibfield  {title} {\bibinfo {title} {Designing flat-band
  tight-binding models with tunable multifold band touching points},\ }\href
  {https://doi.org/10.1103/PhysRevB.104.195128} {\bibfield  {journal} {\bibinfo
   {journal} {Phys. Rev. B}\ }\textbf {\bibinfo {volume} {104}},\ \bibinfo
  {pages} {195128} (\bibinfo {year} {2021})}\BibitemShut {NoStop}%
\bibitem [{\citenamefont {Hwang}\ \emph {et~al.}(2021)\citenamefont {Hwang},
  \citenamefont {Rhim},\ and\ \citenamefont {Yang}}]{hwang21prb}%
  \BibitemOpen
  \bibfield  {author} {\bibinfo {author} {\bibfnamefont {Y.}~\bibnamefont
  {Hwang}}, \bibinfo {author} {\bibfnamefont {J.-W.}\ \bibnamefont {Rhim}},\
  and\ \bibinfo {author} {\bibfnamefont {B.-J.}\ \bibnamefont {Yang}},\
  }\bibfield  {title} {\bibinfo {title} {General construction of flat bands
  with and without band crossings based on wave function singularity},\ }\href
  {https://doi.org/10.1103/PhysRevB.104.085144} {\bibfield  {journal} {\bibinfo
   {journal} {Phys. Rev. B}\ }\textbf {\bibinfo {volume} {104}},\ \bibinfo
  {pages} {085144} (\bibinfo {year} {2021})}\BibitemShut {NoStop}%
\bibitem [{\citenamefont {Neves}\ \emph {et~al.}(2024)\citenamefont {Neves},
  \citenamefont {Wakefield}, \citenamefont {Fang}, \citenamefont {Nguyen},
  \citenamefont {Ye},\ and\ \citenamefont {Checkelsky}}]{neves24npjcm}%
  \BibitemOpen
  \bibfield  {author} {\bibinfo {author} {\bibfnamefont {P.~M.}\ \bibnamefont
  {Neves}}, \bibinfo {author} {\bibfnamefont {J.~P.}\ \bibnamefont
  {Wakefield}}, \bibinfo {author} {\bibfnamefont {S.}~\bibnamefont {Fang}},
  \bibinfo {author} {\bibfnamefont {H.}~\bibnamefont {Nguyen}}, \bibinfo
  {author} {\bibfnamefont {L.}~\bibnamefont {Ye}},\ and\ \bibinfo {author}
  {\bibfnamefont {J.~G.}\ \bibnamefont {Checkelsky}},\ }\bibfield  {title}
  {\bibinfo {title} {Crystal net catalog of model flat band materials},\ }\href
  {https://doi.org/10.1038/s41524-024-01220-x} {\bibfield  {journal} {\bibinfo
  {journal} {npj Comput. Mater.}\ }\textbf {\bibinfo {volume} {10}},\ \bibinfo
  {pages} {39} (\bibinfo {year} {2024})}\BibitemShut {NoStop}%
\bibitem [{\citenamefont {Kang}\ \emph
  {et~al.}(2020{\natexlab{a}})\citenamefont {Kang}, \citenamefont {Ye},
  \citenamefont {Fang}, \citenamefont {You}, \citenamefont {Levitan},
  \citenamefont {Han}, \citenamefont {Facio}, \citenamefont {Jozwiak},
  \citenamefont {Bostwick}, \citenamefont {Rotenberg}, \citenamefont {Chan},
  \citenamefont {McDonald}, \citenamefont {Graf}, \citenamefont {Kaznatcheev},
  \citenamefont {Vescovo}, \citenamefont {Bell}, \citenamefont {Kaxiras},
  \citenamefont {van~den Brink}, \citenamefont {Richter}, \citenamefont
  {Prasad~Ghimire}, \citenamefont {Checkelsky},\ and\ \citenamefont
  {Comin}}]{kang20nm}%
  \BibitemOpen
  \bibfield  {author} {\bibinfo {author} {\bibfnamefont {M.}~\bibnamefont
  {Kang}}, \bibinfo {author} {\bibfnamefont {L.}~\bibnamefont {Ye}}, \bibinfo
  {author} {\bibfnamefont {S.}~\bibnamefont {Fang}}, \bibinfo {author}
  {\bibfnamefont {J.-S.}\ \bibnamefont {You}}, \bibinfo {author} {\bibfnamefont
  {A.}~\bibnamefont {Levitan}}, \bibinfo {author} {\bibfnamefont
  {M.}~\bibnamefont {Han}}, \bibinfo {author} {\bibfnamefont {J.~I.}\
  \bibnamefont {Facio}}, \bibinfo {author} {\bibfnamefont {C.}~\bibnamefont
  {Jozwiak}}, \bibinfo {author} {\bibfnamefont {A.}~\bibnamefont {Bostwick}},
  \bibinfo {author} {\bibfnamefont {E.}~\bibnamefont {Rotenberg}}, \bibinfo
  {author} {\bibfnamefont {M.~K.}\ \bibnamefont {Chan}}, \bibinfo {author}
  {\bibfnamefont {R.~D.}\ \bibnamefont {McDonald}}, \bibinfo {author}
  {\bibfnamefont {D.}~\bibnamefont {Graf}}, \bibinfo {author} {\bibfnamefont
  {K.}~\bibnamefont {Kaznatcheev}}, \bibinfo {author} {\bibfnamefont
  {E.}~\bibnamefont {Vescovo}}, \bibinfo {author} {\bibfnamefont {D.~C.}\
  \bibnamefont {Bell}}, \bibinfo {author} {\bibfnamefont {E.}~\bibnamefont
  {Kaxiras}}, \bibinfo {author} {\bibfnamefont {J.}~\bibnamefont {van~den
  Brink}}, \bibinfo {author} {\bibfnamefont {M.}~\bibnamefont {Richter}},
  \bibinfo {author} {\bibfnamefont {M.}~\bibnamefont {Prasad~Ghimire}},
  \bibinfo {author} {\bibfnamefont {J.~G.}\ \bibnamefont {Checkelsky}},\ and\
  \bibinfo {author} {\bibfnamefont {R.}~\bibnamefont {Comin}},\ }\bibfield
  {title} {\bibinfo {title} {{Dirac fermions and flat bands in the ideal kagome
  metal FeSn}},\ }\href {https://doi.org/10.1038/s41563-019-0531-0} {\bibfield
  {journal} {\bibinfo  {journal} {Nat. Mater.}\ }\textbf {\bibinfo {volume}
  {19}},\ \bibinfo {pages} {163} (\bibinfo {year}
  {2020}{\natexlab{a}})}\BibitemShut {NoStop}%
\bibitem [{\citenamefont {Kang}\ \emph
  {et~al.}(2020{\natexlab{b}})\citenamefont {Kang}, \citenamefont {Fang},
  \citenamefont {Ye}, \citenamefont {Po}, \citenamefont {Denlinger},
  \citenamefont {Jozwiak}, \citenamefont {Bostwick}, \citenamefont {Rotenberg},
  \citenamefont {Kaxiras}, \citenamefont {Checkelsky},\ and\ \citenamefont
  {Comin}}]{kang20nc}%
  \BibitemOpen
  \bibfield  {author} {\bibinfo {author} {\bibfnamefont {M.}~\bibnamefont
  {Kang}}, \bibinfo {author} {\bibfnamefont {S.}~\bibnamefont {Fang}}, \bibinfo
  {author} {\bibfnamefont {L.}~\bibnamefont {Ye}}, \bibinfo {author}
  {\bibfnamefont {H.~C.}\ \bibnamefont {Po}}, \bibinfo {author} {\bibfnamefont
  {J.}~\bibnamefont {Denlinger}}, \bibinfo {author} {\bibfnamefont
  {C.}~\bibnamefont {Jozwiak}}, \bibinfo {author} {\bibfnamefont
  {A.}~\bibnamefont {Bostwick}}, \bibinfo {author} {\bibfnamefont
  {E.}~\bibnamefont {Rotenberg}}, \bibinfo {author} {\bibfnamefont
  {E.}~\bibnamefont {Kaxiras}}, \bibinfo {author} {\bibfnamefont {J.~G.}\
  \bibnamefont {Checkelsky}},\ and\ \bibinfo {author} {\bibfnamefont
  {R.}~\bibnamefont {Comin}},\ }\bibfield  {title} {\bibinfo {title}
  {{Topological flat bands in frustrated kagome lattice CoSn}},\ }\href
  {https://doi.org/10.1038/s41467-020-17465-1} {\bibfield  {journal} {\bibinfo
  {journal} {Nat. Commun.}\ }\textbf {\bibinfo {volume} {11}},\ \bibinfo
  {pages} {4004} (\bibinfo {year} {2020}{\natexlab{b}})}\BibitemShut {NoStop}%
\bibitem [{\citenamefont {Teng}\ \emph {et~al.}(2022)\citenamefont {Teng},
  \citenamefont {Chen}, \citenamefont {Ye}, \citenamefont {Rosenberg},
  \citenamefont {Liu}, \citenamefont {Yin}, \citenamefont {Jiang},
  \citenamefont {Oh}, \citenamefont {Hasan}, \citenamefont {Neubauer},
  \citenamefont {Gao}, \citenamefont {Xie}, \citenamefont {Hashimoto},
  \citenamefont {Lu}, \citenamefont {Jozwiak}, \citenamefont {Bostwick},
  \citenamefont {Rotenberg}, \citenamefont {Birgeneau}, \citenamefont {Chu}, ,
  \citenamefont {Yi},\ and\ \citenamefont {Dai}}]{teng22n}%
  \BibitemOpen
  \bibfield  {author} {\bibinfo {author} {\bibfnamefont {X.}~\bibnamefont
  {Teng}}, \bibinfo {author} {\bibfnamefont {L.}~\bibnamefont {Chen}}, \bibinfo
  {author} {\bibfnamefont {F.}~\bibnamefont {Ye}}, \bibinfo {author}
  {\bibfnamefont {E.}~\bibnamefont {Rosenberg}}, \bibinfo {author}
  {\bibfnamefont {Z.}~\bibnamefont {Liu}}, \bibinfo {author} {\bibfnamefont
  {J.-X.}\ \bibnamefont {Yin}}, \bibinfo {author} {\bibfnamefont {Y.-X.}\
  \bibnamefont {Jiang}}, \bibinfo {author} {\bibfnamefont {J.~S.}\ \bibnamefont
  {Oh}}, \bibinfo {author} {\bibfnamefont {M.~Z.}\ \bibnamefont {Hasan}},
  \bibinfo {author} {\bibfnamefont {K.~J.}\ \bibnamefont {Neubauer}}, \bibinfo
  {author} {\bibfnamefont {B.}~\bibnamefont {Gao}}, \bibinfo {author}
  {\bibfnamefont {Y.}~\bibnamefont {Xie}}, \bibinfo {author} {\bibfnamefont
  {M.}~\bibnamefont {Hashimoto}}, \bibinfo {author} {\bibfnamefont
  {D.}~\bibnamefont {Lu}}, \bibinfo {author} {\bibfnamefont {C.}~\bibnamefont
  {Jozwiak}}, \bibinfo {author} {\bibfnamefont {A.}~\bibnamefont {Bostwick}},
  \bibinfo {author} {\bibfnamefont {E.}~\bibnamefont {Rotenberg}}, \bibinfo
  {author} {\bibfnamefont {R.~J.}\ \bibnamefont {Birgeneau}}, \bibinfo {author}
  {\bibfnamefont {J.-H.}\ \bibnamefont {Chu}}, , \bibinfo {author}
  {\bibfnamefont {M.}~\bibnamefont {Yi}},\ and\ \bibinfo {author}
  {\bibfnamefont {P.}~\bibnamefont {Dai}},\ }\bibfield  {title} {\bibinfo
  {title} {Discovery of charge density wave in a kagome lattice
  antiferromagnet},\ }\href {https://doi.org/10.1038/s41586-022-05034-z}
  {\bibfield  {journal} {\bibinfo  {journal} {Nature}\ }\textbf {\bibinfo
  {volume} {609}},\ \bibinfo {pages} {490} (\bibinfo {year}
  {2022})}\BibitemShut {NoStop}%
\bibitem [{\citenamefont {Yin}\ \emph {et~al.}(2022)\citenamefont {Yin},
  \citenamefont {Jiang}, \citenamefont {Teng}, \citenamefont {Hossain},
  \citenamefont {Mardanya}, \citenamefont {Chang}, \citenamefont {Ye},
  \citenamefont {Xu}, \citenamefont {Denner}, \citenamefont {Neupert},
  \citenamefont {Lienhard}, \citenamefont {Deng}, \citenamefont {Setty},
  \citenamefont {Si}, \citenamefont {Chang}, \citenamefont {Guguchia},
  \citenamefont {Gao}, \citenamefont {Shumiya}, \citenamefont {Zhang},
  \citenamefont {Cochran}, \citenamefont {Multer}, \citenamefont {Yi},
  \citenamefont {Dai},\ and\ \citenamefont {Hasan}}]{yin22prl}%
  \BibitemOpen
  \bibfield  {author} {\bibinfo {author} {\bibfnamefont {J.-X.}\ \bibnamefont
  {Yin}}, \bibinfo {author} {\bibfnamefont {Y.-X.}\ \bibnamefont {Jiang}},
  \bibinfo {author} {\bibfnamefont {X.}~\bibnamefont {Teng}}, \bibinfo {author}
  {\bibfnamefont {M.~S.}\ \bibnamefont {Hossain}}, \bibinfo {author}
  {\bibfnamefont {S.}~\bibnamefont {Mardanya}}, \bibinfo {author}
  {\bibfnamefont {T.-R.}\ \bibnamefont {Chang}}, \bibinfo {author}
  {\bibfnamefont {Z.}~\bibnamefont {Ye}}, \bibinfo {author} {\bibfnamefont
  {G.}~\bibnamefont {Xu}}, \bibinfo {author} {\bibfnamefont {M.~M.}\
  \bibnamefont {Denner}}, \bibinfo {author} {\bibfnamefont {T.}~\bibnamefont
  {Neupert}}, \bibinfo {author} {\bibfnamefont {B.}~\bibnamefont {Lienhard}},
  \bibinfo {author} {\bibfnamefont {H.-B.}\ \bibnamefont {Deng}}, \bibinfo
  {author} {\bibfnamefont {C.}~\bibnamefont {Setty}}, \bibinfo {author}
  {\bibfnamefont {Q.}~\bibnamefont {Si}}, \bibinfo {author} {\bibfnamefont
  {G.}~\bibnamefont {Chang}}, \bibinfo {author} {\bibfnamefont
  {Z.}~\bibnamefont {Guguchia}}, \bibinfo {author} {\bibfnamefont
  {B.}~\bibnamefont {Gao}}, \bibinfo {author} {\bibfnamefont {N.}~\bibnamefont
  {Shumiya}}, \bibinfo {author} {\bibfnamefont {Q.}~\bibnamefont {Zhang}},
  \bibinfo {author} {\bibfnamefont {T.~A.}\ \bibnamefont {Cochran}}, \bibinfo
  {author} {\bibfnamefont {D.}~\bibnamefont {Multer}}, \bibinfo {author}
  {\bibfnamefont {M.}~\bibnamefont {Yi}}, \bibinfo {author} {\bibfnamefont
  {P.}~\bibnamefont {Dai}},\ and\ \bibinfo {author} {\bibfnamefont {M.~Z.}\
  \bibnamefont {Hasan}},\ }\bibfield  {title} {\bibinfo {title} {{Discovery of
  Charge Order and Corresponding Edge State in Kagome Magnet FeGe}},\ }\href
  {https://doi.org/10.1103/PhysRevLett.129.166401} {\bibfield  {journal}
  {\bibinfo  {journal} {Phys. Rev. Lett.}\ }\textbf {\bibinfo {volume} {129}},\
  \bibinfo {pages} {166401} (\bibinfo {year} {2022})}\BibitemShut {NoStop}%
\bibitem [{\citenamefont {Teng}\ \emph {et~al.}(2023)\citenamefont {Teng},
  \citenamefont {Oh}, \citenamefont {Tan}, \citenamefont {Chen}, \citenamefont
  {Huang}, \citenamefont {Gao}, \citenamefont {Yin}, \citenamefont {Chu},
  \citenamefont {Hashimoto}, \citenamefont {Lu}, \citenamefont {Jozwiak},
  \citenamefont {Bostwick}, \citenamefont {Rotenberg}, \citenamefont
  {Granroth}, \citenamefont {Yan}, \citenamefont {Birgeneau}, \citenamefont
  {Dai},\ and\ \citenamefont {Yi}}]{teng23np}%
  \BibitemOpen
  \bibfield  {author} {\bibinfo {author} {\bibfnamefont {X.}~\bibnamefont
  {Teng}}, \bibinfo {author} {\bibfnamefont {J.~S.}\ \bibnamefont {Oh}},
  \bibinfo {author} {\bibfnamefont {H.}~\bibnamefont {Tan}}, \bibinfo {author}
  {\bibfnamefont {L.}~\bibnamefont {Chen}}, \bibinfo {author} {\bibfnamefont
  {J.}~\bibnamefont {Huang}}, \bibinfo {author} {\bibfnamefont
  {B.}~\bibnamefont {Gao}}, \bibinfo {author} {\bibfnamefont {J.-X.}\
  \bibnamefont {Yin}}, \bibinfo {author} {\bibfnamefont {J.-H.}\ \bibnamefont
  {Chu}}, \bibinfo {author} {\bibfnamefont {M.}~\bibnamefont {Hashimoto}},
  \bibinfo {author} {\bibfnamefont {D.}~\bibnamefont {Lu}}, \bibinfo {author}
  {\bibfnamefont {C.}~\bibnamefont {Jozwiak}}, \bibinfo {author} {\bibfnamefont
  {A.}~\bibnamefont {Bostwick}}, \bibinfo {author} {\bibfnamefont
  {E.}~\bibnamefont {Rotenberg}}, \bibinfo {author} {\bibfnamefont {G.~E.}\
  \bibnamefont {Granroth}}, \bibinfo {author} {\bibfnamefont {B.}~\bibnamefont
  {Yan}}, \bibinfo {author} {\bibfnamefont {R.~J.}\ \bibnamefont {Birgeneau}},
  \bibinfo {author} {\bibfnamefont {P.}~\bibnamefont {Dai}},\ and\ \bibinfo
  {author} {\bibfnamefont {M.}~\bibnamefont {Yi}},\ }\bibfield  {title}
  {\bibinfo {title} {{Magnetism and charge density wave order in kagome
  FeGe}},\ }\href {https://doi.org/10.1038/s41567-023-01985-w} {\bibfield
  {journal} {\bibinfo  {journal} {Nat. Phys.}\ }\textbf {\bibinfo {volume}
  {609}},\ \bibinfo {pages} {490} (\bibinfo {year} {2023})}\BibitemShut
  {NoStop}%
\bibitem [{\citenamefont {Stoner}(1938)}]{stoner38prsa}%
  \BibitemOpen
  \bibfield  {author} {\bibinfo {author} {\bibfnamefont {E.~C.}\ \bibnamefont
  {Stoner}},\ }\bibfield  {title} {\bibinfo {title} {Collective electron
  ferromagnetism},\ }\href {https://doi.org/10.1098/rspa.1938.0066} {\bibfield
  {journal} {\bibinfo  {journal} {Proc. R. Soc. A}\ }\textbf {\bibinfo {volume}
  {165}},\ \bibinfo {pages} {372} (\bibinfo {year} {1938})}\BibitemShut
  {NoStop}%
\bibitem [{\citenamefont {Xie}\ \emph {et~al.}(2021)\citenamefont {Xie},
  \citenamefont {Chen}, \citenamefont {Chen}, \citenamefont {Wang},
  \citenamefont {Yin}, \citenamefont {Stewart}, \citenamefont {Stone},
  \citenamefont {Daemen}, \citenamefont {Feng}, \citenamefont {Cao},
  \citenamefont {Lei}, \citenamefont {Yin}, \citenamefont {MacDonald},\ and\
  \citenamefont {Dai}}]{xie21cp}%
  \BibitemOpen
  \bibfield  {author} {\bibinfo {author} {\bibfnamefont {Y.}~\bibnamefont
  {Xie}}, \bibinfo {author} {\bibfnamefont {L.}~\bibnamefont {Chen}}, \bibinfo
  {author} {\bibfnamefont {T.}~\bibnamefont {Chen}}, \bibinfo {author}
  {\bibfnamefont {Q.}~\bibnamefont {Wang}}, \bibinfo {author} {\bibfnamefont
  {Q.}~\bibnamefont {Yin}}, \bibinfo {author} {\bibfnamefont {J.~R.}\
  \bibnamefont {Stewart}}, \bibinfo {author} {\bibfnamefont {M.~B.}\
  \bibnamefont {Stone}}, \bibinfo {author} {\bibfnamefont {L.~L.}\ \bibnamefont
  {Daemen}}, \bibinfo {author} {\bibfnamefont {E.}~\bibnamefont {Feng}},
  \bibinfo {author} {\bibfnamefont {H.}~\bibnamefont {Cao}}, \bibinfo {author}
  {\bibfnamefont {H.}~\bibnamefont {Lei}}, \bibinfo {author} {\bibfnamefont
  {Z.}~\bibnamefont {Yin}}, \bibinfo {author} {\bibfnamefont {A.~H.}\
  \bibnamefont {MacDonald}},\ and\ \bibinfo {author} {\bibfnamefont
  {P.}~\bibnamefont {Dai}},\ }\bibfield  {title} {\bibinfo {title} {{Spin
  excitations in metallic kagome lattice FeSn and CoSn}},\ }\href
  {https://doi.org/10.1038/s42005-021-00736-8} {\bibfield  {journal} {\bibinfo
  {journal} {Commun. Phys.}\ }\textbf {\bibinfo {volume} {4}},\ \bibinfo
  {pages} {240} (\bibinfo {year} {2021})}\BibitemShut {NoStop}%
\bibitem [{\citenamefont {{Jiang}}\ \emph {et~al.}(2021)\citenamefont
  {{Jiang}}, \citenamefont {{Yin}}, \citenamefont {{Denner}}, \citenamefont
  {{Shumiya}}, \citenamefont {{Ortiz}}, \citenamefont {{Xu}}, \citenamefont
  {{Guguchia}}, \citenamefont {{He}}, \citenamefont {{Hossain}}, \citenamefont
  {{Liu}}, \citenamefont {{Ruff}}, \citenamefont {{Kautzsch}}, \citenamefont
  {{Zhang}}, \citenamefont {{Chang}}, \citenamefont {{Belopolski}},
  \citenamefont {{Zhang}}, \citenamefont {{Cochran}}, \citenamefont {{Multer}},
  \citenamefont {{Litskevich}}, \citenamefont {{Cheng}}, \citenamefont
  {{Yang}}, \citenamefont {{Wang}}, \citenamefont {{Thomale}}, \citenamefont
  {{Neupert}}, \citenamefont {{Wilson}},\ and\ \citenamefont {{Zahid
  Hasan}}}]{jiang21nm}%
  \BibitemOpen
  \bibfield  {author} {\bibinfo {author} {\bibfnamefont {Y.-X.}\ \bibnamefont
  {{Jiang}}}, \bibinfo {author} {\bibfnamefont {J.-X.}\ \bibnamefont {{Yin}}},
  \bibinfo {author} {\bibfnamefont {M.~M.}\ \bibnamefont {{Denner}}}, \bibinfo
  {author} {\bibfnamefont {N.}~\bibnamefont {{Shumiya}}}, \bibinfo {author}
  {\bibfnamefont {B.~R.}\ \bibnamefont {{Ortiz}}}, \bibinfo {author}
  {\bibfnamefont {G.}~\bibnamefont {{Xu}}}, \bibinfo {author} {\bibfnamefont
  {Z.}~\bibnamefont {{Guguchia}}}, \bibinfo {author} {\bibfnamefont
  {J.}~\bibnamefont {{He}}}, \bibinfo {author} {\bibfnamefont {M.~S.}\
  \bibnamefont {{Hossain}}}, \bibinfo {author} {\bibfnamefont {X.}~\bibnamefont
  {{Liu}}}, \bibinfo {author} {\bibfnamefont {J.}~\bibnamefont {{Ruff}}},
  \bibinfo {author} {\bibfnamefont {L.}~\bibnamefont {{Kautzsch}}}, \bibinfo
  {author} {\bibfnamefont {S.~S.}\ \bibnamefont {{Zhang}}}, \bibinfo {author}
  {\bibfnamefont {G.}~\bibnamefont {{Chang}}}, \bibinfo {author} {\bibfnamefont
  {I.}~\bibnamefont {{Belopolski}}}, \bibinfo {author} {\bibfnamefont
  {Q.}~\bibnamefont {{Zhang}}}, \bibinfo {author} {\bibfnamefont {T.~A.}\
  \bibnamefont {{Cochran}}}, \bibinfo {author} {\bibfnamefont {D.}~\bibnamefont
  {{Multer}}}, \bibinfo {author} {\bibfnamefont {M.}~\bibnamefont
  {{Litskevich}}}, \bibinfo {author} {\bibfnamefont {Z.-J.}\ \bibnamefont
  {{Cheng}}}, \bibinfo {author} {\bibfnamefont {X.~P.}\ \bibnamefont {{Yang}}},
  \bibinfo {author} {\bibfnamefont {Z.}~\bibnamefont {{Wang}}}, \bibinfo
  {author} {\bibfnamefont {R.}~\bibnamefont {{Thomale}}}, \bibinfo {author}
  {\bibfnamefont {T.}~\bibnamefont {{Neupert}}}, \bibinfo {author}
  {\bibfnamefont {S.~D.}\ \bibnamefont {{Wilson}}},\ and\ \bibinfo {author}
  {\bibfnamefont {M.}~\bibnamefont {{Zahid Hasan}}},\ }\bibfield  {title}
  {\bibinfo {title} {{Unconventional chiral charge order in kagome
  superconductor KV$_3$Sb$_5$}},\ }\href
  {https://doi.org/10.1038/s41563-021-01034-y} {\bibfield  {journal} {\bibinfo
  {journal} {Nat. Mater.}\ }\textbf {\bibinfo {volume} {20}},\ \bibinfo {pages}
  {1353} (\bibinfo {year} {2021})}\BibitemShut {NoStop}%
\bibitem [{\citenamefont {{Zhao}}\ \emph {et~al.}(2021)\citenamefont {{Zhao}},
  \citenamefont {{Li}}, \citenamefont {{Ortiz}}, \citenamefont {{Teicher}},
  \citenamefont {{Park}}, \citenamefont {{Ye}}, \citenamefont {{Wang}},
  \citenamefont {{Balents}}, \citenamefont {{Wilson}},\ and\ \citenamefont
  {{Zeljkovic}}}]{zhao21n}%
  \BibitemOpen
  \bibfield  {author} {\bibinfo {author} {\bibfnamefont {H.}~\bibnamefont
  {{Zhao}}}, \bibinfo {author} {\bibfnamefont {H.}~\bibnamefont {{Li}}},
  \bibinfo {author} {\bibfnamefont {B.~R.}\ \bibnamefont {{Ortiz}}}, \bibinfo
  {author} {\bibfnamefont {S.~M.~L.}\ \bibnamefont {{Teicher}}}, \bibinfo
  {author} {\bibfnamefont {T.}~\bibnamefont {{Park}}}, \bibinfo {author}
  {\bibfnamefont {M.}~\bibnamefont {{Ye}}}, \bibinfo {author} {\bibfnamefont
  {Z.}~\bibnamefont {{Wang}}}, \bibinfo {author} {\bibfnamefont
  {L.}~\bibnamefont {{Balents}}}, \bibinfo {author} {\bibfnamefont {S.~D.}\
  \bibnamefont {{Wilson}}},\ and\ \bibinfo {author} {\bibfnamefont
  {I.}~\bibnamefont {{Zeljkovic}}},\ }\bibfield  {title} {\bibinfo {title}
  {{Cascade of correlated electron states in a kagome superconductor
  CsV$_3$Sb$_5$}},\ }\href {https://doi.org/10.1038/s41586-021-03946-w}
  {\bibfield  {journal} {\bibinfo  {journal} {Nature}\ }\textbf {\bibinfo
  {volume} {599}},\ \bibinfo {pages} {216} (\bibinfo {year}
  {2021})}\BibitemShut {NoStop}%
\bibitem [{\citenamefont {Mielke}\ and\ \citenamefont
  {Tasaki}(1993)}]{mielke93cmp}%
  \BibitemOpen
  \bibfield  {author} {\bibinfo {author} {\bibfnamefont {A.}~\bibnamefont
  {Mielke}}\ and\ \bibinfo {author} {\bibfnamefont {H.}~\bibnamefont
  {Tasaki}},\ }\bibfield  {title} {\bibinfo {title} {Ferromagnetism in the
  hubbard model},\ }\href {https://doi.org/10.1007/BF02108079} {\bibfield
  {journal} {\bibinfo  {journal} {Commun. Math. Phys.}\ }\textbf {\bibinfo
  {volume} {158}},\ \bibinfo {pages} {341} (\bibinfo {year}
  {1993})}\BibitemShut {NoStop}%
\bibitem [{\citenamefont {Hanisch}\ \emph {et~al.}(1995)\citenamefont
  {Hanisch}, \citenamefont {Kleine}, \citenamefont {Ritzl},\ and\ \citenamefont
  {Müller-Hartmann}}]{hanisch95annp}%
  \BibitemOpen
  \bibfield  {author} {\bibinfo {author} {\bibfnamefont {T.}~\bibnamefont
  {Hanisch}}, \bibinfo {author} {\bibfnamefont {B.}~\bibnamefont {Kleine}},
  \bibinfo {author} {\bibfnamefont {A.}~\bibnamefont {Ritzl}},\ and\ \bibinfo
  {author} {\bibfnamefont {E.}~\bibnamefont {Müller-Hartmann}},\ }\bibfield
  {title} {\bibinfo {title} {Ferromagnetism in the hubbard model: instability
  of the nagaoka state on the triangular, honeycomb and kagome lattices},\
  }\href {https://doi.org/https://doi.org/10.1002/andp.19955070405} {\bibfield
  {journal} {\bibinfo  {journal} {Ann. Phys.}\ }\textbf {\bibinfo {volume}
  {507}},\ \bibinfo {pages} {303} (\bibinfo {year} {1995})}\BibitemShut
  {NoStop}%
\bibitem [{\citenamefont {Pollmann}\ \emph {et~al.}(2008)\citenamefont
  {Pollmann}, \citenamefont {Fulde},\ and\ \citenamefont
  {Shtengel}}]{pollmann08prl}%
  \BibitemOpen
  \bibfield  {author} {\bibinfo {author} {\bibfnamefont {F.}~\bibnamefont
  {Pollmann}}, \bibinfo {author} {\bibfnamefont {P.}~\bibnamefont {Fulde}},\
  and\ \bibinfo {author} {\bibfnamefont {K.}~\bibnamefont {Shtengel}},\
  }\bibfield  {title} {\bibinfo {title} {Kinetic ferromagnetism on a kagome
  lattice},\ }\href {https://doi.org/10.1103/PhysRevLett.100.136404} {\bibfield
   {journal} {\bibinfo  {journal} {Phys. Rev. Lett.}\ }\textbf {\bibinfo
  {volume} {100}},\ \bibinfo {pages} {136404} (\bibinfo {year}
  {2008})}\BibitemShut {NoStop}%
\bibitem [{\citenamefont {Ren}\ \emph {et~al.}(2021)\citenamefont {Ren},
  \citenamefont {Jiang}, \citenamefont {Qiao},\ and\ \citenamefont
  {Sheng}}]{ren21prl}%
  \BibitemOpen
  \bibfield  {author} {\bibinfo {author} {\bibfnamefont {Y.}~\bibnamefont
  {Ren}}, \bibinfo {author} {\bibfnamefont {H.-C.}\ \bibnamefont {Jiang}},
  \bibinfo {author} {\bibfnamefont {Z.}~\bibnamefont {Qiao}},\ and\ \bibinfo
  {author} {\bibfnamefont {D.~N.}\ \bibnamefont {Sheng}},\ }\bibfield  {title}
  {\bibinfo {title} {Orbital chern insulator and quantum phase diagram of a
  kagome electron system with half-filled flat bands},\ }\href
  {https://doi.org/10.1103/PhysRevLett.126.117602} {\bibfield  {journal}
  {\bibinfo  {journal} {Phys. Rev. Lett.}\ }\textbf {\bibinfo {volume} {126}},\
  \bibinfo {pages} {117602} (\bibinfo {year} {2021})}\BibitemShut {NoStop}%
\bibitem [{\citenamefont {Nishimoto}\ \emph {et~al.}(2010)\citenamefont
  {Nishimoto}, \citenamefont {Nakamura}, \citenamefont {O'Brien},\ and\
  \citenamefont {Fulde}}]{nishimoto10prl}%
  \BibitemOpen
  \bibfield  {author} {\bibinfo {author} {\bibfnamefont {S.}~\bibnamefont
  {Nishimoto}}, \bibinfo {author} {\bibfnamefont {M.}~\bibnamefont {Nakamura}},
  \bibinfo {author} {\bibfnamefont {A.}~\bibnamefont {O'Brien}},\ and\ \bibinfo
  {author} {\bibfnamefont {P.}~\bibnamefont {Fulde}},\ }\bibfield  {title}
  {\bibinfo {title} {Metal-insulator transition of fermions on a kagome lattice
  at 1/3 filling},\ }\href {https://doi.org/10.1103/PhysRevLett.104.196401}
  {\bibfield  {journal} {\bibinfo  {journal} {Phys. Rev. Lett.}\ }\textbf
  {\bibinfo {volume} {104}},\ \bibinfo {pages} {196401} (\bibinfo {year}
  {2010})}\BibitemShut {NoStop}%
\bibitem [{\citenamefont {O'Brien}\ \emph {et~al.}(2010)\citenamefont
  {O'Brien}, \citenamefont {Pollmann},\ and\ \citenamefont
  {Fulde}}]{obrien10prb}%
  \BibitemOpen
  \bibfield  {author} {\bibinfo {author} {\bibfnamefont {A.}~\bibnamefont
  {O'Brien}}, \bibinfo {author} {\bibfnamefont {F.}~\bibnamefont {Pollmann}},\
  and\ \bibinfo {author} {\bibfnamefont {P.}~\bibnamefont {Fulde}},\ }\bibfield
   {title} {\bibinfo {title} {Strongly correlated fermions on a kagome
  lattice},\ }\href {https://doi.org/10.1103/PhysRevB.81.235115} {\bibfield
  {journal} {\bibinfo  {journal} {Phys. Rev. B}\ }\textbf {\bibinfo {volume}
  {81}},\ \bibinfo {pages} {235115} (\bibinfo {year} {2010})}\BibitemShut
  {NoStop}%
\bibitem [{\citenamefont {Wen}\ \emph {et~al.}(2010)\citenamefont {Wen},
  \citenamefont {R\"uegg}, \citenamefont {Wang},\ and\ \citenamefont
  {Fiete}}]{wen10prb}%
  \BibitemOpen
  \bibfield  {author} {\bibinfo {author} {\bibfnamefont {J.}~\bibnamefont
  {Wen}}, \bibinfo {author} {\bibfnamefont {A.}~\bibnamefont {R\"uegg}},
  \bibinfo {author} {\bibfnamefont {C.-C.~J.}\ \bibnamefont {Wang}},\ and\
  \bibinfo {author} {\bibfnamefont {G.~A.}\ \bibnamefont {Fiete}},\ }\bibfield
  {title} {\bibinfo {title} {Interaction-driven topological insulators on the
  kagome and the decorated honeycomb lattices},\ }\href
  {https://doi.org/10.1103/PhysRevB.82.075125} {\bibfield  {journal} {\bibinfo
  {journal} {Phys. Rev. B}\ }\textbf {\bibinfo {volume} {82}},\ \bibinfo
  {pages} {075125} (\bibinfo {year} {2010})}\BibitemShut {NoStop}%
\bibitem [{\citenamefont {Zhu}\ \emph {et~al.}(2016)\citenamefont {Zhu},
  \citenamefont {Gong}, \citenamefont {Zeng}, \citenamefont {Fu},\ and\
  \citenamefont {Sheng}}]{zhu16prl}%
  \BibitemOpen
  \bibfield  {author} {\bibinfo {author} {\bibfnamefont {W.}~\bibnamefont
  {Zhu}}, \bibinfo {author} {\bibfnamefont {S.-S.}\ \bibnamefont {Gong}},
  \bibinfo {author} {\bibfnamefont {T.-S.}\ \bibnamefont {Zeng}}, \bibinfo
  {author} {\bibfnamefont {L.}~\bibnamefont {Fu}},\ and\ \bibinfo {author}
  {\bibfnamefont {D.~N.}\ \bibnamefont {Sheng}},\ }\bibfield  {title} {\bibinfo
  {title} {Interaction-driven spontaneous quantum hall effect on a kagome
  lattice},\ }\href {https://doi.org/10.1103/PhysRevLett.117.096402} {\bibfield
   {journal} {\bibinfo  {journal} {Phys. Rev. Lett.}\ }\textbf {\bibinfo
  {volume} {117}},\ \bibinfo {pages} {096402} (\bibinfo {year}
  {2016})}\BibitemShut {NoStop}%
\bibitem [{\citenamefont {Ren}\ \emph {et~al.}(2018)\citenamefont {Ren},
  \citenamefont {Zeng}, \citenamefont {Zhu},\ and\ \citenamefont
  {Sheng}}]{ren18prb}%
  \BibitemOpen
  \bibfield  {author} {\bibinfo {author} {\bibfnamefont {Y.}~\bibnamefont
  {Ren}}, \bibinfo {author} {\bibfnamefont {T.-S.}\ \bibnamefont {Zeng}},
  \bibinfo {author} {\bibfnamefont {W.}~\bibnamefont {Zhu}},\ and\ \bibinfo
  {author} {\bibfnamefont {D.~N.}\ \bibnamefont {Sheng}},\ }\bibfield  {title}
  {\bibinfo {title} {Quantum anomalous hall phase stabilized via realistic
  interactions on a kagome lattice},\ }\href
  {https://doi.org/10.1103/PhysRevB.98.205146} {\bibfield  {journal} {\bibinfo
  {journal} {Phys. Rev. B}\ }\textbf {\bibinfo {volume} {98}},\ \bibinfo
  {pages} {205146} (\bibinfo {year} {2018})}\BibitemShut {NoStop}%
\bibitem [{\citenamefont {Liu}\ \emph {et~al.}(2010)\citenamefont {Liu},
  \citenamefont {Yao},\ and\ \citenamefont {Ma}}]{liu10prb}%
  \BibitemOpen
  \bibfield  {author} {\bibinfo {author} {\bibfnamefont {Q.}~\bibnamefont
  {Liu}}, \bibinfo {author} {\bibfnamefont {H.}~\bibnamefont {Yao}},\ and\
  \bibinfo {author} {\bibfnamefont {T.}~\bibnamefont {Ma}},\ }\bibfield
  {title} {\bibinfo {title} {Spontaneous symmetry breaking in a two-dimensional
  kagome lattice},\ }\href {https://doi.org/10.1103/PhysRevB.82.045102}
  {\bibfield  {journal} {\bibinfo  {journal} {Phys. Rev. B}\ }\textbf {\bibinfo
  {volume} {82}},\ \bibinfo {pages} {045102} (\bibinfo {year}
  {2010})}\BibitemShut {NoStop}%
\bibitem [{\citenamefont {Pollmann}\ \emph {et~al.}(2014)\citenamefont
  {Pollmann}, \citenamefont {Roychowdhury}, \citenamefont {Hotta},\ and\
  \citenamefont {Penc}}]{pollmann14prb}%
  \BibitemOpen
  \bibfield  {author} {\bibinfo {author} {\bibfnamefont {F.}~\bibnamefont
  {Pollmann}}, \bibinfo {author} {\bibfnamefont {K.}~\bibnamefont
  {Roychowdhury}}, \bibinfo {author} {\bibfnamefont {C.}~\bibnamefont
  {Hotta}},\ and\ \bibinfo {author} {\bibfnamefont {K.}~\bibnamefont {Penc}},\
  }\bibfield  {title} {\bibinfo {title} {Interplay of charge and spin
  fluctuations of strongly interacting electrons on the kagome lattice},\
  }\href {https://doi.org/10.1103/PhysRevB.90.035118} {\bibfield  {journal}
  {\bibinfo  {journal} {Phys. Rev. B}\ }\textbf {\bibinfo {volume} {90}},\
  \bibinfo {pages} {035118} (\bibinfo {year} {2014})}\BibitemShut {NoStop}%
\bibitem [{\citenamefont {Okamoto}\ \emph {et~al.}(2022)\citenamefont
  {Okamoto}, \citenamefont {Mohanta}, \citenamefont {Dagotto},\ and\
  \citenamefont {Sheng}}]{okamoto22cp}%
  \BibitemOpen
  \bibfield  {author} {\bibinfo {author} {\bibfnamefont {S.}~\bibnamefont
  {Okamoto}}, \bibinfo {author} {\bibfnamefont {N.}~\bibnamefont {Mohanta}},
  \bibinfo {author} {\bibfnamefont {E.}~\bibnamefont {Dagotto}},\ and\ \bibinfo
  {author} {\bibfnamefont {D.~N.}\ \bibnamefont {Sheng}},\ }\bibfield  {title}
  {\bibinfo {title} {Topological flat bands in a kagome lattice multiorbital
  system},\ }\href {https://doi.org/10.1038/s42005-022-00969-1} {\bibfield
  {journal} {\bibinfo  {journal} {Commun. Phys.}\ }\textbf {\bibinfo {volume}
  {5}},\ \bibinfo {pages} {198} (\bibinfo {year} {2022})}\BibitemShut {NoStop}%
\bibitem [{\citenamefont {{Setty}}\ \emph {et~al.}(2022)\citenamefont
  {{Setty}}, \citenamefont {{Lane}}, \citenamefont {{Chen}}, \citenamefont
  {{Hu}}, \citenamefont {{Zhu}},\ and\ \citenamefont {{Si}}}]{setty22ax}%
  \BibitemOpen
  \bibfield  {author} {\bibinfo {author} {\bibfnamefont {C.}~\bibnamefont
  {{Setty}}}, \bibinfo {author} {\bibfnamefont {C.~A.}\ \bibnamefont {{Lane}}},
  \bibinfo {author} {\bibfnamefont {L.}~\bibnamefont {{Chen}}}, \bibinfo
  {author} {\bibfnamefont {H.}~\bibnamefont {{Hu}}}, \bibinfo {author}
  {\bibfnamefont {J.-X.}\ \bibnamefont {{Zhu}}},\ and\ \bibinfo {author}
  {\bibfnamefont {Q.}~\bibnamefont {{Si}}},\ }\bibfield  {title} {\bibinfo
  {title} {{Electron correlations and charge density wave in the topological
  kagome metal FeGe}},\ }\href {https://doi.org/10.48550/arXiv.2203.01930}
  {\bibfield  {journal} {\bibinfo  {journal} {arXiv e-prints}\ ,\ \bibinfo
  {eid} {arXiv:2203.01930}} (\bibinfo {year} {2022})},\ \Eprint
  {https://arxiv.org/abs/2203.01930} {arXiv:2203.01930 [cond-mat.str-el]}
  \BibitemShut {NoStop}%
\bibitem [{\citenamefont {Arovas}\ \emph {et~al.}(2022)\citenamefont {Arovas},
  \citenamefont {Berg}, \citenamefont {Kivelson},\ and\ \citenamefont
  {Raghu}}]{arovas22arcmp}%
  \BibitemOpen
  \bibfield  {author} {\bibinfo {author} {\bibfnamefont {D.~P.}\ \bibnamefont
  {Arovas}}, \bibinfo {author} {\bibfnamefont {E.}~\bibnamefont {Berg}},
  \bibinfo {author} {\bibfnamefont {S.~A.}\ \bibnamefont {Kivelson}},\ and\
  \bibinfo {author} {\bibfnamefont {S.}~\bibnamefont {Raghu}},\ }\bibfield
  {title} {\bibinfo {title} {The hubbard model},\ }\href
  {https://doi.org/10.1146/annurev-conmatphys-031620-102024} {\bibfield
  {journal} {\bibinfo  {journal} {Annu. Rev. Condens. Matter Phys.}\ }\textbf
  {\bibinfo {volume} {13}},\ \bibinfo {pages} {239} (\bibinfo {year}
  {2022})}\BibitemShut {NoStop}%
\bibitem [{\citenamefont {Qin}\ \emph {et~al.}(2022)\citenamefont {Qin},
  \citenamefont {Sch\"{a}fer}, \citenamefont {Andergassen}, \citenamefont
  {Corboz},\ and\ \citenamefont {Gull}}]{qin22arcmp}%
  \BibitemOpen
  \bibfield  {author} {\bibinfo {author} {\bibfnamefont {M.}~\bibnamefont
  {Qin}}, \bibinfo {author} {\bibfnamefont {T.}~\bibnamefont {Sch\"{a}fer}},
  \bibinfo {author} {\bibfnamefont {S.}~\bibnamefont {Andergassen}}, \bibinfo
  {author} {\bibfnamefont {P.}~\bibnamefont {Corboz}},\ and\ \bibinfo {author}
  {\bibfnamefont {E.}~\bibnamefont {Gull}},\ }\bibfield  {title} {\bibinfo
  {title} {The hubbard model: A computational perspective},\ }\href
  {https://doi.org/10.1146/annurev-conmatphys-090921-033948} {\bibfield
  {journal} {\bibinfo  {journal} {Annu. Rev. Condens. Matter Phys.}\ }\textbf
  {\bibinfo {volume} {13}},\ \bibinfo {pages} {275} (\bibinfo {year}
  {2022})}\BibitemShut {NoStop}%
\bibitem [{sup()}]{supp}%
  \BibitemOpen
  \bibinfo {note} {{See Supplemental Material at [URL] for the details of
  Hartree-Fock analysis, miscellaneous ground states in pure Hubbard model,
  effect of extended repulsion, mean-field renormalized band structure and band
  topology, and mean-field analysis of magnetic orders with PSG, which includes
  Refs.~\cite{kudin02jcp,fukui05jpsp,sun09prl,szasz21prb,cookmeyer21prl,disante23prr,rossi23prb}}}\BibitemShut
  {NoStop}%
\bibitem [{\citenamefont {Huse}\ and\ \citenamefont {Elser}(1988)}]{huse88prl}%
  \BibitemOpen
  \bibfield  {author} {\bibinfo {author} {\bibfnamefont {D.~A.}\ \bibnamefont
  {Huse}}\ and\ \bibinfo {author} {\bibfnamefont {V.}~\bibnamefont {Elser}},\
  }\bibfield  {title} {\bibinfo {title} {Simple variational wave functions for
  two-dimensional heisenberg spin-\textonehalf{} antiferromagnets},\ }\href
  {https://doi.org/10.1103/PhysRevLett.60.2531} {\bibfield  {journal} {\bibinfo
   {journal} {Phys. Rev. Lett.}\ }\textbf {\bibinfo {volume} {60}},\ \bibinfo
  {pages} {2531} (\bibinfo {year} {1988})}\BibitemShut {NoStop}%
\bibitem [{\citenamefont {Domenge}\ \emph {et~al.}(2005)\citenamefont
  {Domenge}, \citenamefont {Sindzingre}, \citenamefont {Lhuillier},\ and\
  \citenamefont {Pierre}}]{domenge05prb}%
  \BibitemOpen
  \bibfield  {author} {\bibinfo {author} {\bibfnamefont {J.-C.}\ \bibnamefont
  {Domenge}}, \bibinfo {author} {\bibfnamefont {P.}~\bibnamefont {Sindzingre}},
  \bibinfo {author} {\bibfnamefont {C.}~\bibnamefont {Lhuillier}},\ and\
  \bibinfo {author} {\bibfnamefont {L.}~\bibnamefont {Pierre}},\ }\bibfield
  {title} {\bibinfo {title} {Twelve sublattice ordered phase in the
  ${J}_{1}\ensuremath{-}{J}_{2}$ model on the kagom\'e lattice},\ }\href
  {https://doi.org/10.1103/PhysRevB.72.024433} {\bibfield  {journal} {\bibinfo
  {journal} {Phys. Rev. B}\ }\textbf {\bibinfo {volume} {72}},\ \bibinfo
  {pages} {024433} (\bibinfo {year} {2005})}\BibitemShut {NoStop}%
\bibitem [{\citenamefont {Domenge}\ \emph {et~al.}(2008)\citenamefont
  {Domenge}, \citenamefont {Lhuillier}, \citenamefont {Messio}, \citenamefont
  {Pierre},\ and\ \citenamefont {Viot}}]{domenge08prb}%
  \BibitemOpen
  \bibfield  {author} {\bibinfo {author} {\bibfnamefont {J.-C.}\ \bibnamefont
  {Domenge}}, \bibinfo {author} {\bibfnamefont {C.}~\bibnamefont {Lhuillier}},
  \bibinfo {author} {\bibfnamefont {L.}~\bibnamefont {Messio}}, \bibinfo
  {author} {\bibfnamefont {L.}~\bibnamefont {Pierre}},\ and\ \bibinfo {author}
  {\bibfnamefont {P.}~\bibnamefont {Viot}},\ }\bibfield  {title} {\bibinfo
  {title} {Chirality and ${\mathbb{z}}_{2}$ vortices in a heisenberg spin model
  on the kagome lattice},\ }\href {https://doi.org/10.1103/PhysRevB.77.172413}
  {\bibfield  {journal} {\bibinfo  {journal} {Phys. Rev. B}\ }\textbf {\bibinfo
  {volume} {77}},\ \bibinfo {pages} {172413} (\bibinfo {year}
  {2008})}\BibitemShut {NoStop}%
\bibitem [{\citenamefont {Messio}\ \emph {et~al.}(2011)\citenamefont {Messio},
  \citenamefont {Lhuillier},\ and\ \citenamefont {Misguich}}]{messio11prb}%
  \BibitemOpen
  \bibfield  {author} {\bibinfo {author} {\bibfnamefont {L.}~\bibnamefont
  {Messio}}, \bibinfo {author} {\bibfnamefont {C.}~\bibnamefont {Lhuillier}},\
  and\ \bibinfo {author} {\bibfnamefont {G.}~\bibnamefont {Misguich}},\
  }\bibfield  {title} {\bibinfo {title} {Lattice symmetries and regular
  magnetic orders in classical frustrated antiferromagnets},\ }\href
  {https://doi.org/10.1103/PhysRevB.83.184401} {\bibfield  {journal} {\bibinfo
  {journal} {Phys. Rev. B}\ }\textbf {\bibinfo {volume} {83}},\ \bibinfo
  {pages} {184401} (\bibinfo {year} {2011})}\BibitemShut {NoStop}%
\bibitem [{\citenamefont {Ohgushi}\ \emph {et~al.}(2000)\citenamefont
  {Ohgushi}, \citenamefont {Murakami},\ and\ \citenamefont
  {Nagaosa}}]{ohgushi00prb}%
  \BibitemOpen
  \bibfield  {author} {\bibinfo {author} {\bibfnamefont {K.}~\bibnamefont
  {Ohgushi}}, \bibinfo {author} {\bibfnamefont {S.}~\bibnamefont {Murakami}},\
  and\ \bibinfo {author} {\bibfnamefont {N.}~\bibnamefont {Nagaosa}},\
  }\bibfield  {title} {\bibinfo {title} {Spin anisotropy and quantum hall
  effect in the kagom\'e lattice: Chiral spin state based on a ferromagnet},\
  }\href {https://doi.org/10.1103/PhysRevB.62.R6065} {\bibfield  {journal}
  {\bibinfo  {journal} {Phys. Rev. B}\ }\textbf {\bibinfo {volume} {62}},\
  \bibinfo {pages} {R6065} (\bibinfo {year} {2000})}\BibitemShut {NoStop}%
\bibitem [{\citenamefont {Taguchi}\ \emph {et~al.}(2001)\citenamefont
  {Taguchi}, \citenamefont {Oohara}, \citenamefont {Yoshizawa}, \citenamefont
  {Nagaosa},\ and\ \citenamefont {Tokura}}]{taguchi01sc}%
  \BibitemOpen
  \bibfield  {author} {\bibinfo {author} {\bibfnamefont {Y.}~\bibnamefont
  {Taguchi}}, \bibinfo {author} {\bibfnamefont {Y.}~\bibnamefont {Oohara}},
  \bibinfo {author} {\bibfnamefont {H.}~\bibnamefont {Yoshizawa}}, \bibinfo
  {author} {\bibfnamefont {N.}~\bibnamefont {Nagaosa}},\ and\ \bibinfo {author}
  {\bibfnamefont {Y.}~\bibnamefont {Tokura}},\ }\bibfield  {title} {\bibinfo
  {title} {Spin chirality, berry phase, and anomalous hall effect in a
  frustrated ferromagnet},\ }\href {https://doi.org/10.1126/science.1058161}
  {\bibfield  {journal} {\bibinfo  {journal} {Science}\ }\textbf {\bibinfo
  {volume} {291}},\ \bibinfo {pages} {2573} (\bibinfo {year}
  {2001})}\BibitemShut {NoStop}%
\bibitem [{\citenamefont {Martin}\ and\ \citenamefont
  {Batista}(2008)}]{martin08prl}%
  \BibitemOpen
  \bibfield  {author} {\bibinfo {author} {\bibfnamefont {I.}~\bibnamefont
  {Martin}}\ and\ \bibinfo {author} {\bibfnamefont {C.~D.}\ \bibnamefont
  {Batista}},\ }\bibfield  {title} {\bibinfo {title} {Itinerant electron-driven
  chiral magnetic ordering and spontaneous quantum hall effect in triangular
  lattice models},\ }\href {https://doi.org/10.1103/PhysRevLett.101.156402}
  {\bibfield  {journal} {\bibinfo  {journal} {Phys. Rev. Lett.}\ }\textbf
  {\bibinfo {volume} {101}},\ \bibinfo {pages} {156402} (\bibinfo {year}
  {2008})}\BibitemShut {NoStop}%
\bibitem [{\citenamefont {Barros}\ \emph {et~al.}(2014)\citenamefont {Barros},
  \citenamefont {Venderbos}, \citenamefont {Chern},\ and\ \citenamefont
  {Batista}}]{barros14prb}%
  \BibitemOpen
  \bibfield  {author} {\bibinfo {author} {\bibfnamefont {K.}~\bibnamefont
  {Barros}}, \bibinfo {author} {\bibfnamefont {J.~W.~F.}\ \bibnamefont
  {Venderbos}}, \bibinfo {author} {\bibfnamefont {G.-W.}\ \bibnamefont
  {Chern}},\ and\ \bibinfo {author} {\bibfnamefont {C.~D.}\ \bibnamefont
  {Batista}},\ }\bibfield  {title} {\bibinfo {title} {Exotic magnetic orderings
  in the kagome kondo-lattice model},\ }\href
  {https://doi.org/10.1103/PhysRevB.90.245119} {\bibfield  {journal} {\bibinfo
  {journal} {Phys. Rev. B}\ }\textbf {\bibinfo {volume} {90}},\ \bibinfo
  {pages} {245119} (\bibinfo {year} {2014})}\BibitemShut {NoStop}%
\bibitem [{\citenamefont {Emery}\ \emph {et~al.}(1990)\citenamefont {Emery},
  \citenamefont {Kivelson},\ and\ \citenamefont {Lin}}]{emery90prl}%
  \BibitemOpen
  \bibfield  {author} {\bibinfo {author} {\bibfnamefont {V.~J.}\ \bibnamefont
  {Emery}}, \bibinfo {author} {\bibfnamefont {S.~A.}\ \bibnamefont
  {Kivelson}},\ and\ \bibinfo {author} {\bibfnamefont {H.~Q.}\ \bibnamefont
  {Lin}},\ }\bibfield  {title} {\bibinfo {title} {Phase separation in the t-j
  model},\ }\href {https://doi.org/10.1103/PhysRevLett.64.475} {\bibfield
  {journal} {\bibinfo  {journal} {Phys. Rev. Lett.}\ }\textbf {\bibinfo
  {volume} {64}},\ \bibinfo {pages} {475} (\bibinfo {year} {1990})}\BibitemShut
  {NoStop}%
\bibitem [{\citenamefont {Qin}(2022)}]{qin22jpcm}%
  \BibitemOpen
  \bibfield  {author} {\bibinfo {author} {\bibfnamefont {M.}~\bibnamefont
  {Qin}},\ }\bibfield  {title} {\bibinfo {title} {Effect of hole doping on the
  120 degree order in the triangular lattice hubbard model: a hartree–fock
  revisit},\ }\href {https://doi.org/10.1088/1361-648X/ac5e79} {\bibfield
  {journal} {\bibinfo  {journal} {J. Phys.: Condens. Matter}\ }\textbf
  {\bibinfo {volume} {34}},\ \bibinfo {pages} {235603} (\bibinfo {year}
  {2022})}\BibitemShut {NoStop}%
\bibitem [{exp()}]{explanation}%
  \BibitemOpen
  \bibinfo {note} {{Here we are assuming that the (symmetry-breaking)
  commensurate orders are weak enough, and that the (possibly folded) bands
  develop a charge gap due to strong interactions.}}\BibitemShut {Stop}%
\bibitem [{\citenamefont {Balents}\ \emph {et~al.}(2002)\citenamefont
  {Balents}, \citenamefont {Fisher},\ and\ \citenamefont
  {Girvin}}]{PhysRevB.65.224412}%
  \BibitemOpen
  \bibfield  {author} {\bibinfo {author} {\bibfnamefont {L.}~\bibnamefont
  {Balents}}, \bibinfo {author} {\bibfnamefont {M.~P.~A.}\ \bibnamefont
  {Fisher}},\ and\ \bibinfo {author} {\bibfnamefont {S.~M.}\ \bibnamefont
  {Girvin}},\ }\bibfield  {title} {\bibinfo {title} {Fractionalization in an
  easy-axis kagome antiferromagnet},\ }\href
  {https://doi.org/10.1103/PhysRevB.65.224412} {\bibfield  {journal} {\bibinfo
  {journal} {Phys. Rev. B}\ }\textbf {\bibinfo {volume} {65}},\ \bibinfo
  {pages} {224412} (\bibinfo {year} {2002})}\BibitemShut {NoStop}%
\bibitem [{\citenamefont {Yan}\ \emph {et~al.}(2011)\citenamefont {Yan},
  \citenamefont {Huse},\ and\ \citenamefont {White}}]{yan2011spin}%
  \BibitemOpen
  \bibfield  {author} {\bibinfo {author} {\bibfnamefont {S.}~\bibnamefont
  {Yan}}, \bibinfo {author} {\bibfnamefont {D.~A.}\ \bibnamefont {Huse}},\ and\
  \bibinfo {author} {\bibfnamefont {S.~R.}\ \bibnamefont {White}},\ }\bibfield
  {title} {\bibinfo {title} {Spin-liquid ground state of the $s = 1/2$ kagome
  heisenberg antiferromagnet},\ }\href
  {https://doi.org/10.1126/science.1201080} {\bibfield  {journal} {\bibinfo
  {journal} {Science}\ }\textbf {\bibinfo {volume} {332}},\ \bibinfo {pages}
  {1173} (\bibinfo {year} {2011})}\BibitemShut {NoStop}%
\bibitem [{\citenamefont {Nishimoto}\ \emph {et~al.}(2013)\citenamefont
  {Nishimoto}, \citenamefont {Shibata},\ and\ \citenamefont
  {Hotta}}]{nishimoto2013controlling}%
  \BibitemOpen
  \bibfield  {author} {\bibinfo {author} {\bibfnamefont {S.}~\bibnamefont
  {Nishimoto}}, \bibinfo {author} {\bibfnamefont {N.}~\bibnamefont {Shibata}},\
  and\ \bibinfo {author} {\bibfnamefont {C.}~\bibnamefont {Hotta}},\ }\bibfield
   {title} {\bibinfo {title} {Controlling frustrated liquids and solids with an
  applied field in a kagome heisenberg antiferromagnet},\ }\href
  {https://doi.org/10.1038/ncomms3287} {\bibfield  {journal} {\bibinfo
  {journal} {Nat. Commun.}\ }\textbf {\bibinfo {volume} {4}},\ \bibinfo {pages}
  {2287} (\bibinfo {year} {2013})}\BibitemShut {NoStop}%
\bibitem [{\citenamefont {Jiang}\ \emph {et~al.}(2012)\citenamefont {Jiang},
  \citenamefont {Wang},\ and\ \citenamefont {Balents}}]{jiang2012identifying}%
  \BibitemOpen
  \bibfield  {author} {\bibinfo {author} {\bibfnamefont {H.-C.}\ \bibnamefont
  {Jiang}}, \bibinfo {author} {\bibfnamefont {Z.}~\bibnamefont {Wang}},\ and\
  \bibinfo {author} {\bibfnamefont {L.}~\bibnamefont {Balents}},\ }\bibfield
  {title} {\bibinfo {title} {Identifying topological order by entanglement
  entropy},\ }\href {https://doi.org/10.1038/nphys2465} {\bibfield  {journal}
  {\bibinfo  {journal} {Nat. Phys.}\ }\textbf {\bibinfo {volume} {8}},\
  \bibinfo {pages} {902} (\bibinfo {year} {2012})}\BibitemShut {NoStop}%
\bibitem [{\citenamefont {Messio}\ \emph {et~al.}(2012)\citenamefont {Messio},
  \citenamefont {Bernu},\ and\ \citenamefont {Lhuillier}}]{messio12prl}%
  \BibitemOpen
  \bibfield  {author} {\bibinfo {author} {\bibfnamefont {L.}~\bibnamefont
  {Messio}}, \bibinfo {author} {\bibfnamefont {B.}~\bibnamefont {Bernu}},\ and\
  \bibinfo {author} {\bibfnamefont {C.}~\bibnamefont {Lhuillier}},\ }\bibfield
  {title} {\bibinfo {title} {Kagome antiferromagnet: A chiral topological spin
  liquid?},\ }\href {https://doi.org/10.1103/PhysRevLett.108.207204} {\bibfield
   {journal} {\bibinfo  {journal} {Phys. Rev. Lett.}\ }\textbf {\bibinfo
  {volume} {108}},\ \bibinfo {pages} {207204} (\bibinfo {year}
  {2012})}\BibitemShut {NoStop}%
\bibitem [{\citenamefont {Wang}\ and\ \citenamefont
  {Vishwanath}(2006)}]{PhysRevB.74.174423}%
  \BibitemOpen
  \bibfield  {author} {\bibinfo {author} {\bibfnamefont {F.}~\bibnamefont
  {Wang}}\ and\ \bibinfo {author} {\bibfnamefont {A.}~\bibnamefont
  {Vishwanath}},\ }\bibfield  {title} {\bibinfo {title} {Spin-liquid states on
  the triangular and kagom\'e lattices: A projective-symmetry-group analysis of
  schwinger boson states},\ }\href {https://doi.org/10.1103/PhysRevB.74.174423}
  {\bibfield  {journal} {\bibinfo  {journal} {Phys. Rev. B}\ }\textbf {\bibinfo
  {volume} {74}},\ \bibinfo {pages} {174423} (\bibinfo {year}
  {2006})}\BibitemShut {NoStop}%
\bibitem [{not()}]{note1}%
  \BibitemOpen
  \bibinfo {note} {{We note that a phase diagram for self-consistent mean field
  ansatze obtained from PSG has been obtained in several works
  \cite{messio12prl,rossi23prb}, where the SS order is absent. Here we follow a
  different interpretation for the ansatze and the SS order appears as a
  leading instability towards magnetic order -- see \cite{supp} for a detailed
  discussion.}}\BibitemShut {Stop}%
\bibitem [{\citenamefont {Tan}\ \emph {et~al.}(2021)\citenamefont {Tan},
  \citenamefont {Liu}, \citenamefont {Wang},\ and\ \citenamefont
  {Yan}}]{tan21prl}%
  \BibitemOpen
  \bibfield  {author} {\bibinfo {author} {\bibfnamefont {H.}~\bibnamefont
  {Tan}}, \bibinfo {author} {\bibfnamefont {Y.}~\bibnamefont {Liu}}, \bibinfo
  {author} {\bibfnamefont {Z.}~\bibnamefont {Wang}},\ and\ \bibinfo {author}
  {\bibfnamefont {B.}~\bibnamefont {Yan}},\ }\bibfield  {title} {\bibinfo
  {title} {Charge density waves and electronic properties of superconducting
  kagome metals},\ }\href {https://doi.org/10.1103/PhysRevLett.127.046401}
  {\bibfield  {journal} {\bibinfo  {journal} {Phys. Rev. Lett.}\ }\textbf
  {\bibinfo {volume} {127}},\ \bibinfo {pages} {046401} (\bibinfo {year}
  {2021})}\BibitemShut {NoStop}%
\bibitem [{\citenamefont {Lin}\ and\ \citenamefont
  {Nandkishore}(2021)}]{lin21prb}%
  \BibitemOpen
  \bibfield  {author} {\bibinfo {author} {\bibfnamefont {Y.-P.}\ \bibnamefont
  {Lin}}\ and\ \bibinfo {author} {\bibfnamefont {R.~M.}\ \bibnamefont
  {Nandkishore}},\ }\bibfield  {title} {\bibinfo {title} {Complex charge
  density waves at van hove singularity on hexagonal lattices: Haldane-model
  phase diagram and potential realization in the kagome metals
  $a{V}_{3}{\mathrm{sb}}_{5}$ ($a$=k, rb, cs)},\ }\href
  {https://doi.org/10.1103/PhysRevB.104.045122} {\bibfield  {journal} {\bibinfo
   {journal} {Phys. Rev. B}\ }\textbf {\bibinfo {volume} {104}},\ \bibinfo
  {pages} {045122} (\bibinfo {year} {2021})}\BibitemShut {NoStop}%
\bibitem [{\citenamefont {Park}\ \emph {et~al.}(2021)\citenamefont {Park},
  \citenamefont {Ye},\ and\ \citenamefont {Balents}}]{park21prb}%
  \BibitemOpen
  \bibfield  {author} {\bibinfo {author} {\bibfnamefont {T.}~\bibnamefont
  {Park}}, \bibinfo {author} {\bibfnamefont {M.}~\bibnamefont {Ye}},\ and\
  \bibinfo {author} {\bibfnamefont {L.}~\bibnamefont {Balents}},\ }\bibfield
  {title} {\bibinfo {title} {Electronic instabilities of kagome metals: Saddle
  points and landau theory},\ }\href
  {https://doi.org/10.1103/PhysRevB.104.035142} {\bibfield  {journal} {\bibinfo
   {journal} {Phys. Rev. B}\ }\textbf {\bibinfo {volume} {104}},\ \bibinfo
  {pages} {035142} (\bibinfo {year} {2021})}\BibitemShut {NoStop}%
\bibitem [{\citenamefont {Feng}\ \emph {et~al.}(2021)\citenamefont {Feng},
  \citenamefont {Zhang}, \citenamefont {Jiang},\ and\ \citenamefont
  {Hu}}]{feng21prb}%
  \BibitemOpen
  \bibfield  {author} {\bibinfo {author} {\bibfnamefont {X.}~\bibnamefont
  {Feng}}, \bibinfo {author} {\bibfnamefont {Y.}~\bibnamefont {Zhang}},
  \bibinfo {author} {\bibfnamefont {K.}~\bibnamefont {Jiang}},\ and\ \bibinfo
  {author} {\bibfnamefont {J.}~\bibnamefont {Hu}},\ }\bibfield  {title}
  {\bibinfo {title} {Low-energy effective theory and symmetry classification of
  flux phases on the kagome lattice},\ }\href
  {https://doi.org/10.1103/PhysRevB.104.165136} {\bibfield  {journal} {\bibinfo
   {journal} {Phys. Rev. B}\ }\textbf {\bibinfo {volume} {104}},\ \bibinfo
  {pages} {165136} (\bibinfo {year} {2021})}\BibitemShut {NoStop}%
\bibitem [{\citenamefont {Christensen}\ \emph {et~al.}(2021)\citenamefont
  {Christensen}, \citenamefont {Birol}, \citenamefont {Andersen},\ and\
  \citenamefont {Fernandes}}]{christensen21prb}%
  \BibitemOpen
  \bibfield  {author} {\bibinfo {author} {\bibfnamefont {M.~H.}\ \bibnamefont
  {Christensen}}, \bibinfo {author} {\bibfnamefont {T.}~\bibnamefont {Birol}},
  \bibinfo {author} {\bibfnamefont {B.~M.}\ \bibnamefont {Andersen}},\ and\
  \bibinfo {author} {\bibfnamefont {R.~M.}\ \bibnamefont {Fernandes}},\
  }\bibfield  {title} {\bibinfo {title} {Theory of the charge density wave in
  $a{\mathrm{v}}_{3}{\mathrm{sb}}_{5}$ kagome metals},\ }\href
  {https://doi.org/10.1103/PhysRevB.104.214513} {\bibfield  {journal} {\bibinfo
   {journal} {Phys. Rev. B}\ }\textbf {\bibinfo {volume} {104}},\ \bibinfo
  {pages} {214513} (\bibinfo {year} {2021})}\BibitemShut {NoStop}%
\bibitem [{\citenamefont {Jiang}\ \emph {et~al.}(2022)\citenamefont {Jiang},
  \citenamefont {Yu}, \citenamefont {Wang}, \citenamefont {Lu}, \citenamefont
  {Meng}, \citenamefont {Jiang},\ and\ \citenamefont {Liu}}]{jiang22cpl}%
  \BibitemOpen
  \bibfield  {author} {\bibinfo {author} {\bibfnamefont {Y.}~\bibnamefont
  {Jiang}}, \bibinfo {author} {\bibfnamefont {Z.}~\bibnamefont {Yu}}, \bibinfo
  {author} {\bibfnamefont {Y.}~\bibnamefont {Wang}}, \bibinfo {author}
  {\bibfnamefont {T.}~\bibnamefont {Lu}}, \bibinfo {author} {\bibfnamefont
  {S.}~\bibnamefont {Meng}}, \bibinfo {author} {\bibfnamefont {K.}~\bibnamefont
  {Jiang}},\ and\ \bibinfo {author} {\bibfnamefont {M.}~\bibnamefont {Liu}},\
  }\bibfield  {title} {\bibinfo {title} {{Screening Promising
  CsV$_{3}$Sb$_{5}$-Like Kagome Materials from Systematic First-Principles
  Evaluation}},\ }\href {https://doi.org/10.1088/0256-307X/39/4/047402}
  {\bibfield  {journal} {\bibinfo  {journal} {Chin. Phys. Lett.}\ }\textbf
  {\bibinfo {volume} {39}},\ \bibinfo {pages} {047402} (\bibinfo {year}
  {2022})}\BibitemShut {NoStop}%
\bibitem [{\citenamefont {Jo}\ \emph {et~al.}(2012)\citenamefont {Jo},
  \citenamefont {Guzman}, \citenamefont {Thomas}, \citenamefont {Hosur},
  \citenamefont {Vishwanath},\ and\ \citenamefont {Stamper-Kurn}}]{jo12prl}%
  \BibitemOpen
  \bibfield  {author} {\bibinfo {author} {\bibfnamefont {G.-B.}\ \bibnamefont
  {Jo}}, \bibinfo {author} {\bibfnamefont {J.}~\bibnamefont {Guzman}}, \bibinfo
  {author} {\bibfnamefont {C.~K.}\ \bibnamefont {Thomas}}, \bibinfo {author}
  {\bibfnamefont {P.}~\bibnamefont {Hosur}}, \bibinfo {author} {\bibfnamefont
  {A.}~\bibnamefont {Vishwanath}},\ and\ \bibinfo {author} {\bibfnamefont
  {D.~M.}\ \bibnamefont {Stamper-Kurn}},\ }\bibfield  {title} {\bibinfo {title}
  {Ultracold atoms in a tunable optical kagome lattice},\ }\href
  {https://doi.org/10.1103/PhysRevLett.108.045305} {\bibfield  {journal}
  {\bibinfo  {journal} {Phys. Rev. Lett.}\ }\textbf {\bibinfo {volume} {108}},\
  \bibinfo {pages} {045305} (\bibinfo {year} {2012})}\BibitemShut {NoStop}%
\bibitem [{\citenamefont {Mazurenko}\ \emph {et~al.}(2017)\citenamefont
  {Mazurenko}, \citenamefont {Chiu}, \citenamefont {Ji}, \citenamefont
  {Parsons}, \citenamefont {Kan\'asz-Nagy}, \citenamefont {Schmidt},
  \citenamefont {Grusdt}, \citenamefont {Demler}, \citenamefont {Greif},\ and\
  \citenamefont {Greiner}}]{mazurenko17n}%
  \BibitemOpen
  \bibfield  {author} {\bibinfo {author} {\bibfnamefont {A.}~\bibnamefont
  {Mazurenko}}, \bibinfo {author} {\bibfnamefont {C.~S.}\ \bibnamefont {Chiu}},
  \bibinfo {author} {\bibfnamefont {G.}~\bibnamefont {Ji}}, \bibinfo {author}
  {\bibfnamefont {M.~F.}\ \bibnamefont {Parsons}}, \bibinfo {author}
  {\bibfnamefont {M.}~\bibnamefont {Kan\'asz-Nagy}}, \bibinfo {author}
  {\bibfnamefont {R.}~\bibnamefont {Schmidt}}, \bibinfo {author} {\bibfnamefont
  {F.}~\bibnamefont {Grusdt}}, \bibinfo {author} {\bibfnamefont
  {E.}~\bibnamefont {Demler}}, \bibinfo {author} {\bibfnamefont
  {D.}~\bibnamefont {Greif}},\ and\ \bibinfo {author} {\bibfnamefont
  {M.}~\bibnamefont {Greiner}},\ }\bibfield  {title} {\bibinfo {title} {A
  cold-atom fermi–hubbard antiferromagnet},\ }\href
  {https://doi.org/10.1038/nature22362} {\bibfield  {journal} {\bibinfo
  {journal} {Nature}\ }\textbf {\bibinfo {volume} {545}},\ \bibinfo {pages}
  {462} (\bibinfo {year} {2017})}\BibitemShut {NoStop}%
\bibitem [{\citenamefont {Watanabe}\ and\ \citenamefont
  {Ogata}(2005)}]{watanabe05jpsj}%
  \BibitemOpen
  \bibfield  {author} {\bibinfo {author} {\bibfnamefont {H.}~\bibnamefont
  {Watanabe}}\ and\ \bibinfo {author} {\bibfnamefont {M.}~\bibnamefont
  {Ogata}},\ }\bibfield  {title} {\bibinfo {title} {Charge order and
  superconductivity in two-dimensional triangular lattice at n=2/3},\ }\href
  {https://doi.org/10.1143/JPSJ.74.2901} {\bibfield  {journal} {\bibinfo
  {journal} {J. Phys. Soc. Jpn.}\ }\textbf {\bibinfo {volume} {74}},\ \bibinfo
  {pages} {2901} (\bibinfo {year} {2005})}\BibitemShut {NoStop}%
\bibitem [{\citenamefont {Song}\ \emph {et~al.}(2021)\citenamefont {Song},
  \citenamefont {Vishwanath},\ and\ \citenamefont {Zhang}}]{song21prb}%
  \BibitemOpen
  \bibfield  {author} {\bibinfo {author} {\bibfnamefont {X.-Y.}\ \bibnamefont
  {Song}}, \bibinfo {author} {\bibfnamefont {A.}~\bibnamefont {Vishwanath}},\
  and\ \bibinfo {author} {\bibfnamefont {Y.-H.}\ \bibnamefont {Zhang}},\
  }\bibfield  {title} {\bibinfo {title} {Doping the chiral spin liquid:
  Topological superconductor or chiral metal},\ }\href
  {https://doi.org/10.1103/PhysRevB.103.165138} {\bibfield  {journal} {\bibinfo
   {journal} {Phys. Rev. B}\ }\textbf {\bibinfo {volume} {103}},\ \bibinfo
  {pages} {165138} (\bibinfo {year} {2021})}\BibitemShut {NoStop}%
\bibitem [{\citenamefont {Peng}\ \emph {et~al.}(2021)\citenamefont {Peng},
  \citenamefont {Jiang}, \citenamefont {Wang},\ and\ \citenamefont
  {Jiang}}]{peng21njp}%
  \BibitemOpen
  \bibfield  {author} {\bibinfo {author} {\bibfnamefont {C.}~\bibnamefont
  {Peng}}, \bibinfo {author} {\bibfnamefont {Y.-F.}\ \bibnamefont {Jiang}},
  \bibinfo {author} {\bibfnamefont {Y.}~\bibnamefont {Wang}},\ and\ \bibinfo
  {author} {\bibfnamefont {H.-C.}\ \bibnamefont {Jiang}},\ }\bibfield  {title}
  {\bibinfo {title} {Gapless spin liquid and pair density wave of the hubbard
  model on three-leg triangular cylinders},\ }\href
  {https://doi.org/10.1088/1367-2630/ac3a83} {\bibfield  {journal} {\bibinfo
  {journal} {New J. Phys.}\ }\textbf {\bibinfo {volume} {23}},\ \bibinfo
  {pages} {123004} (\bibinfo {year} {2021})}\BibitemShut {NoStop}%
\bibitem [{\citenamefont {Zhu}\ \emph {et~al.}(2022)\citenamefont {Zhu},
  \citenamefont {Sheng},\ and\ \citenamefont {Vishwanath}}]{zhu22prb}%
  \BibitemOpen
  \bibfield  {author} {\bibinfo {author} {\bibfnamefont {Z.}~\bibnamefont
  {Zhu}}, \bibinfo {author} {\bibfnamefont {D.~N.}\ \bibnamefont {Sheng}},\
  and\ \bibinfo {author} {\bibfnamefont {A.}~\bibnamefont {Vishwanath}},\
  }\bibfield  {title} {\bibinfo {title} {Doped mott insulators in the
  triangular-lattice hubbard model},\ }\href
  {https://doi.org/10.1103/PhysRevB.105.205110} {\bibfield  {journal} {\bibinfo
   {journal} {Phys. Rev. B}\ }\textbf {\bibinfo {volume} {105}},\ \bibinfo
  {pages} {205110} (\bibinfo {year} {2022})}\BibitemShut {NoStop}%
\bibitem [{\citenamefont {Huang}\ \emph {et~al.}(2023)\citenamefont {Huang},
  \citenamefont {Gong},\ and\ \citenamefont {Sheng}}]{huang23prl}%
  \BibitemOpen
  \bibfield  {author} {\bibinfo {author} {\bibfnamefont {Y.}~\bibnamefont
  {Huang}}, \bibinfo {author} {\bibfnamefont {S.-S.}\ \bibnamefont {Gong}},\
  and\ \bibinfo {author} {\bibfnamefont {D.~N.}\ \bibnamefont {Sheng}},\
  }\bibfield  {title} {\bibinfo {title} {Quantum phase diagram and
  spontaneously emergent topological chiral superconductivity in doped
  triangular-lattice mott insulators},\ }\href
  {https://doi.org/10.1103/PhysRevLett.130.136003} {\bibfield  {journal}
  {\bibinfo  {journal} {Phys. Rev. Lett.}\ }\textbf {\bibinfo {volume} {130}},\
  \bibinfo {pages} {136003} (\bibinfo {year} {2023})}\BibitemShut {NoStop}%
\bibitem [{\citenamefont {Zhu}\ and\ \citenamefont {Chen}(2023)}]{zhu23prb}%
  \BibitemOpen
  \bibfield  {author} {\bibinfo {author} {\bibfnamefont {Z.}~\bibnamefont
  {Zhu}}\ and\ \bibinfo {author} {\bibfnamefont {Q.}~\bibnamefont {Chen}},\
  }\bibfield  {title} {\bibinfo {title} {Superconductivity in doped triangular
  mott insulators: The roles of parent spin backgrounds and charge kinetic
  energy},\ }\href {https://doi.org/10.1103/PhysRevB.107.L220502} {\bibfield
  {journal} {\bibinfo  {journal} {Phys. Rev. B}\ }\textbf {\bibinfo {volume}
  {107}},\ \bibinfo {pages} {L220502} (\bibinfo {year} {2023})}\BibitemShut
  {NoStop}%
\bibitem [{\citenamefont {Cheng}\ \emph {et~al.}(2010)\citenamefont {Cheng},
  \citenamefont {Sun}, \citenamefont {Galitski},\ and\ \citenamefont
  {Das~Sarma}}]{cheng10prb}%
  \BibitemOpen
  \bibfield  {author} {\bibinfo {author} {\bibfnamefont {M.}~\bibnamefont
  {Cheng}}, \bibinfo {author} {\bibfnamefont {K.}~\bibnamefont {Sun}}, \bibinfo
  {author} {\bibfnamefont {V.}~\bibnamefont {Galitski}},\ and\ \bibinfo
  {author} {\bibfnamefont {S.}~\bibnamefont {Das~Sarma}},\ }\bibfield  {title}
  {\bibinfo {title} {Stable topological superconductivity in a family of
  two-dimensional fermion models},\ }\href
  {https://doi.org/10.1103/PhysRevB.81.024504} {\bibfield  {journal} {\bibinfo
  {journal} {Phys. Rev. B}\ }\textbf {\bibinfo {volume} {81}},\ \bibinfo
  {pages} {024504} (\bibinfo {year} {2010})}\BibitemShut {NoStop}%
\bibitem [{\citenamefont {Gneist}\ \emph {et~al.}(2022)\citenamefont {Gneist},
  \citenamefont {Kiese}, \citenamefont {Henkel}, \citenamefont {Thomale},
  \citenamefont {Classen},\ and\ \citenamefont {Scherer}}]{gneist22epjb}%
  \BibitemOpen
  \bibfield  {author} {\bibinfo {author} {\bibfnamefont {N.}~\bibnamefont
  {Gneist}}, \bibinfo {author} {\bibfnamefont {D.}~\bibnamefont {Kiese}},
  \bibinfo {author} {\bibfnamefont {R.}~\bibnamefont {Henkel}}, \bibinfo
  {author} {\bibfnamefont {R.}~\bibnamefont {Thomale}}, \bibinfo {author}
  {\bibfnamefont {L.}~\bibnamefont {Classen}},\ and\ \bibinfo {author}
  {\bibfnamefont {M.~M.}\ \bibnamefont {Scherer}},\ }\bibfield  {title}
  {\bibinfo {title} {Functional renormalization of spinless triangular-lattice
  fermions: N-patch vs. truncated-unity scheme},\ }\href
  {https://doi.org/10.1140/epjb/s10051-022-00395-w} {\bibfield  {journal}
  {\bibinfo  {journal} {Eur. Phys. J. B}\ }\textbf {\bibinfo {volume} {95}},\
  \bibinfo {pages} {157} (\bibinfo {year} {2022})}\BibitemShut {NoStop}%
\bibitem [{\citenamefont {He}\ \emph {et~al.}(2023)\citenamefont {He},
  \citenamefont {Yang}, \citenamefont {Profe}, \citenamefont {Bergholtz},\ and\
  \citenamefont {Kennes}}]{he23prr}%
  \BibitemOpen
  \bibfield  {author} {\bibinfo {author} {\bibfnamefont {Y.}~\bibnamefont
  {He}}, \bibinfo {author} {\bibfnamefont {K.}~\bibnamefont {Yang}}, \bibinfo
  {author} {\bibfnamefont {J.~B.}\ \bibnamefont {Profe}}, \bibinfo {author}
  {\bibfnamefont {E.~J.}\ \bibnamefont {Bergholtz}},\ and\ \bibinfo {author}
  {\bibfnamefont {D.~M.}\ \bibnamefont {Kennes}},\ }\bibfield  {title}
  {\bibinfo {title} {Superconductivity of repulsive spinless fermions with
  sublattice potentials},\ }\href
  {https://doi.org/10.1103/PhysRevResearch.5.L012009} {\bibfield  {journal}
  {\bibinfo  {journal} {Phys. Rev. Res.}\ }\textbf {\bibinfo {volume} {5}},\
  \bibinfo {pages} {L012009} (\bibinfo {year} {2023})}\BibitemShut {NoStop}%
\bibitem [{\citenamefont {Savary}\ and\ \citenamefont
  {Balents}(2016)}]{savary16rpp}%
  \BibitemOpen
  \bibfield  {author} {\bibinfo {author} {\bibfnamefont {L.}~\bibnamefont
  {Savary}}\ and\ \bibinfo {author} {\bibfnamefont {L.}~\bibnamefont
  {Balents}},\ }\bibfield  {title} {\bibinfo {title} {Quantum spin liquids: a
  review},\ }\href {https://doi.org/10.1088/0034-4885/80/1/016502} {\bibfield
  {journal} {\bibinfo  {journal} {Rep. Prog. Phys.}\ }\textbf {\bibinfo
  {volume} {80}},\ \bibinfo {pages} {016502} (\bibinfo {year}
  {2016})}\BibitemShut {NoStop}%
\bibitem [{\citenamefont {Liu}\ and\ \citenamefont
  {Bergholtz}(2023)}]{liu23bk}%
  \BibitemOpen
  \bibfield  {author} {\bibinfo {author} {\bibfnamefont {Z.}~\bibnamefont
  {Liu}}\ and\ \bibinfo {author} {\bibfnamefont {E.~J.}\ \bibnamefont
  {Bergholtz}},\ }\bibfield  {title} {\bibinfo {title} {Recent developments in
  fractional chern insulators},\ }in\ \href
  {https://doi.org/https://doi.org/10.1016/B978-0-323-90800-9.00136-0} {\emph
  {\bibinfo {booktitle} {Reference Module in Materials Science and Materials
  Engineering}}}\ (\bibinfo  {publisher} {Elsevier},\ \bibinfo {year}
  {2023})\BibitemShut {NoStop}%
\bibitem [{\citenamefont {{Liu}}\ \emph {et~al.}(2023)\citenamefont {{Liu}},
  \citenamefont {{Liu}}, \citenamefont {{Bao}}, \citenamefont {{Yang}},
  \citenamefont {{Ji}}, \citenamefont {{Liu}}, \citenamefont {{Xu}},
  \citenamefont {{Yang}}, \citenamefont {{Chai}}, \citenamefont {{Lu}},
  \citenamefont {{Liu}}, \citenamefont {{Wang}}, \citenamefont {{Jiang}},
  \citenamefont {{Tao}}, \citenamefont {{Ren}}, \citenamefont {{Xu}},
  \citenamefont {{Cao}}, \citenamefont {{Xu}}, \citenamefont {{Cheng}},\ and\
  \citenamefont {{Cao}}}]{liu23ax}%
  \BibitemOpen
  \bibfield  {author} {\bibinfo {author} {\bibfnamefont {Y.}~\bibnamefont
  {{Liu}}}, \bibinfo {author} {\bibfnamefont {Z.-Y.}\ \bibnamefont {{Liu}}},
  \bibinfo {author} {\bibfnamefont {J.-K.}\ \bibnamefont {{Bao}}}, \bibinfo
  {author} {\bibfnamefont {P.-T.}\ \bibnamefont {{Yang}}}, \bibinfo {author}
  {\bibfnamefont {L.-W.}\ \bibnamefont {{Ji}}}, \bibinfo {author}
  {\bibfnamefont {J.-Y.}\ \bibnamefont {{Liu}}}, \bibinfo {author}
  {\bibfnamefont {C.-C.}\ \bibnamefont {{Xu}}}, \bibinfo {author}
  {\bibfnamefont {W.-Z.}\ \bibnamefont {{Yang}}}, \bibinfo {author}
  {\bibfnamefont {W.-L.}\ \bibnamefont {{Chai}}}, \bibinfo {author}
  {\bibfnamefont {J.-Y.}\ \bibnamefont {{Lu}}}, \bibinfo {author}
  {\bibfnamefont {C.-C.}\ \bibnamefont {{Liu}}}, \bibinfo {author}
  {\bibfnamefont {B.-S.}\ \bibnamefont {{Wang}}}, \bibinfo {author}
  {\bibfnamefont {H.}~\bibnamefont {{Jiang}}}, \bibinfo {author} {\bibfnamefont
  {Q.}~\bibnamefont {{Tao}}}, \bibinfo {author} {\bibfnamefont
  {Z.}~\bibnamefont {{Ren}}}, \bibinfo {author} {\bibfnamefont {X.-F.}\
  \bibnamefont {{Xu}}}, \bibinfo {author} {\bibfnamefont {C.}~\bibnamefont
  {{Cao}}}, \bibinfo {author} {\bibfnamefont {Z.-A.}\ \bibnamefont {{Xu}}},
  \bibinfo {author} {\bibfnamefont {J.-G.}\ \bibnamefont {{Cheng}}},\ and\
  \bibinfo {author} {\bibfnamefont {G.-H.}\ \bibnamefont {{Cao}}},\ }\bibfield
  {title} {\bibinfo {title} {{Superconductivity emerged from density-wave order
  in a kagome bad metal}},\ }\href {https://doi.org/10.48550/arXiv.2309.13514}
  {\bibfield  {journal} {\bibinfo  {journal} {arXiv e-prints}\ ,\ \bibinfo
  {eid} {arXiv:2309.13514}} (\bibinfo {year} {2023})},\ \Eprint
  {https://arxiv.org/abs/2309.13514} {arXiv:2309.13514 [cond-mat.supr-con]}
  \BibitemShut {NoStop}%
\bibitem [{\citenamefont {Kudin}\ \emph {et~al.}(2002)\citenamefont {Kudin},
  \citenamefont {Scuseria},\ and\ \citenamefont {Cancès}}]{kudin02jcp}%
  \BibitemOpen
  \bibfield  {author} {\bibinfo {author} {\bibfnamefont {K.~N.}\ \bibnamefont
  {Kudin}}, \bibinfo {author} {\bibfnamefont {G.~E.}\ \bibnamefont
  {Scuseria}},\ and\ \bibinfo {author} {\bibfnamefont {E.}~\bibnamefont
  {Cancès}},\ }\bibfield  {title} {\bibinfo {title} {{A black-box
  self-consistent field convergence algorithm: One step closer}},\ }\href
  {https://doi.org/10.1063/1.1470195} {\bibfield  {journal} {\bibinfo
  {journal} {J. Chem. Phys.}\ }\textbf {\bibinfo {volume} {116}},\ \bibinfo
  {pages} {8255} (\bibinfo {year} {2002})}\BibitemShut {NoStop}%
\bibitem [{\citenamefont {Fukui}\ \emph {et~al.}(2005)\citenamefont {Fukui},
  \citenamefont {Hatsugai},\ and\ \citenamefont {Suzuki}}]{fukui05jpsp}%
  \BibitemOpen
  \bibfield  {author} {\bibinfo {author} {\bibfnamefont {T.}~\bibnamefont
  {Fukui}}, \bibinfo {author} {\bibfnamefont {Y.}~\bibnamefont {Hatsugai}},\
  and\ \bibinfo {author} {\bibfnamefont {H.}~\bibnamefont {Suzuki}},\
  }\bibfield  {title} {\bibinfo {title} {Chern numbers in discretized brillouin
  zone: Efficient method of computing (spin) hall conductances},\ }\href
  {https://doi.org/10.1143/JPSJ.74.1674} {\bibfield  {journal} {\bibinfo
  {journal} {J. Phys. Soc. Jpn.}\ }\textbf {\bibinfo {volume} {74}},\ \bibinfo
  {pages} {1674} (\bibinfo {year} {2005})}\BibitemShut {NoStop}%
\bibitem [{\citenamefont {Sun}\ \emph {et~al.}(2009)\citenamefont {Sun},
  \citenamefont {Yao}, \citenamefont {Fradkin},\ and\ \citenamefont
  {Kivelson}}]{sun09prl}%
  \BibitemOpen
  \bibfield  {author} {\bibinfo {author} {\bibfnamefont {K.}~\bibnamefont
  {Sun}}, \bibinfo {author} {\bibfnamefont {H.}~\bibnamefont {Yao}}, \bibinfo
  {author} {\bibfnamefont {E.}~\bibnamefont {Fradkin}},\ and\ \bibinfo {author}
  {\bibfnamefont {S.~A.}\ \bibnamefont {Kivelson}},\ }\bibfield  {title}
  {\bibinfo {title} {Topological insulators and nematic phases from spontaneous
  symmetry breaking in 2d fermi systems with a quadratic band crossing},\
  }\href {https://doi.org/10.1103/PhysRevLett.103.046811} {\bibfield  {journal}
  {\bibinfo  {journal} {Phys. Rev. Lett.}\ }\textbf {\bibinfo {volume} {103}},\
  \bibinfo {pages} {046811} (\bibinfo {year} {2009})}\BibitemShut {NoStop}%
\bibitem [{\citenamefont {Szasz}\ and\ \citenamefont
  {Motruk}(2021)}]{szasz21prb}%
  \BibitemOpen
  \bibfield  {author} {\bibinfo {author} {\bibfnamefont {A.}~\bibnamefont
  {Szasz}}\ and\ \bibinfo {author} {\bibfnamefont {J.}~\bibnamefont {Motruk}},\
  }\bibfield  {title} {\bibinfo {title} {Phase diagram of the anisotropic
  triangular lattice hubbard model},\ }\href
  {https://doi.org/10.1103/PhysRevB.103.235132} {\bibfield  {journal} {\bibinfo
   {journal} {Phys. Rev. B}\ }\textbf {\bibinfo {volume} {103}},\ \bibinfo
  {pages} {235132} (\bibinfo {year} {2021})}\BibitemShut {NoStop}%
\bibitem [{\citenamefont {Cookmeyer}\ \emph {et~al.}(2021)\citenamefont
  {Cookmeyer}, \citenamefont {Motruk},\ and\ \citenamefont
  {Moore}}]{cookmeyer21prl}%
  \BibitemOpen
  \bibfield  {author} {\bibinfo {author} {\bibfnamefont {T.}~\bibnamefont
  {Cookmeyer}}, \bibinfo {author} {\bibfnamefont {J.}~\bibnamefont {Motruk}},\
  and\ \bibinfo {author} {\bibfnamefont {J.~E.}\ \bibnamefont {Moore}},\
  }\bibfield  {title} {\bibinfo {title} {Four-spin terms and the origin of the
  chiral spin liquid in mott insulators on the triangular lattice},\ }\href
  {https://doi.org/10.1103/PhysRevLett.127.087201} {\bibfield  {journal}
  {\bibinfo  {journal} {Phys. Rev. Lett.}\ }\textbf {\bibinfo {volume} {127}},\
  \bibinfo {pages} {087201} (\bibinfo {year} {2021})}\BibitemShut {NoStop}%
\bibitem [{\citenamefont {Di~Sante}\ \emph {et~al.}(2023)\citenamefont
  {Di~Sante}, \citenamefont {Kim}, \citenamefont {Hanke}, \citenamefont
  {Wehling}, \citenamefont {Franchini}, \citenamefont {Thomale},\ and\
  \citenamefont {Sangiovanni}}]{disante23prr}%
  \BibitemOpen
  \bibfield  {author} {\bibinfo {author} {\bibfnamefont {D.}~\bibnamefont
  {Di~Sante}}, \bibinfo {author} {\bibfnamefont {B.}~\bibnamefont {Kim}},
  \bibinfo {author} {\bibfnamefont {W.}~\bibnamefont {Hanke}}, \bibinfo
  {author} {\bibfnamefont {T.}~\bibnamefont {Wehling}}, \bibinfo {author}
  {\bibfnamefont {C.}~\bibnamefont {Franchini}}, \bibinfo {author}
  {\bibfnamefont {R.}~\bibnamefont {Thomale}},\ and\ \bibinfo {author}
  {\bibfnamefont {G.}~\bibnamefont {Sangiovanni}},\ }\bibfield  {title}
  {\bibinfo {title} {Electronic correlations and universal long-range scaling
  in kagome metals},\ }\href
  {https://doi.org/10.1103/PhysRevResearch.5.L012008} {\bibfield  {journal}
  {\bibinfo  {journal} {Phys. Rev. Res.}\ }\textbf {\bibinfo {volume} {5}},\
  \bibinfo {pages} {L012008} (\bibinfo {year} {2023})}\BibitemShut {NoStop}%
\bibitem [{\citenamefont {Rossi}\ \emph {et~al.}(2023)\citenamefont {Rossi},
  \citenamefont {Motruk}, \citenamefont {Rademaker},\ and\ \citenamefont
  {Abanin}}]{rossi23prb}%
  \BibitemOpen
  \bibfield  {author} {\bibinfo {author} {\bibfnamefont {D.}~\bibnamefont
  {Rossi}}, \bibinfo {author} {\bibfnamefont {J.}~\bibnamefont {Motruk}},
  \bibinfo {author} {\bibfnamefont {L.}~\bibnamefont {Rademaker}},\ and\
  \bibinfo {author} {\bibfnamefont {D.~A.}\ \bibnamefont {Abanin}},\ }\bibfield
   {title} {\bibinfo {title} {Schwinger boson study of the
  ${J}_{1}\text{\ensuremath{-}}{J}_{2}\text{\ensuremath{-}}{J}_{3}$ kagome
  heisenberg antiferromagnet with dzyaloshinskii-moriya interactions},\ }\href
  {https://doi.org/10.1103/PhysRevB.108.144406} {\bibfield  {journal} {\bibinfo
   {journal} {Phys. Rev. B}\ }\textbf {\bibinfo {volume} {108}},\ \bibinfo
  {pages} {144406} (\bibinfo {year} {2023})}\BibitemShut {NoStop}%
\end{thebibliography}%

\clearpage
\onecolumngrid

\begin{center}{\large\bf
Supplemental Material for\\``Complex magnetic and spatial symmetry breaking from correlations in kagome flat bands"
}\end{center}

\setcounter{secnumdepth}{3}
\setcounter{equation}{0}
\setcounter{figure}{0}
\renewcommand{\theequation}{S\arabic{equation}}
\renewcommand{\thefigure}{S\arabic{figure}}
\newcommand\Scite[1]{[S\citealp{#1}]}
\makeatletter \renewcommand\@biblabel[1]{[S#1]} \makeatother

\tableofcontents

\section{Hartree-Fock analysis}

In this section, we introduce the Hartree-Fock (HF) analysis used in our work. The central concept is to approximate the interacting model by a mean-field noninteracting theory. In this mean-field theory, the fermions are treated in terms of the environments of individual fermions.  The structures of these fermionic environments are set by the ground state, where various symmetry breaking orders may occur. Each fermion experiences the background potential from its environment, known as the Hartree potential. Meanwhile, the fermion can also immerse into the environment, followed by the ejection of a new fermion as a replacement. This process involves the Fermi exchange and leads to the Fock potential. The mean-field theory is described by the Hartree-Fock Hamiltonian, where the Hartree and Fock potentials replace the interaction. The ground state can be solved by diagonalizing and updating the Hartree-Fock Hamiltonian self-consistently under energy minimization.

Now we lay out the explicit formalism. For a given lattice, we consider the Hubbard model
\begin{equation}
H=-\sum_{ii'\tau\tau'}\sum_\sigma t_{ii'\tau\tau'}c_{i\tau\sigma}^\dagger c_{i'\tau'\sigma}+\frac{1}{2}\sum_{ii'\tau\tau'}\sum_{\sigma\sigma'}U_{ii'\tau\tau'}c_{i\tau\sigma}^\dagger c_{i'\tau'\sigma'}^\dagger c_{i'\tau'\sigma'}c_{i\tau\sigma}
\end{equation}
with Bravais lattice site $i$, sublattice $\tau$, and spin $\sigma$. Any wavefunction can be represented by its density matrix
\begin{equation}
P_{ii'\tau\tau'\sigma\sigma'}=\epvl{c_{i'\tau'\sigma'}^\dagger c_{i\tau\sigma}},
\end{equation}
which is a collection of the particle-hole condensates. The Hartree-Fock Hamiltonian is derived as
\begin{equation}
H^\text{HF}[P]=\sum_{ii'\tau\tau'\sigma\sigma'}c_{i\tau\sigma}^\dagger H^{HF}_{ii'\tau\tau'\sigma\sigma'}c_{i'\tau'\sigma'}=T+H^\text{H}[P]+H^\text{F}[P],
\end{equation}
where the three terms are
\begin{equation}
\begin{aligned}
&\text{Noninteracting potential: }T_{ii'\tau\tau'\sigma\sigma'}=-\delta_{\sigma\sigma'}t_{ii'\tau\tau'},\\
&\text{Hartree potential: }H^\text{H}_{ii'\tau\tau'\sigma\sigma'}[P]=\delta_{ii'}\delta_{\tau\tau'}\delta_{\sigma\sigma'}\sum_{i''\tau''\sigma''}U_{ii''\tau\tau''}P_{i''i''\tau''\tau''\sigma''\sigma''},\\
&\text{Fock potential: }H^\text{F}_{ii'\tau\tau'\sigma\sigma'}[P]=-U_{ii'\tau\tau'}P_{ii'\tau\tau'\sigma\sigma'}.
\end{aligned}
\end{equation}
With the density matrix and the Hartree-Fock Hamiltonian, the energy of the wavefunction can be computed
\begin{equation}
E[P]=\epvl{H}=\frac{1}{2}\Tr(P[T+H^\text{HF}]).
\end{equation}

We want to obtain the $N$-particle ground state, where the occupation number $N=n_fN_L$ is defined by the filling $n_f$ and the number of lattice sites $N_L$. This is achieved by variationally solving the density matrix $P$ under iterative energy minimization. The starting point is an initial density matrix $P_m=P_0$, from which the Hartree-Fock Hamiltonian $H^\text{HF}_m=H^\text{HF}[P_m]$ and energy $E_m=(1/2)\Tr(P_m[T+H^\text{HF}_m])$ are obtained. At the $m$-th iteration, the variational update is carried out as follows:
\begin{enumerate}
\item Diagonalize the Hartree-Fock Hamiltonian $H^\text{HF}_m=U EU^\dagger$, where $E=\text{diag}(E_1,E_2,\dots)$ with $E_1<E_2<\dots$ are the eigenvalues and $U=(u_1,u_2,\dots)$ are the eigenstates.
\item Assemble the density matrix $P_{m+1}=U D_NU^\dagger$, where $D_N=\text{diag}(1,1,\dots,1,0,0,\dots)$ selects the $N$ eigenstates with the lowest energies.
\item Compose $H^\text{HF}_{m+1}=H^\text{HF}[P_{m+1}]$.
\item Calculate the energy density $e_{m+1}=E_{m+1}/N_L$.
\item Check convergence: Stop if the error is small enough $|e_{m+1}-e_m|<\delta e$ and $|P_{m+1,ab}-P_{m,ab}|<\delta p$. We choose $\delta e,\delta p=10^{-15}$ in our computation, although the iteration may only achieve $\delta p=10^{-14}$ especially when the system size is large.
\item Adopt the optimal damping algorithm \cite{kudin02jcp} to accelerate the covergence.
\end{enumerate}
Our computation adopts a $12\times12\times3$ lattice with periodic boundary condition.  Larger lattices have also been examined for the sanity check.

With the ground state obtained, the spin-charge (S-C) patterns can be studied by computing the densities
\begin{equation}
s^\nu_{ii'\tau\tau'}
=\sum_{\sigma\sigma'}\epvl{c_{i'\tau'\sigma'}^\dagger\tilde\sigma^\nu_{\sigma'\sigma}c_{i\tau\sigma}}
=\sum_{\sigma\sigma'}\epvl{c_{i'\tau'\sigma'}^\dagger c_{i\tau\sigma}}\tilde\sigma^\nu_{\sigma'\sigma}
=\sum_{\sigma\sigma'}P_{ii'\tau\tau'\sigma\sigma'}\tilde\sigma^\nu_{\sigma'\sigma}
=\Tr(P_{ii'\tau\tau'}\tilde\sigma^\nu).
\end{equation}
Here $P_{ii'\tau\tau'}=(P_{ii'\tau\tau'\sigma\sigma'})$ is a $2\times2$ density matrix in the spin space. The charge and spin densities $s^\nu=(s^0,\mbf s)$ are computed with the $2\times2$ identity and Pauli matrices $\tilde\sigma^\nu=(\sigma^0,\boldsymbol\sigma/2)$. The onsite and intersite densities correspond to the site and bond densities, respectively. While the site densities are real, the bond densities are generally complex. The real parts of the bond densities $\text{Re}(s^\nu_{ii'\tau\tau'})$ correspond to the ordinary charge and spin bond densities. Meanwhile, the imaginary parts $\text{Im}(s^\nu_{ii'\tau\tau'})$ are related to the charge and spin bond currents. To see this relation, we derive the charge bond current $j_{ab}$ from site $a$ to site $b$. The conservation law of site density at $b$ gives
\begin{equation}
\begin{aligned}
\sum_aj_{ab}
&=\frac{d}{dt}\epvl{c_b^\dagger c_b}
=\frac{1}{i}\epvl{[c_b^\dagger c_b,H]}\\
&\approx\frac{1}{i}\sum_{ac}\epvl{[c_b^\dagger c_b,-t_{ac}c_a^\dagger c_c]}\\
&=-\frac{1}{i}\sum_{ac}t_{ac}\epvl{c_a^\dagger[c_b^\dagger c_b,c_c]+[c_b^\dagger c_b,c_a^\dagger]c_c}\\
&=-\frac{1}{i}\sum_{ac}t_{ac}\epvl{c_a^\dagger(c_b^\dagger\{c_b,c_c\}-\{c_b^\dagger,c_c\}c_b)+(c_b^\dagger\{c_b,c_a^\dagger\}-\{c_b^\dagger,c_a^\dagger\}c_b)c_c}\\
&=-\frac{1}{i}\sum_{ac}t_{ac}\epvl{-\delta_{bc}c_a^\dagger c_b+\delta_{ba}c_b^\dagger c_c}
=-\frac{1}{i}\sum_a(-t_{ab}\epvl{c_a^\dagger c_b}+t_{ba}\epvl{c_b^\dagger c_a})\\
&=-\sum_a2\text{Im}(t_{ba}P_{ab}),
\end{aligned}
\end{equation}
implying the relation
\begin{equation}
j_{ab}=-2\text{Im}(t_{ba}s^0_{ab}).
\end{equation}
We have adopted a lowest-order approximation in line 3, where the higher-order term from the interaction is neglected. In fact, this turns out to be exact under the density-density interaction, since the higher-order term vanishes $\text{Im}(\epvl{c_a^\dagger c_b}\epvl{c_b^\dagger c_a})=\text{Im}|\epvl{c_a^\dagger c_b}|^2=0$. With the real hoppings $t_{ba}\in\mathbb R$, the charge bond currents are directly proportional to the imaginary part of the bond density $\text{Im}(s^0_{ab})$. The derivation of spin bond current is analogous. In the figures, we draw the site densities $s^\nu_{ii\tau\tau}$, real bond densities $\text{Re}(s^\nu_{ii'\tau\tau'})$, and imaginary bond densities $\text{Im}(s^\nu_{ii'\tau\tau'})$ to represent the S-C patterns.

The mean-field renormalized band structure can also be obtained. This is simply done by diagonalizing the Hartree-Fock Hamlitonian in the momentum space, where a proper (reduced) Brillouin zone is chosen to match the periodicity. When the periodicity is enlarged, the number of bands is increased in the reduced Brillouin zone. The original flat bands are renormalized and acquire light dispersion under the symmetry-breaking orders. The Berry curvature can be computed in the (reduced) Brillouin zone \cite{fukui05jpsp}, from which the band topology is determined.

\begin{figure}[t]
\centering
\includegraphics[scale=1]{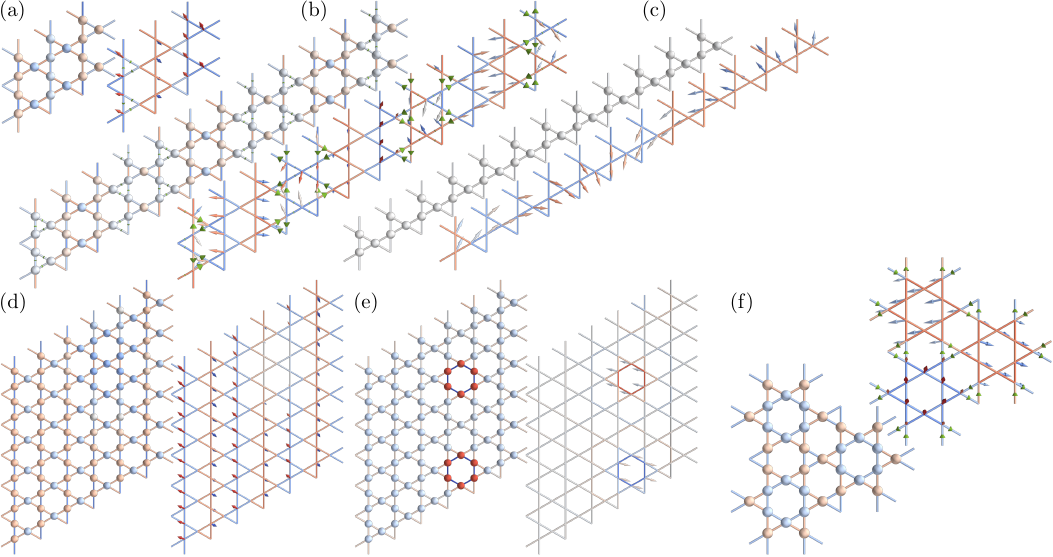}
\caption{\label{suppfig:scp} S-C patterns for some ground states. (a) S-C stripe order at $n_\text{FB}=-2/3$. (b) S-C stripe order at $n_\text{FB}=-3/4$. (c) SS order at $n_\text{FB}=-1$. (d) S-C stripe order at $n_\text{FB}=-1$. The nonuniformity occurs due to the metallic property. (e) DCLS order at $n_\text{FB}=-1$. (f) $120^\circ$SoDAFM at $n_\text{FB}=-1/2$.}
\end{figure}

\section{Miscellaneous ground states in pure Hubbard model}

In the main text, we have presented the primary symmetry-breaking orders of pure Hubbard model in kagome flat band. Here we list some more ground states that appear in the phase diagram (Fig.~\ref{suppfig:scp}). Note that the computation may obtain a nonuniform pattern if the ground state is metallic. This occurs especially when the Fermi level involves a flat segment, as is the case of the AFMSt-CStP1 order.

First, phase separation \cite{emery90prl,hanisch95annp,qin22jpcm} can appear between different phases in the doping phase diagram, where domains of both phases coexist. Second, coplanar spiral spin (SS) orders \cite{hanisch95annp,qin22jpcm} generally occur along the FM phase boundary. Third, we observe a wide dilute CLS (DCLS) phase at weak coupling. Increasing filling raises the CLS density, until it turns into the AFMSt-CStP2 order at $n_\text{FB}=-1/2$. Last, the hole-type DCLS occurs at $n_\text{FB}>1/2$ with macroscopic spin-configuration degeneracy. This phase corresponds to the ``paramagnetism" (PM) under the breakdown of global FM percolation \cite{mielke92jpa}.

On the other hand, we notice a $2\sqrt3\times2\sqrt3$ $120^\circ$ star-of-David (SoD) AFM at $n_\text{FB}=-1/2$ slightly above the ground state. The noncoplanar spin orders, including the TCTS order at $n_\text{FB}=-5/6$ and the THSV order at $n_\text{FB}=-2/3$, can be thought of as the intermediate deformations between the two $120^\circ$AFMs.

\section{Effect of extended repulsion on symmetry-breaking orders}

In this section, we discuss further the effect of extended repulsion on the energetic favorability of ground states.

\subsection{Phase diagram}

\begin{figure}[t]
\centering
\includegraphics[scale=1]{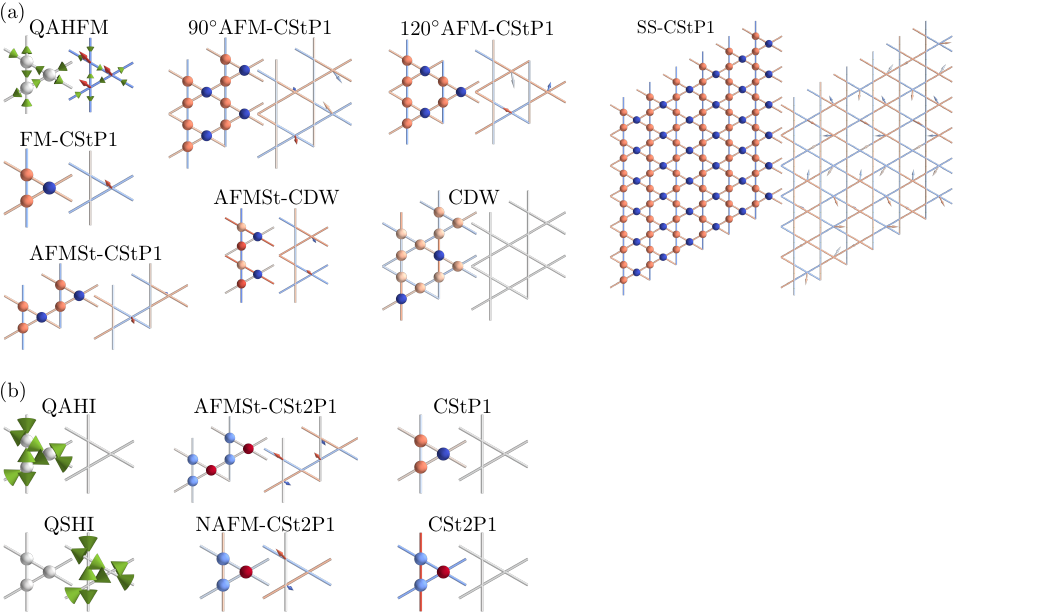}
\caption{\label{suppfig:scnfb0n1} S-C patterns of the ground states at (a) half filling $n_\text{FB}=0$ and (b) empty filling $n_\text{FB}=-1$.}
\end{figure}

To inspect the effects of longer-ranged Coulomb repulsion, we introduce the nearest- and second-neighbor repulsions $U_1=2U_2>0$ to the pure Hubbard model. These repulsions generally favor the charge orders, thereby driving the ground state through a series of phases with different S-C patterns (Fig.~\ref{suppfig:scnfb0n1}).

At half filling $n_\text{FB}=0$, the FM ground state is a QBCP half semimetal. The large-$U_0$ regime \cite{ren21prl} is described by an effective spinless model \cite{zhu16prl,ren18prb}. Under the extended repulsion, the time-reversal-symmetry-breaking loop currents appear as a weak-coupling instability \cite{sun09prl}. This gaps out the QBCP and leads to the QAHFM \cite{ren21prl}, which expands a wide regime of small extended repulsion $U_{1,2}\ll U_0$. At a larger extended repulsion, the charge nematicity induces the FM-CStP1 order. The darker and lighter regions in the phase diagram are gapless and gapped, respectively. The hatch region indicates a phase crossover from QAHFM. Further increasing the extended repulsion leads to a series $90^\circ$AFM $\rightarrow$ $120^\circ$AFM $\rightarrow$ SS $\rightarrow$ AFMSt away from the FM. Note that these magnetic orders occur on the lower-density triangular superlattice in the CStP1 order. The phase series may be related to the anisotropic Hubbard model \cite{szasz21prb} or certain equivalent spin models \cite{cookmeyer21prl} on the triangular lattice. On the other hand, there is an AFMSt-CDW order with a different stripe pattern, as well as a 5-higher-1-lower-density CDW matching the total fermion filling $n_f/2=5/6$.

The empty filling $n_\text{FB}=-1$ also presents a rich phase diagram, where additional phases occur under extended repulsion \cite{liu10prb,wen10prb}. In the large-$U_0$ regime, a large extended repulsion drives the AFM on a nematic CSt2P1 order. The AFM develops on the lower-density square superlattice, where both the N\'eel and stripe (NAFM and AFMSt) orders are observed. In the small-$U_0$ regime, we observe the degenerate quantum anomalous and spin Hall insulators (QAHI and QSHI). There also exists a CSt2P1 order, which transits to the CStP1 order at larger extended repulsion. Due to the occurrence of nonuniform patterns, it is difficult to accurately resolve the phase profile in the small-$U_0$ regime. For example, in the sketched darker blue region, our finite-size computation observes nonuniform S-C stripes but is unable to determine the ground state precisely. On the other hand, we observe an exact transition line $U_0=3U_1=6U_2$ between the CStP1 and NAFM-CSt2P1 phases.

\subsection{Competition of charge orders}

\begin{figure}[t]
\centering
\includegraphics[scale=1]{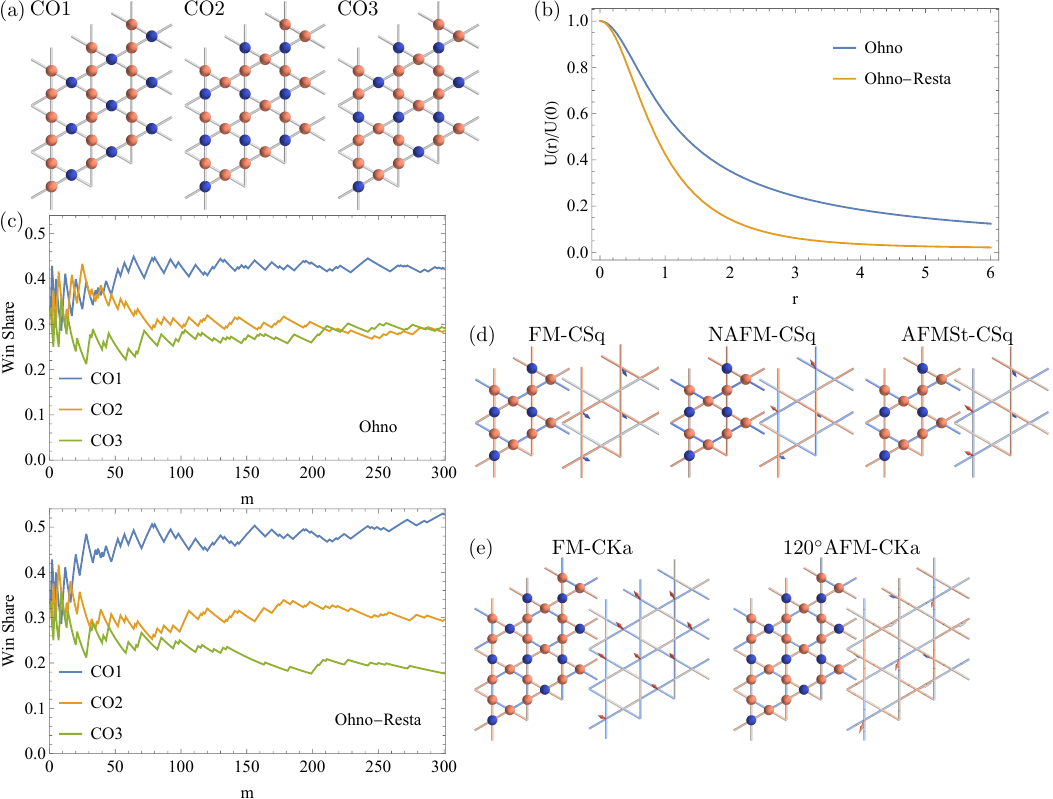}
\caption{\label{suppfig:cos} Effect of extended repulsion on the symmetry-breaking orders. (a) The three charge orders. (b) The scaling of the screened Coulomb repulsion. (c) The competition of the three charge orders under extended repulsion. At each $m$, we compute the interaction energy $E[n]$ up to the $m$-th neighbor. The win share $w_a/W$ of CO$a$ at $m$ counts the number of times it wins for any $\tilde m\leq m$, divided by the total number of counting $W=\sum_{a=1}^3w_a$. (d)(e) The symmetry-breaking orders on the square and kagome superlattices in dominant CO2 and CO3 at the half filling $n_\text{FB}=0$.}
\end{figure}

The primary effect of extended repulsion is the choice between three different types of charge orders (CO1,2,3) [Fig.~\ref{suppfig:cos}(a)]. Under onsite and nearest-neighbor repulsions, the three charge orders are degenerate. However, the degeneracy is lifted when further-neighbor repulsions are involved. For example, the second-, third-, fourth-neighbor repulsions favor CO1,2,3, respectively. This poses the question of how long we should allow the extended repulsion to reach in our computation. To see the interaction-range dependence of energetic favor among the three charge orders, we compute the interaction energy
\begin{equation}
E[n]=\sum_\tau\sum_{i'\tau'}U_{0i'\tau\tau'}n_{0\tau}n_{i'\tau'}
\end{equation}
in the three charge orders. Here the charge densities $n_{i\tau}$ are chosen as 0 and 1 for the lower- and higher-density sites, respectively. For the extended repulsion, we adopt the screening formula from the recent literature [Fig.~\ref{suppfig:cos}(b)] \cite{disante23prr}. The original Coulomb repulsion is inverse to the intersite distance $r$
\begin{equation}
U(r)\sim\frac{1}{r}.
\end{equation}
Under the local screening, the power-law scaling is screened by the typical extent of the Wannier functions $\delta$
\begin{equation}
\begin{aligned}
U_\text{Ohno}(r)\sim\frac{1}{\tilde r},\quad \tilde r=\sqrt{r^2+\delta^2}.
\end{aligned}
\end{equation}
The consideration of nonlocal screening further modifies the scaling to
\begin{equation}
U_\text{Ohno-Resta}(r)\sim
\begin{cases}
1/\frac{q_\text{TF}R}{\sinh[q_\text{TF}(R-\tilde r)]+q_\text{TF}\tilde r}\tilde r,&\tilde r<R,\\
1/\tilde r,&\tilde r\geq R.
\end{cases}
\end{equation}
Here $R$ is the ionic radius and $q_\text{TF}$ is the Thomas-Fermi wave vector. Note that the local-screening model only works for the large-gap insulators or semiconductors. Meanwhile, the nonlocal-screening model can describe the small-gap systems and metals by setting a large ionic radius. We compute the screened Coulomb potential in both scenarios and apply them to the interaction energy of the three charge orders [Fig.~\ref{suppfig:cos}(c)]. Here we adopt the parameters $\delta=0.75\text{\r{A}}$, $R=6\text{\r{A}}$, and $q_\text{TF}=0.75\text{\r{A}}^{-1}$, and the lattice constant $a=2.5\text{\r{A}}$. The results show that CO1 is the most frequent winner under the competition. Therefore, we admit the extended repulsion up to second neighbor in our computation, which selects CO1 among the three charge orders.

It is still interesting to examine what symmetry-breaking orders can appear under CO2 and CO3. The change of dominant charge order may affect the phase diagram significantly, especially at the half filling $n_\text{FB}=0$. As discussed in the main text, the dominant charge order determines the effective lower-density superlattice for the spin orders. We have explored the spin orders on the triangular superlattice from CO1. Meanwhile, CO2 and CO3 induce the square and kagome superlattices, respectively, which allow more possible symmetry-breaking orders to occur. Under the dominant CO2, the square superlattice (CSq) hosts the FM, NAFM, and AFMSt orders [Figs.~\ref{suppfig:cos}(d)]. On the other hand, the kagome superlattice in the dominant CO3 (CKa) can support the FM and $120^\circ$AFM [Figs.~\ref{suppfig:cos}(e)].

\section{Mean-field renormalized band structure and band topology}

\begin{figure}[t]
\centering
\includegraphics[scale=1]{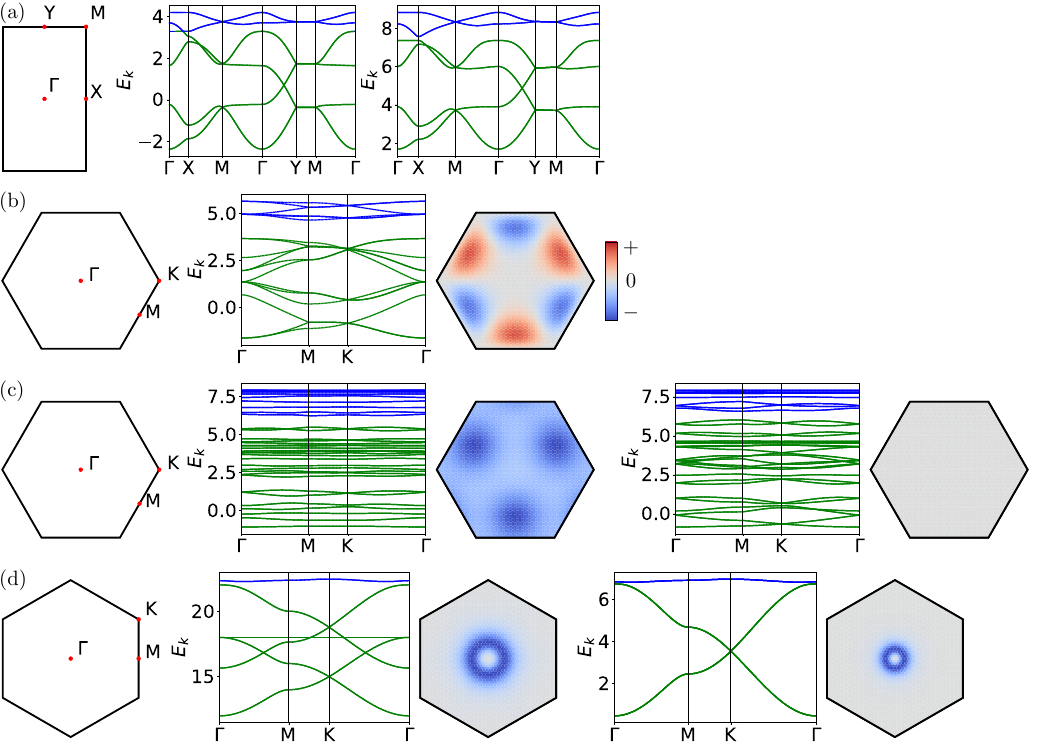}
\caption{\label{suppfig:hf} Mean-field renormalized band structures under symmetry-breaking orders. Here we present the (reduced) Brillouin zones and the band structures along the closed loops. The distribution of Berry curvature is further mapped out in the (reduced) Brillouin zone for nontrivial orders, whose sign and strength are indicated by a color bar. (a) AFMSt-CStP1 order at empty filling $n_\text{FB}=-1$ under (left) $U_0=2.5$ and (right) $U_0=3$, $U_1=2U_2=0.5$. (b) $120^\circ$TAFM with $C=0$ at empty filling $n_\text{FB}=-1$ under $U_0=4$. (c) Noncoplanar spin orders, including (left) TCTS order with $C=-1$ at filling $n_\text{FB}=-5/6$ and (right) THSV order with $C=0$ at filling $n_\text{FB}=-2/3$ under $U_0=6$. (d) Topological insulators under extended repulsion, including (left) QAHFM with $C=-1$ at half filling $n_\text{FB}=0$ under $U_0=12$, $U_1=2U_2=0.8$ and (right) QAHI with $C=-2$ at empty filling $n_\text{FB}=-1$ under $U_0=1$, $U_1=2U_2=0.5$. Note that the QAHI is degenerate with a QSHI with completely vanishing Berry curvature and $C=0$.}
\end{figure}

While the FM band structures have been presented in the main text, here we show the band structures for a few more orders (Figs.~\ref{suppfig:hf}). First, for the AFMSt-CStP1 order at empty filling $n_\text{FB}=-1$ [Figs.~\ref{suppfig:hf}(a)], the $1\times2$ periodicity leads to the $1\times1/2$ reduced Brillouin zone with a doubled number of bands. There is a band crossing on the $\Gamma$-$X$ flat segment of the Fermi surface, which is gapped under larger repulsion. Second, the $120^\circ$TAFM at empty filling $n_\text{FB}=-1$ [Figs.~\ref{suppfig:hf}(b)] has the $\sqrt3\times\sqrt3$ periodicity. The number of bands is tripled in the $1/\sqrt3\times1/\sqrt3$ reduced Brillouin zone. Although the occupied bands carry nonzero Berry curvatures, the Chern number vanishes $C=0$, consistent with the zero scalar chirality under coplanar spin orders \cite{ohgushi00prb,taguchi01sc,martin08prl,barros14prb}. Alternatively, the zero Chern number can be understood from an effective time-reversal symmetry, which is a combination of time-reversal and spin-$\text{C}_3$-rotation symmetries. Third, noncoplanar spin orders occur at fractional fillings [Figs.~\ref{suppfig:hf}(c)]. For the TCTS order at filling $n_\text{FB}=-5/6$ and THSV order at filling $n_\text{FB}=-2/3$, the periodicity becomes $2\sqrt3\times2\sqrt3$. The number of bands is multiplied by 12 in the $1/2\sqrt3\times1/2\sqrt3$ reduced Brillouin zone. For the TCTS order, the Berry curvature distribution leads to the Chern number $C=\pm1$. This nontrivial band topology arises from the noncoplanar spin structures, where nonzero scalar chirality appears under time-reversal symmetry breaking \cite{ohgushi00prb,taguchi01sc,martin08prl,barros14prb}. Meanwhile, the Berry curvature completely vanishes for the THSV order, and the Chern number is $C=0$. This trivial band topology results from the `reality' under combined parity-time-reversal $PT$ symmetry, which can be observed by rotating the lattice with respect to a hexagon center and reversing the spins. Finally, topological insulators can appear under extended repulsion [Figs.~\ref{suppfig:hf}(d)]. For the QAHFM at half filling $n_\text{FB}=0$, the QBCP is gapped only for the minority spin, since the loop currents only develop in this branch. The Berry curvature distribution leads to the Chern number $C=\pm1$. Meanwhile, the degenerate QAHI and QSHI can develop at empty filling $n_\text{FB}=-1$. For QAHI (QSHI), the same (opposite) Berry curvature distributions in the two spin branches lead to the Chern number $C=\pm2$ ($0$).

\section{Mean field ansatze from projective symmetry group (PSG)}

\subsection{Symmetry, convention and PSG}

\begin{figure}[!thb]
\centering
\includegraphics[width=0.6\textwidth]{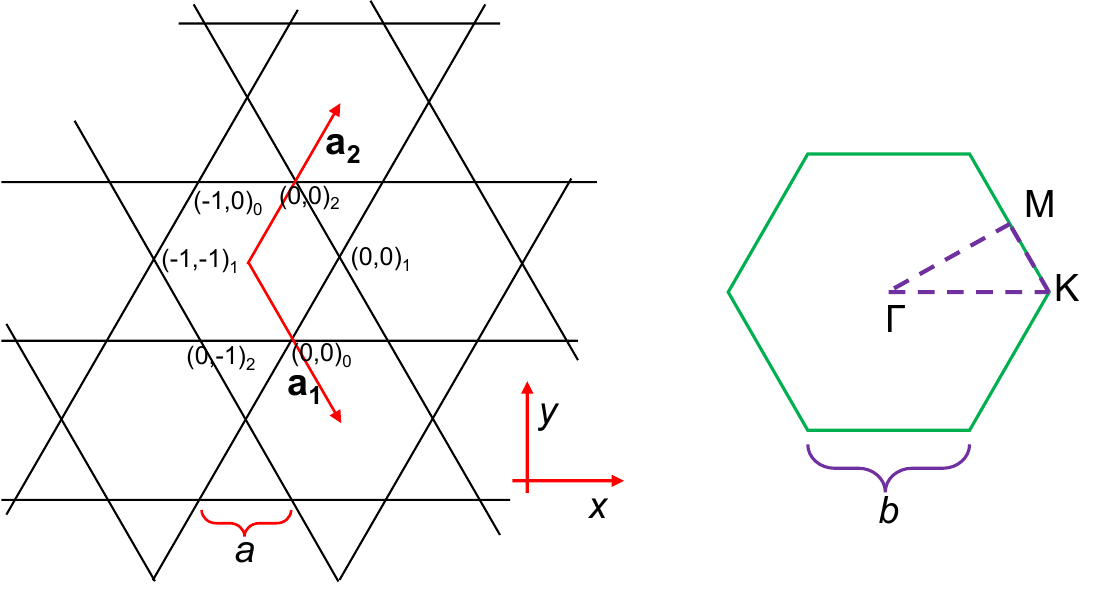}
\caption{The kagome lattice, coordinates, and its Brillouin zone used for the PSG mean field ansatze. Note we have $b = \frac{2\pi}{3a}$.}\label{psg_lattice_setup}
\end{figure}

We set up the coordinate convention for the kagome lattice, see \ref{psg_lattice_setup}. We denote the length of the hexagon edge as $a$. We set up the coordinate system in the following way: the origin is at a hexagon center; the unit translations are $\bm{a}_1 = a\hat{x}-\sqrt{3} a \hat{y}$ and $\bm{a}_2 = a\hat{x}+\sqrt{3} a \hat{y}$. The reciprocal lattice vectors are then $\bm{b}_1 = \pi(1,1/\sqrt{3})$, $\bm{b}_2 = \pi(-1,1/\sqrt{3})$. The three sublattices $\tau=0,1,2$ have their origin at $(0,0,0)_0 = \bm{a}_1/2$, $(0,0,0)_1 = (\bm{a}_1+\bm{a}_2)/2$ and $(0,0,0)_2 = \bm{a}_2/2$. The symmetry group has generators $\{T_1,T_2,D,C_6\}$, where 
\begin{subequations}
\begin{align}
T_1\colon(r_1,r_2)_\tau&\rightarrow (r_1+1,r_2)_\tau\\
T_2\colon (r_1,r_2)_\tau&\rightarrow (r_1,r_2+1)_\tau\\
D \colon (r_1,r_2)_\tau&\rightarrow (r_2,r_1)_{2-\tau}\\
{C}_6\colon (r_1,r_2)_\tau&\rightarrow(
r1-r2-\delta_{\tau=2},r_1)_{\tau+1}
\end{align}
\end{subequations}
The group relations are  $T_1T_2T_1^{-1}T_2^{-1}=1$, ${C}^{-1}_6T_1{C}_6T_2=1$, ${C}^{-1}_6T_2{C}_6T_1^{-1}T_2^{-1}=1$, $DT_1DT_2^{-1}=1$, $DT_2DT_1^{-1}=1$, $({C}^{-1}_6D)^2=1$,  ${C}_6^6 =D^2=1$. 

The projective symmetry group (PSG) classification for $\mathbb{Z}_2$ spin liquids using Schwinger bosons on the kagome lattice has been done by Wang and Vishwanath in Ref.~\cite{PhysRevB.74.174423}. We summarize their results:
\begin{equation}\label{psgk}
\begin{aligned}
\phi_{T_1}(\vec{r}_\tau)&=0,\\
\phi_{T_2}(\vec{r}_\tau) &=n_1r_1\pi,\\
\phi_{{C}_6}(\vec{r}_\tau) &= \frac{n_{{C}_6}}{2}\pi
+n_1\left[r_1r_2+\frac{(r_1-1)r_1}{2}\right]\pi,\\
\phi_D(\vec{r}_\tau) &= \frac{n_D}{2}\pi + n_1 r_1r_2\pi,\\
\phi_{\mathcal{T}}(\vec{r}_\tau) & = 0,
\end{aligned}
\end{equation}
where the phase $\phi_{\mathcal{O}}(\vec{r}_\tau)$ for a symmetry operation $\mathcal{O}\in \{T_1,T_2,D,C_6\}$ appears in the projective symmetry transformation of Schwinger bosons:
\begin{equation}
\widetilde{\mathcal{O}}\colon
b_{\vec{r}_\tau}\rightarrow
e^{i\phi_{\mathcal{O}}(\mathcal{O}(\vec{r}_\tau))} U^\dag_{\mathcal{O}} b_{\mathcal{O}(\vec{r}_\tau)},
\end{equation}
here $b_{\vec{r}_\tau} = \Big(\begin{smallmatrix} b_{\vec{r}_\tau,\uparrow}\\b_{\vec{r}_\tau,\downarrow}\end{smallmatrix}\Big)$ is the two-component Schwinger boson annihilation operator,
$U_{\mathcal{O}}$ is the unitary (anti-unitary) matrix associated to a spatial (time-reversal) symmetry operation $\mathcal{O}$ that acts on the spin operator: $\mathcal{O}\colon \bm{S}\rightarrow U_{\mathcal{O}}\bm{S}U_{\mathcal{O}}^\dag$ or in the case of time-reversal operation $\mathcal{T}$, we have $\mathcal{T}\colon \bm{S}\rightarrow \mathcal{K}U_{\mathcal{T}}\bm{S}U_{\mathcal{T}}^\dag\mathcal{K}$ with $\mathcal{K}$ the complex conjugation operator.

\subsection{PSG to parton mean field ansatze}

The mean-field Hamiltonian symmetric under the projective operations $\widetilde{\mathcal{O}}$ is called a mean-field ansatz. Write the most general form of this Hamiltonian as
\begin{equation}
H = \sum_{i,j}
b^\dag_i u^h_{ij} b_j + b^\dag_i u^p_{ij} (b^\dag_j)^T +h.c.,
\end{equation}
where $i,j$ denote lattice sites that include both the unit cell and sublattice indices.
We assume the $(01)$ bond (see Fig.~\ref{psg_lattice_setup} for its definition) to be 
\begin{subequations}\label{uhup}
\begin{eqnarray}
u^h_{01} &=&  a\sigma^0 + i(b\sigma^1+c \sigma^2+d\sigma^3),\\
u^p_{01} &=& \left(a'\sigma^0 + i (b'\sigma^1+c'\sigma^2+d'\sigma^3)\right)\cdot i\sigma^2,
\end{eqnarray}
\end{subequations}
where $a,b,c,d$ and $a',b',c',d'$ are real due to time-reversal symmetry $\mathcal{T}$.

The order-two operation $D$ has well defined operation on the lattice sites, however its spin rotation has yet to be defined. To do this, we assume the kagome lattice  is a 2D reduction of a layered kagome lattice having space group P6/mmm and point group $D_{6h}\cong Dih_3\times \mathbb{Z}_2^2$. The P6/mmm is the symmetry for many kagome lattices, including the kagome compounds $A$V$_3$Sb$_5$ with $A=$ K, Rb, and Cs \cite{jiang21nm,zhao21n}.
The point group symmetry contains a six-fold axis $C_6$ along $z$ that goes through the hexagon center (the origin), a two-fold axis that is parallel to the origin and $(0,0,0)_{0,1,2}$, and also a mirror in the plane spanned by $\hat{z}$ and the origin and $(0,0,0)_{0,1,2}$. Since the V atom sits on the 3g Wyckoff position whose location is $(1/2,0,1/2)$, $(0,1/2,1/2)$ and $(1/2,1/2,1/2)$, the symmetry $D$ should be interpreted as a mirror reflection and not two-fold rotation (the two-fold rotation would map between layers at $z=1/2$ and $z=-1/2$ so it is not a symmetry of a single kagome layer). 

Now we write done the constraints on the bond parameters from the projective symmetry transformation. For nearest-neighbor (NN) bonds: using the fact that the $01$ bond is mapped back to itself under ${C}^{-1}_6D$, we get
\begin{subequations}
\begin{align}
\left(a,-\left(-\frac{b}{2}+\frac{\sqrt{3} c}{2}\right),-\left(\frac{\sqrt{3} b}{2}+\frac{c}{2}\right),-(-d)\right) &= (a,b,c,d),\\
(-1)^{n_{{C}_6}+n_D}\left(-a',-\frac{b'}{2}+\frac{\sqrt{3} c'}{2},\frac{\sqrt{3}b'}{2}+\frac{c'}{2},-d'\right)
&=(a',b',c',d'),
\end{align}
\end{subequations}

For next-nearest-neighbor (NNN) bonds: there are six independent bonds not related by translations. Setting the bond $(0,0)_2\leftrightarrow (0,0)_0$ to have the bond parameters $(A,B,C,D)$ and $(A',B',C',D')$, this bond is mapped back (to its hermitian conjugate) by $D$. We have the condition
\begin{subequations}
\begin{align}
(A,B,C,D) &= (A,B,-C,D),\\
(A',B',C',D') & = (-1)^{n_D}(-A',-B',C',-D').
\end{align}
\end{subequations}

These are obtained from the transformation rule for the bonds. The solutions are given in Table \ref{table:hoppair}.

\begin{table}[!thb]
\centering
\caption{Hopping and pairing parameters. The onsite bond is for the site $(0,0,0)_1$, and parameters shown are the \emph{allowed} ones; the NN bond is for the bond $(0,0,0)_1\leftarrow (0,0,0)_0$, and the relations shown are the \emph{constraints}; the NNN bond is for the bond $(0,0,0)_2\leftarrow (0,0,0)_0$ and the relations shown are the \emph{constraints}.}\label{table:hoppair}
\begin{tabular}{c|c|l|l}
\hline\hline
$n_1\pi$-$(n_{{C}_6},n_D)$ & Onsite & NN & NNN\\
\hline
0-(0,0)&$\mu$,$-$&$c=-\frac{b}{\sqrt{3}}$, $a'=0,c'=\sqrt{3}b',d'=0$&$C=0$, $A'=B'=D'=0$\\
0-(0,1)&$\mu$,$\delta'$&$c=-\frac{b}{\sqrt{3}}$, $c'=-\frac{b'}{\sqrt{3}}$&$C=0$, $C'=0$\\
0-(1,0)&$\mu$,$\gamma'$&$c=-\frac{b}{\sqrt{3}}$, $c'=-\frac{b'}{\sqrt{3}}$&$C=0$, $A'=B'=D'=0$\\
0-(1,1)&$\mu$,$\beta'$&$c=-\frac{b}{\sqrt{3}}$, $a'=0,c'=\sqrt{3}b',d'=0$&$C=0$, $C'=0$\\
$\pi$-(0,0)&$\mu$,$-$&$c=-\frac{b}{\sqrt{3}}$, $a'=0,c'=\sqrt{3}b',d'=0$&$C=0$, $A'=B'=D'=0$\\
$\pi$-(0,1)&$\mu$,$\delta'$&$c=-\frac{b}{\sqrt{3}}$, $c'=-\frac{b'}{\sqrt{3}}$&$C=0$, $C'=0$\\
$\pi$-(1,0)&$\mu$,$\gamma'$&$c=-\frac{b}{\sqrt{3}}$, $c'=-\frac{b'}{\sqrt{3}}$&$C=0$, $A'=B'=D'=0$\\
$\pi$-(1,1)&$\mu$,$\beta'$&$c=-\frac{b}{\sqrt{3}}$, $a'=0,c'= \sqrt{3}b',d'=0$&$C=0$, $C'=0$\\
\hline
\hline
\end{tabular}
\end{table}

Using this information one can write down the tight binding model for the partons symmetric under PSG. 
In the following we restrict ourselves to the zero-flux states, i.e. $n_1=0$.

Recall that the bosonic Bogoliubov--de Gennes (BdG) Hamiltonian are of the generic form
\begin{equation}\label{Hh}
H = \sum_{\bm{k}} B^\dag_{\bm{k}}\mathcal{H}(\bm{k}) B_{\bm{k}},
\end{equation}
where we used the basis $B_{\bm{k}} = (b_{\bm{k},0},b_{\bm{k},1},b_{\bm{k},2},b^\dag_{-\bm{k},0},b^\dag_{-\bm{k},1},b^\dag_{-\bm{k},2})^T$, where each $b_{\bm{k},\tau}$ has two components labeled by spin $\sigma=\uparrow,\downarrow$. The Hamiltonian matrix $\mathcal{H}(\bm{k})$ reads

\begin{equation}\label{Hamiltonianden}
\mathcal{H}(\bm{k}) = \left(\begin{array}{cc} H_h(\bm{k}) & H_p(\bm{k})\\
H^\dag_p(\bm{k}) & H^T_h(-\bm{k})\end{array}\right),
\end{equation}
where $H_h(\bm{k}) = H^\dag_h(\bm{k})$ and $H_p(\bm{k}) = H^T_p(-\bm{k})$. 

We have, for $n_1=0$,
\begin{subequations}
\begin{align}
U^\dag_{{C}_6}\mathcal{H}(\bm{k}) U_{{C}_6}&=
\mathcal{H}({C}_6(\bm{k})),\\
U^\dag_{D}\mathcal{H}(\bm{k}) U_{D}&=
\mathcal{H}(D(\bm{k})),\\
U^\dag_{\mathcal{T}} \mathcal{H}(\bm{k}) U_{\mathcal{T}}&=\mathcal{H}^*(-\bm{k}),\\
U^\dag_{\mathcal{C}}\mathcal{H}(\bm{k}) U_{\mathcal{C}}&=\mathcal{H}^*(-\bm{k}),
\end{align}
\end{subequations}
where $C_6(\bm{k}) = ( \frac{1}{2}k_x -\frac{\sqrt{3}}{2}k_y,\frac{\sqrt{3}}{2}k_x + \frac{1}{2}k_y)$ and $D(\bm{k}) = (k_x,-k_y)$, and
\begin{equation}
U_{{C}_6} = \left(\begin{array}{cc} e^{-i\frac{\pi}{2} n_{{C}_6}} V_{{C}_6}&0\\0& e^{i\frac{\pi}{2} n_{{C}_6}} V^*_{{C}_6}\end{array}\right),\qquad
U_{D} = \left(\begin{array}{cc} e^{i\frac{\pi}{2} n_{D}} V_{D}&0\\0& e^{-i\frac{\pi}{2} n_{D}} V^*_{D}\end{array}\right),
\end{equation}
with
\begin{equation}
V_{{C}_6}= \left(\begin{array}{ccc}&1&\\&&1\\1&&\end{array}\right)
\otimes \left(e^{-i\frac{\pi}{6} \sigma^3}\right)^\dag,\qquad
V_{D}= \left(\begin{array}{ccc}&&1\\&1&\\1&&\end{array}\right)
\otimes \left(e^{-i\frac{\pi}{2} \sigma^2}\right)^\dag,
\end{equation}
and for the time reversal symmetry and bosonic particle-hole symmetry we have
\begin{equation}
U_{\mathcal{T}} = 1_{6\times 6}\otimes i \sigma^2,\qquad
U_{\mathcal{C}} = \left(\begin{array}{cc} & 1_{6\times 6}\\1_{6\times 6}\end{array}\right).
\end{equation}

Since there are too many parameters, let's restrict to only nonzero singlet terms and set all the triplet terms to be zero: $\beta=\gamma=\delta=\beta'=\gamma'=\delta'=b=c=d=b'=c'=d'=0$. Under this simplification, the four $0$-flux ansatze reduce to just two: $0-(0,1)$ and $0-(1,0)$ (the $0-(0,0)$ and $0-(1,1)$ classes are included in them). 

As an example, let us present the class $0-(0,0)$. This class reduces to a U(1) PSG class when restricting to a spin-singlet ansatz, i.e. $H_p(\bm{k}) = 0$, and we have the hopping part
\begin{equation}\label{hopping_00}
H_h(\bm{k})=
\begin{pmatrix}
 \mu  & a \cos \frac{k_x+\sqrt{3} k_y}{2}+ A \cos \frac{3k_x-\sqrt{3} k_y}{2} & a \cos k_x+A\cos \sqrt{3} k_y \\
 a \cos \frac{k_x+\sqrt{3} k_y}{2}+A \cos\frac{3 k_x - \sqrt{3}k_y}{2} & \mu  & a \cos\frac{k_x-\sqrt{3} k_y}{2}+A\cos \frac{3k_x+\sqrt{3}k_y}{2} \\
 a \cos k_x +A\cos \sqrt{3}k_y& a \cos \frac{k_x-\sqrt{3} k_y}{2}+A \cos \frac{3k_x+\sqrt{3}k_y}{2} & \mu  \\
\end{pmatrix}\otimes \sigma^0.
\end{equation}
Note that the NN hopping part of the Hamiltonians for the $0-(0,1)$ and $0-(1,0)$ classes, $H_h(\bm{k})$ is identical to the one in Eq.~\eqref{hopping_00}. Below we present the pairing part of the Hamiltonian. For the 0-(0,1) class,  the pairing part is
\begin{equation}
\begin{aligned}
&H_p(\bm{k})=\\
& \begin{pmatrix}
 0 & a' \cos \frac{k_x+\sqrt{3} k_y}{2} -A'  \cos \frac{3k_x-\sqrt{3}k_y}{2}& -a'\cos k_x+A'\cos \sqrt{3}k_y \\
 -a'\cos \frac{k_x+\sqrt{3}k_y}{2} + A' \cos \frac{3k_x-\sqrt{3}k_y}{2} & 0 & a' \cos \frac{k_x-\sqrt{3} k_y}{2}-A' \cos \frac{3k_x+\sqrt{3}k_y}{2} \\
 a' \cos k_x-A'\cos \sqrt{3}k_y & -a'\cos \frac{k_x-\sqrt{3}k_y}{2}+A'\cos \frac{3k_x+\sqrt{3}k_y}{2} & 0 \\
\end{pmatrix}\otimes i \sigma^2.
\end{aligned}
\end{equation}

For the class 0-(1,0),  the pairing part is
\begin{equation}
H_p(\bm{k}) = 
\begin{pmatrix}
 0 & ia' \sin \frac{k_x+\sqrt{3} k_y}{2} & -ia'\sin k_x \\
 ia'\sin \frac{k_x+\sqrt{3}k_y}{2} & 0 & ia' \sin \frac{k_x-\sqrt{3} k_y}{2} \\
 -ia' \sin k_x & ia'\sin \frac{k_x-\sqrt{3}k_y}{2} & 0 \\
\end{pmatrix}\otimes i \sigma^2.
\end{equation}

\begin{figure}
\centering
\includegraphics[width=0.85\textwidth]{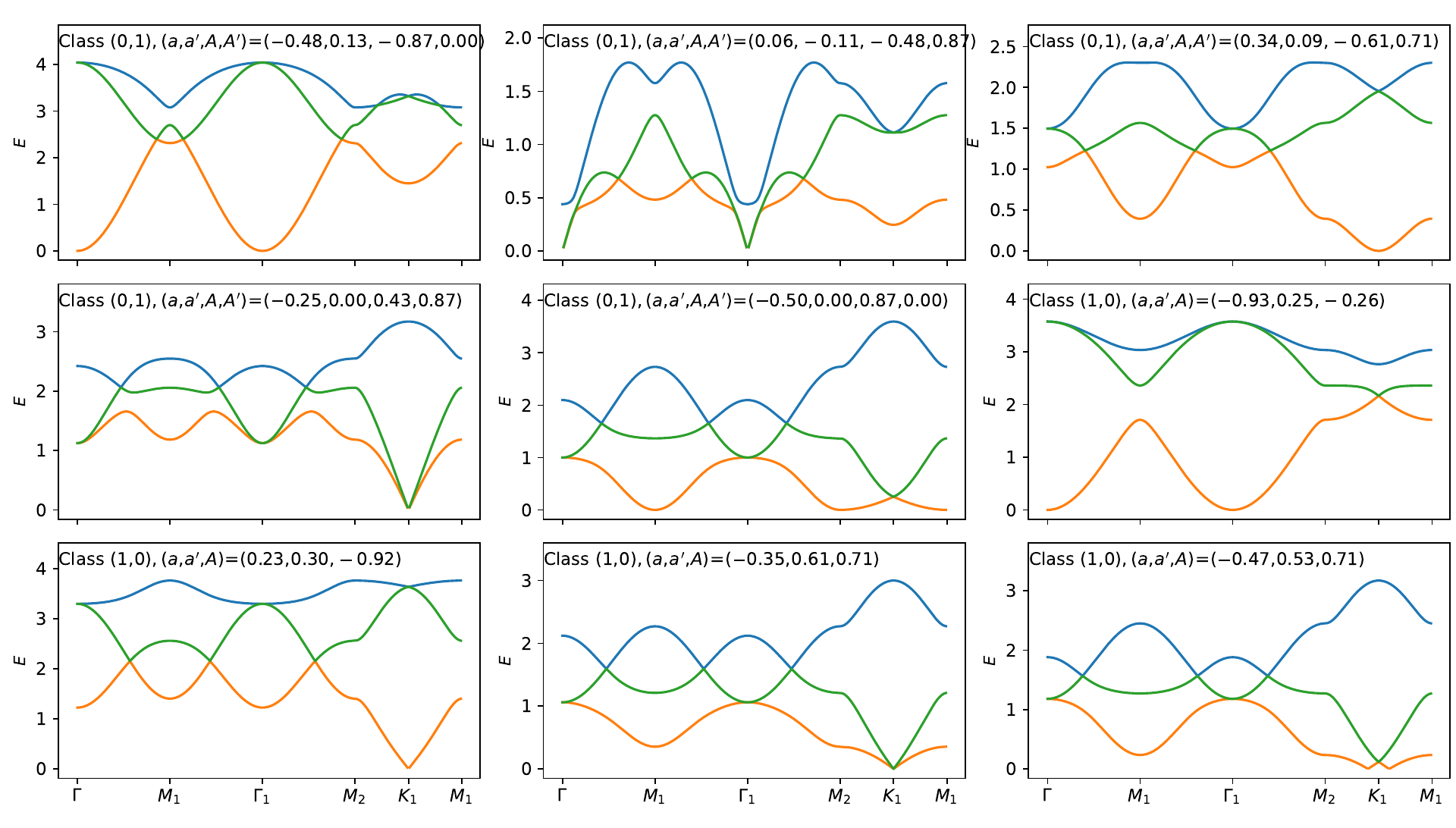}
\caption{Bosonic spinon dispersion for PSG classes $(n_{{C}_6},n_D) = (0,1)$ cases 1-5 and $(1,0)$ case 1-3. For $(1,0)$ case 3, two parameter values are shown: one with double linear dispersion at $\mathrm{K}$ and the other has linear dispersion along the $\overline{\mathrm{MK}}$ line. In all cases, chemical potential $\mu$ is tuned to critical value such that the spinons band becomes gapless. The gapless band is either linear ($z=1$) or quadratic ($z=2$), which is related to the diagonalizability of the bosonic BdG Hamiltonian (see analysis in Sec.~\ref{supp:psg:case_4}).}\label{dispersionsPSG}
\end{figure}

\begin{figure}
\centering
\includegraphics[width=0.8\textwidth]{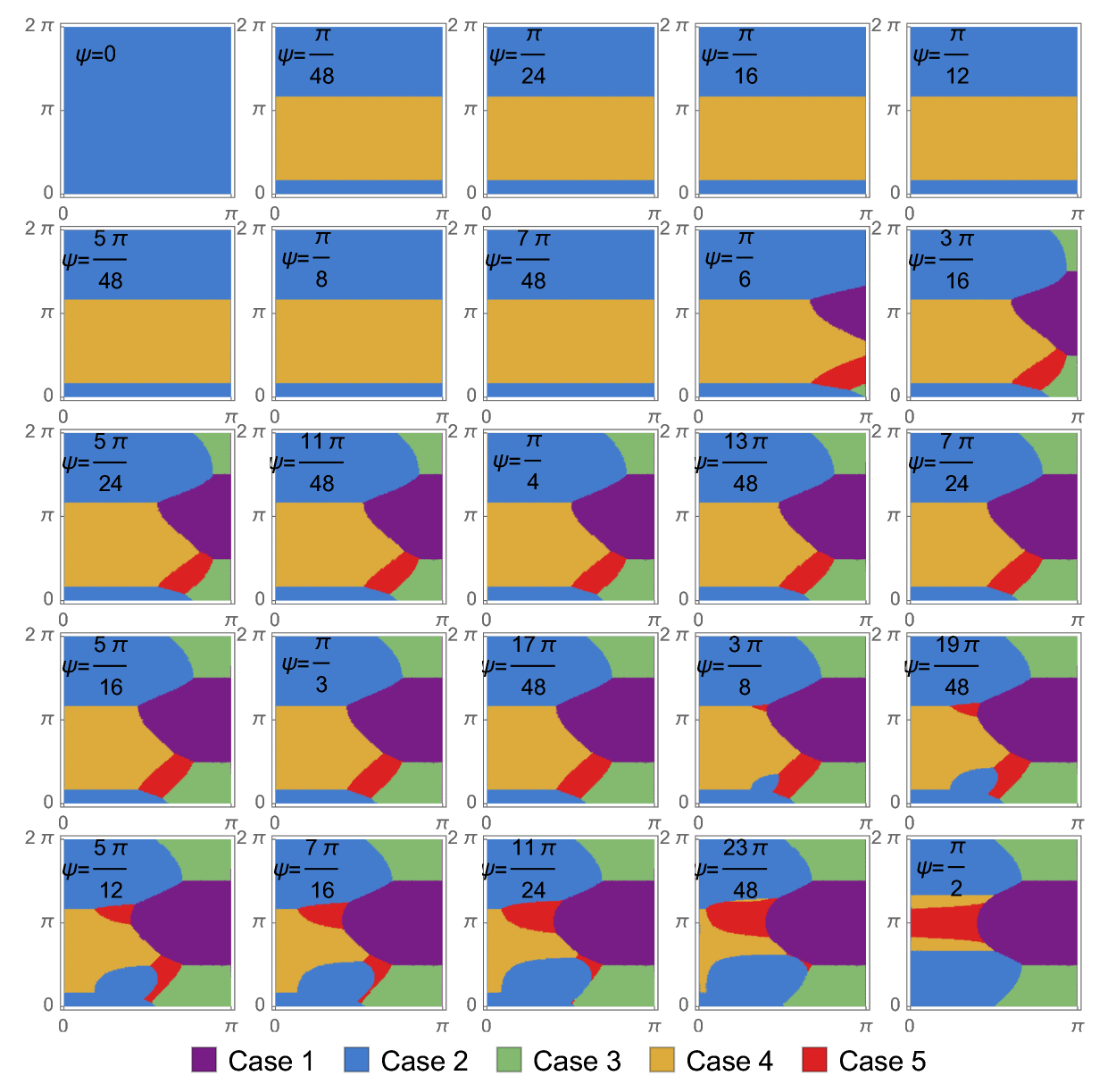}
\caption{Magnetic phase diagram for the spin-singlet ansatz of the PSG class $(n_{{C}_6},n_D) =(0,1)$ up to NNN. In each plot the horizontal (vertical) axis is for the parameter $\theta$ ($\phi$). The relation between $\theta, \phi,\psi$ and the hopping and pairing parameters $a,a',A,A'$ is given in Eq.~\eqref{class01params}.}\label{fig:phase_diagram_01_NNN}
\end{figure}

\begin{figure}
\centering
\includegraphics[width=0.6\textwidth]{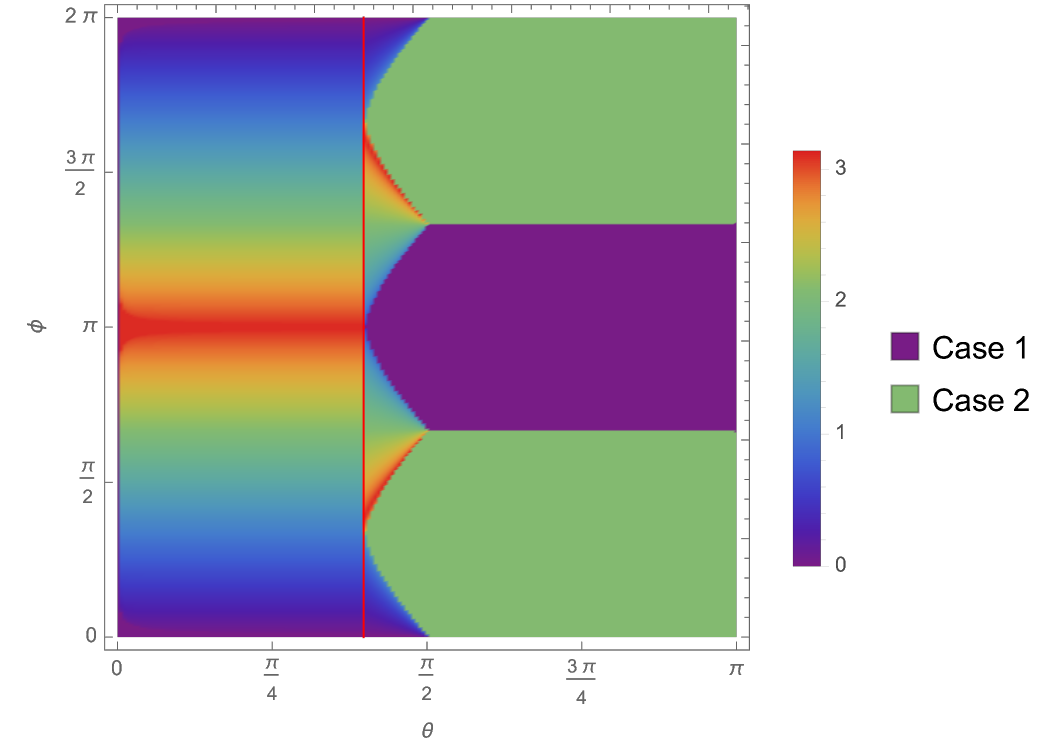}
\caption{Magnetic phase diagram for the spin-singlet ansatz of the PSG class $(n_{{C}_6},n_D) =(1,0)$ up to NNN. The color represents the norm of the condensation momentum $|\bm{k}|$ which distinguishes cases 1-5: purple represents case 1 ($\Gamma$ condensation), green represents case 2 ($\mathrm{K}$ condensation), and the colorbar represents cases 3-5 which support incommensurate spiral orders (continuous condensation momenta along $\Gamma\rightarrow\mathrm{K}\rightarrow\mathrm{M}$, with red representing $\mathrm{M}$). The relation between the parameterization $(\theta,\phi)$ and the mean field parameters $a,a',A$ is given in Eq.~\eqref{class10params}; the red vertical line is $\theta= \tan^{-1}3$.}\label{fig:phase_diagram_10_NNN}
\end{figure}

Below we will analyze the magnetic orders that border these spin liquids. To study them, we bring down the bosonic spinon band by tuning the chemical potential until the bosonic spinons at the band bottom start to condense. The condensation order parameter for the bosonic spinons will determine the type of magnetic order that borders the spin liquid phase. Our focus will be on the $0-(0,1)$ and $0-(1,0)$ classes. 
The phase diagrams 
are shown in Figs.~\ref{fig:phase_diagram_01_NNN} and \ref{fig:phase_diagram_10_NNN}. The magnetic orders that appear in the phase diagram are summarized in Table \ref{Tab:mag_order_psg} in the main text. The spinon dispersion for representative ansatze are given in Fig.~\ref{dispersionsPSG}.

\section{Magnetic phase diagram from parton mean field ansatze}

Before diving into the detailed analysis of the magnetic orders, we first present two different interpretations and treatments possible for the parton mean field ansatze.

In the usual treatment of Schwinger boson mean field theory for spin liquids, one assumes a spin model (typically of Heisenberg type with full SO(3) spin symmetry) whose mean-field decomposition gives the Schwinger boson mean field Hamiltonian. This has been done for the kagome $J_1$-$J_2$ model, see Refs.~\cite{messio12prl,rossi23prb}. Here, instead of following this treatment, we interpret the Schwinger boson mean field Hamiltonian as an \emph{effective} Hamiltonian for spinons. We do not specify the underlying spin Hamiltonian, and do not require the spinon Hamiltonian to be self-consistently determined; rather, the single spin occupancy is to be enforced by applying Gutzwiller projection to the spinon states, and we make the crucial assumption that the magnetic order and spin liquid phases of the unprojected states persist after the projection. In this treatment, the spinon hoppings and pairings are taken as free parameters, and we focus on the magnetic phase diagram spanned by these parameters, which is obtained by bring down the spinon gap to zero at each point in this hopping and pairing parameter phase space. We note that this treatment allows us to discover many more magnetic orders from the Schwinger boson mean field theory than the self-consistent treatment used in previous works \cite{messio12prl,rossi23prb}, and further reveal a distinction between the two $\mathbb{Z}_2$ spin liquid classes --- the $0-(0,1)$ class and the $0-(1,0)$ class, through the magnetic orders that border the spin liquid phase.

\subsection{Class $0-(0,1)$}

We start from a large chemical potential $\mu$ such that the bosonic parton spectrum is gapped, and then lower $\mu$ until the spectrum becomes gapless and the bosonic parton begins to condense. 

If we only consider NN bonds, we find that $\Gamma=(0,0)$ in the Brillouin zone is always among the condensation momenta. At fine tuned parameter points (e.g. $(a,a')=(1,0)$ or $(a,a') = (0,1)$) there are other condensation momenta (the condensation manifold is the whole Brillouin zone for the former and the line $k_x=0$, $k_x = \pm \sqrt{3}k_y$ for the latter). The critical chemical potential is 
\begin{equation}\label{case1case2}
\mu = \max (a+\sqrt{3}|a'|,-2a) = \left\{\begin{array}{ll}-2a,&a <  -\frac{|a'|}{\sqrt{3}},\\a+\sqrt{3}|a'|,&a> -\frac{|a'|}{\sqrt{3}}\end{array}\right..
\end{equation}

If we include NNN bonds, we find the condensations can be $\Gamma$, $\pm\mathrm{K}$, and $\mathrm{M}_{1,2,3}$. The critical chemical potential is
\begin{equation}
\mu = \max\Big(\underbrace{-2(a+A),a+A+\sqrt{3}|a'-A'|}_{\Gamma},
\underbrace{a-2A,\frac{1}{2}(-a+2A+\sqrt{3}|a'+2A'|)}_{\pm\mathrm{K}},
\underbrace{\sqrt{(a-A)^2+(a'+A')^2}}_{\mathrm{M}_{1,2,3}}\Big),
\end{equation}
the corresponding parton condensation momenta are also given. We will call the above cases 1,2,3,4,5, where cases 1 and 2 have condensation momentum at $\Gamma$, cases 3 and 4 at $\pm \mathrm{K}$, and case 5 at $\mathrm{M}_{1,2,3}$. We will also use the parameterization 
\begin{equation}\label{class01params}
a=\sin\psi \sin \theta \cos \phi,\quad a'=\sin \psi\sin \theta \sin\phi, \quad A=\sin \psi\cos \theta,\quad A'=\cos\psi,
\end{equation}
and using this parameterization the condensation diagram is given in Fig.~\ref{fig:phase_diagram_01_NNN}.

\begin{figure}
\centering
\includegraphics[width=0.7\textwidth]{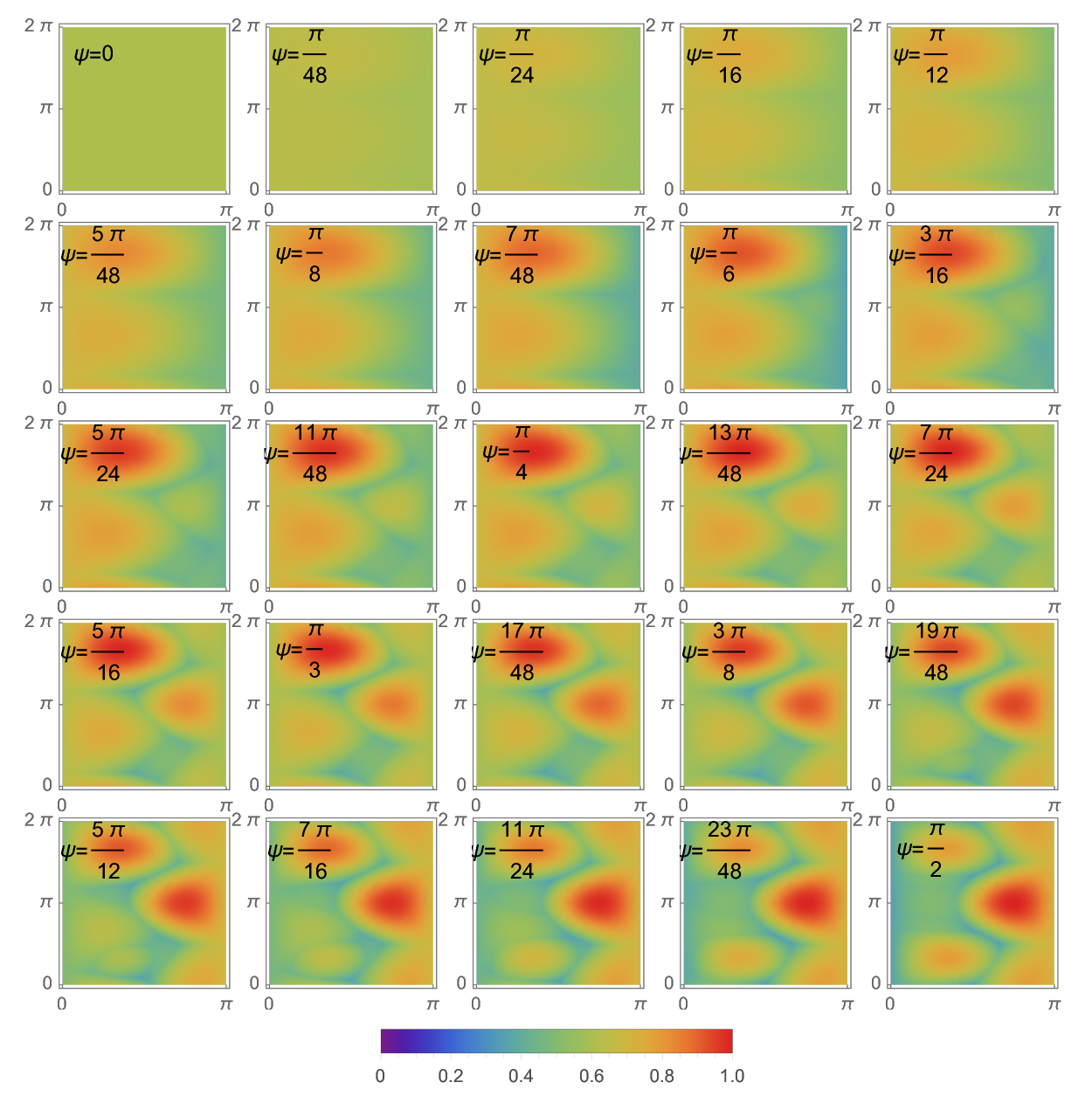}
\caption{The critical chemical potential, corresponding to Fig.~\ref{fig:phase_diagram_01_NNN}. Color denotes the relative size of $\mu_c$ in arbitrary units.}\label{fig:phase_diagram_01_NNN_critical_mu}
\end{figure}

If we keep the parameterization of the bond parameters as normalized in Eq.~\eqref{class01params}, the critical chemical potential for the whole phase diagram is shown in Fig.~\ref{fig:phase_diagram_01_NNN_critical_mu}. The critical chemical potential gives us a rough sense of which magnetic order in the condensation diagram Fig.~\ref{fig:phase_diagram_01_NNN} starts to form the first: keeping a global magnitude of the bond parameters  normalized to Eq.~\eqref{class01params}, the global maximum for the critical chemical potential determines the magnetic order. As we see, by comparing Fig.~\ref{fig:phase_diagram_01_NNN} and Fig.~\ref{fig:phase_diagram_01_NNN_critical_mu}, the largest critical chemical potential (the red spots on the right plots) always corresponds to the purple or blue phases, which are case 1 (collinear FM order) and case 2 ($120^\circ$ coplanar order). Then conceptually this means these two orders tend to form before the rest orders (yellow, red and green) do, and in this sense we can replace the yellow, red and green phases by the parent spin liquid class (0,1) to arrive at a phase diagram with both magnetic order and the parent spin liquid.

The above analysis should be treated with caveats. First, the normalization \eqref{class01params} is only a convenient way to plot the phase diagram and should not be taken for a realistic constraint. A global maximum critical chemical potential is meaningful only when this normalization constraint is retained even when varying the interaction strength. Second, the phase diagram itself is not a real phase diagram, as the bond parameters should in principle be determined self consistently.

For bosonic BdG type Hamiltonian, it is known that the eigenvalues and eigenvectors are obtained by diagonalizing $J\mathcal{H}(\bm{k})$, where $J = \left(\begin{smallmatrix}
1_{6\times 6}&\\&-1_{6\times 6}\end{smallmatrix}\right)$. The zero energy levels are at least two-fold degenerate due to time reversal symmetry $\mathcal{T}$. The bosonic particle-hole $\mathcal{C}$ symmetry may or may not contribute an extra double degeneracy, and depending on these two scenarios the low energy field theory can be different. As we will see, case 1 and case 2 correspond to these two different scenarios.

\subsubsection{Case 1: $\Gamma$ condensation}

At $\Gamma$ the zero-energy eigenstates are four-fold degenerate due to time reversal symmetry $\mathcal{T}$ and particle-hole symmetry $\mathcal{C}$:
\begin{equation}
\mathbf{a},\quad U_{\mathcal{T}}\mathbf{a}^*,\quad U_{\mathcal{C}}\mathbf{a}^*,\quad U_{\mathcal{C}}U_{\mathcal{T}}\mathbf{a},
\end{equation}
where 
\begin{equation}\label{simplesta}
\mathbf{a} = (1,0,1,0,1,0,0,0,0,0,0,0),
\end{equation}
note that this belongs to the scenario where the bosonic particle-hole symmetry leads to extra double degeneracy.

Generally, the spin configuration is computed from the parton condensation as
\begin{equation}\label{spinConf}
S^i_{\tau} \propto \left(\begin{array}{cc} \langle b^\dag_{\tau\uparrow}\rangle & \langle b^\dag_{\tau\uparrow}\rangle \end{array}\right) \sigma^i \left(\begin{array}{c} \langle b_{\tau\uparrow}\rangle\\\langle b_{\tau\downarrow}\rangle\end{array}\right),
\end{equation}
where $i=x,y,z$ are the spin components, and $\tau=0,1,2$ denote the three sublattices. For each $\tau$, we have $ \langle b_\tau\rangle = (c_1\mathbf{a}+c_2U_{\mathcal{T}}\mathbf{a}^*)_{2\tau:2\tau+1}$ where $c_1$ and $c_2$ are arbitrary complex number and the notation $2\tau:2\tau+1$ denotes the extraction of the $2\tau$ and $2\tau+1$ components of the vector. We can then write the spin as (suppressing $\tau$)
\begin{equation}\label{spinconinspecialcase}
\bm{S} = \left(|c_1|^2-|c_2|^2\right) \bm{n}_1+2\mathrm{Re}(c^*_1c_2) \bm{n}_2-2\mathrm{Im}(c^*_1c_2) \bm{n}_3,
\end{equation}
where 
\begin{equation}\label{triad}
\begin{aligned}
\bm{n}_1 &= (2\mathrm{Re}(a^*b),2\mathrm{Im}(a^*b), |a|^2-|b|^2),\\
\bm{n}_2 &= (\mathrm{Re}(-a^{*2}+b^{*2}),-\mathrm{Im}(a^{*2}+b^{*2}),2\mathrm{Re}(a^*b^*)),\\
\bm{n}_3 &= (\mathrm{Im}(-a^{*2}+b^{*2}),\mathrm{Re}(a^{*2}+b^{*2}),2\mathrm{Im}(a^*b^*)).
\end{aligned}
\end{equation}
where $\bm{n}_1,\bm{n}_2$ and $\bm{n}_3$ are mutually orthogonal, and that $|\bm{n}_{1,2,3}|^2=(|a|^2+|b|^2)^2$, 
so we have $|\bm{S}|^2=(|a|^2+|b|^2)^2(|c_1|^2+|c_2|^2)^2$. Apply this to Case 1, we see that it gives the collinear ferromagnetic order.

\subsubsection{Case 2: $\Gamma$ condensation} 

Let us first consider the case of $a'-A'>0$. Then at $\Gamma$ point there are again four eigenstates
\begin{equation}
\mathbf{a}_1,\quad U_{\mathcal{T}}\mathbf{a}_1^*,\quad \mathbf{a}_2,\quad U_{\mathcal{T}}\mathbf{a}_2^*,
\end{equation}
where
\begin{equation}\label{a1a2form}
\mathbf{a}_1 = \left(\frac{\sqrt{3}}{2},-\frac{1}{2},-\frac{\sqrt{3}}{2},-\frac{1}{2},0,1,\frac{\sqrt{3}}{2},-\frac{1}{2},-\frac{\sqrt{3}}{2},-\frac{1}{2},0,1\right),\quad
\mathbf{a}_2 = \left(\frac{i\sqrt{3}}{2},\frac{i}{2},-\frac{i\sqrt{3}}{2} ,\frac{i}{2},0,-i,-\frac{i\sqrt{3}}{2},-\frac{i}{2},\frac{i\sqrt{3}}{2},-\frac{i}{2},0,i\right).
\end{equation}
Note that now $U_{\mathcal{C}}\mathbf{a}^*_{1,2} = \mathbf{a}^*_{1,2}$, so this belongs to the ``nondiagonalizable'' case, whose low energy dispersion is linear ($z=1$). We have
\begin{equation}
\langle b \rangle = r_1 \mathbf{a}_1 + r_2 U_{\mathcal{T}}\mathbf{a}_1^* + r_3 \mathbf{a}_2 + r_4 U_{\mathcal{T}}\mathbf{a}_2^*,
\end{equation}
where $r_{1,2,3,4}$ are all real. Using Eq.~\eqref{spinConf} to get the spin configuration, we get
\begin{subequations}
\begin{align}
\bm{S}_1&=
\Big(\frac{\sqrt{3} (-r_1^2+r_2^2+r_3^2-r_4^2)-2 (r_1 r_2-r_3r_4)}{4},\frac{ \sqrt{3}(r_1r_3-r_2r_4)+r_1r_4+r_2r_3}{2},\frac{r_1^2-r_2^2+r_3^2-r_4^2-2 \sqrt{3}(r_1 r_2+ r_3 r_4)}{4}\Big),\\
\bm{S}_2&= \Big(\frac{\sqrt{3} (r_1^2-r_2^2-r_3^2+r_4^2)-2 (r_1 r_2-r_3r_4)}{4},\frac{-\sqrt{3} (r_1r_3-r_2r_4)+r_1 r_4+r_2r_3}{2},\frac{r_1^2-r_2^2+r_3^2-r_4^2 +2 \sqrt{3} (r_1 r_2+ r_3 r_4)}{4}\Big),\\
\bm{S}_3&= \Big(r_1 r_2-r_3 r_4,-r_1 r_4-r_2 r_3,\frac{-r_1^2+r_2^2-r_3^2+r_4^2}{2}\Big)
\end{align}
\end{subequations}
We have $|\bm{S}_{1,2,3}|^2 = \frac{1}{4}(r_1^2 + r_2^2 + r_3^2 + r_4^2)^2$, $\bm{S}_1\cdot \bm{S}_2 = \bm{S}_2 \cdot \bm{S}_3 = \bm{S}_1 \cdot \bm{S}_3 = -\frac{1}{8}(r_1^2 + r_2^2 + r_3^2 + r_4^2)^2$ and $\bm{S}_1\cdot (\bm{S}_2\times \bm{S}_3 ) =0$, so this describes a three sublattice coplanar $120^\circ$ order.

We can similarly consider $a'<0$. In this case, instead of the eigenstates \eqref{a1a2form} we have
\begin{equation}
\mathbf{a}_1 = \Big(-\frac{\sqrt{3}}{2},\frac{1}{2},\frac{\sqrt{3}}{2},\frac{1}{2},0,-1,\frac{\sqrt{3}}{2},-\frac{1}{2},-\frac{\sqrt{3}}{2},-\frac{1}{2},0,1\Big),\quad
\mathbf{a}_2 = \Big(-\frac{i\sqrt{3}}{2},-\frac{i}{2},\frac{i\sqrt{3}}{2} ,-\frac{i}{2},0,i,-\frac{i\sqrt{3}}{2},-\frac{i}{2},\frac{i\sqrt{3}}{2},-\frac{i}{2},0,i\Big),
\end{equation}
i.e. the first six components of each eigenvector have flipped signs. It turns out that this will also give a three sublattice coplanar $120^\circ$ order, but with the \emph{opposite vector spin chirality} to the $a'>0$ case. 

\subsubsection{Case 3, K point}

Both the Hamiltonian $\mathcal{H}(\bm{k})$ and the nonhermitian one $J\mathcal{H}(\bm{k})$ have four zero eigenvalues at $\mathrm{K} = (\frac{2\pi}{3},0)$. This belongs to the ``diagonalizable'' case, whose low energy gapless dispersion is quadratic (having dynamical exponent $z=2$). This has been checked numerically. The null vectors at both momenta $\pm \mathrm{K}$ are
\begin{equation}
\mathbf{b},\quad U_{\mathcal{T}}\mathbf{b}^*,\quad U_{\mathcal{C}}\mathbf{b}^*,\quad U_{\mathcal{C}}U_{\mathcal{T}}\mathbf{b},
\end{equation}
where 
\begin{equation}\label{simplestbb}
\mathbf{b} = (1,0,-1,0,1,0,0,0,0,0,0,0).
\end{equation}
Note that the null vector $U_{\mathcal{T}}\mathbf{b}^*$ is not the time reversal partner of $\mathbf{b}$; $\mathbf{b}$ and $U_{\mathcal{T}}\mathbf{b}^*$ are just treated as two different null vectors. We have
\begin{equation}
\langle b_{\bm{r}}\rangle = e^{i \phi_{\bm{r}} }
(c_1\mathbf{b}_{1:6}+c_2 (U_{\mathcal{T}}\mathbf{b}^*)_{1:6})+e^{-i \phi_{\bm{r}} }
(c_3\mathbf{b}_{1:6}+c_4 (U_{\mathcal{T}}\mathbf{b}^*)_{1:6}),
\end{equation}
where we defined $\phi_{\bm{r}} = \left(\frac{2\pi}{3},0\right)\cdot \bm{r}$, and $c_1=x_1+ix_2,c_2=x_3+ix_4,c_3=x_5+ix_6,c_4=x_7+ix_8$ are arbitrary complex numbers. The spin order is then of the form
\begin{equation}
\bm{S}_{\bm{r}_\tau}
=\bm{n} + \bm{m}_1 \cos \phi_{\bm{r}} + \bm{m}_2 \sin \phi_{\bm{r}},
\end{equation}
where
\begin{equation}
\begin{aligned}
\bm{n} &= \Big(x_{1} x_{3}+x_{2} x_{4}+x_{5} x_{7}+x_{6} x_{8},x_{1} x_{4}-x_{2} x_{3}+x_{5} x_{8}-x_{6} x_{7},\frac{-x_{1}^2-x_{2}^2+x_{3}^2+x_{4}^2-x_{5}^2-x_{6}^2+x_{7}^2+x_{8}^2}{2}\Big),\\
\bm{m}_1 &= \left(x_{1} x_{7}+x_{2} x_{8}+x_{3} x_{5}+x_{4} x_{6},x_{1} x_{8}-x_{2} x_{7}-x_{3} x_{6}+x_{4} x_{5},-x_{1} x_{5}-x_{2} x_{6}+x_{3} x_{7}+x_{4} x_{8}\right),\\
\bm{m}_2 &= \left(x_{1} x_{8}-x_{2} x_{7}+x_{3} x_{6}-x_{4} x_{5},-x_{1} x_{7}-x_{2} x_{8}+x_{3} x_{5}+x_{4} x_{6},-x_{1} x_{6}+x_{2} x_{5}+x_{3} x_{8}-x_{4} x_{7}\right).
\end{aligned}
\end{equation}
The above notation means that for a fixed unit cell say $\bm{r} = (0,0)$, the three spins on the small triangle labeled by $\tau=0,1,2$ are collinear; however, the neighboring triangle belonging to the $\bm{r}=(1,0)$ cell has $\phi_{\bm{r}} = -\frac{2\pi}{3}$, and the three spins on this triangle ($\bm{r}=(1,0)$) are also collinear but the orientation and amplitude are different from those on the $\bm{r}=(0,0)$ triangle. The order contains both a FM part and a $120^\circ$ part.

We call it the $\sqrt{3}\times \sqrt{3}$, nonuniform umbrella, sublattice-uniform order.

\subsubsection{Case 4, K point}\label{supp:psg:case_4}

This belongs to the ``nondiagonalizable'' case with linear dispersion ($z=1$). To obtain the expectation value of the bosonic spinons we following the procedure as follows: for the Hamiltonian
\begin{equation}
H = \left(\begin{array}{cc} b^\dag_{\bm{k}} & b_{-\bm{k}}\end{array}\right)
\mathcal{H}(\bm{k}) \left(\begin{array}{c} b_{\bm{k}} \\ b^\dag_{-\bm{k}}\end{array}\right),
\end{equation}
whenever $z=1$, or equivalently, whenever there are $2m$ zero eigenvalues but the nullspace of $\mathcal{H}$ is $m$-dimensional, then, we use basis transformation to get into the position-momentum operator space of $(x_{\bm{k}},x_{-\bm{k}},p_{\bm{k}},p_{-\bm{k}})$:
\begin{equation}
\mathcal{H}_r = \frac{1}{\sqrt{2}} \left(\left( \begin{array}{cccc} 1 &0 & i &0\\ 0 & 1 & 0 & -i\end{array}\right)\otimes 1_{6\times 6}\right)^T \mathcal{H}
\frac{1}{\sqrt{2}}\left( \begin{array}{cccc} 1 &0 & i &0\\ 0 & 1 & 0 & -i\end{array}\right)\otimes 1_{6\times 6};
\end{equation}
The matrix $\mathcal{H}_r$ has a dimension twice that of $\mathcal{H}$. Then, we solve the nullspace of $\mathcal{H}_r$, which should give precisely $2m$ distinct eigenvectors $\mathbf{a}_{1,...,2m}$, and these are the gapless modes. In the $z=1$ case we expect the conjugate operators of these gapless modes are gapped, which we denote by $\mathbf{b}_{1,...,2m}$. And the conjugate operators can be obtained by $\mathcal{E} \mathcal{H}_r \mathbf{b}_i = \lambda_i \mathbf{a}_i$, $i = 1,2,...,2m$, but this is irrelevant to the expectation value of bosonic spinons. 

Since the $\mathbf{b}_i$'s are gapped, we assume they correspond to modes that having a zero expectation value in those operators; but the conjugate operators corresponding to $\mathbf{a}_i$'s are maximally fluctuating since they are gapless, Therefore they acquire some expectation value (indefinite), say $c_{1,...,2m}$. The expectation value for the $b$ operators are then
\begin{equation}
\left(\begin{array}{c}\langle b_{\bm{k}}\rangle  \\ \langle b^\dag_{-\bm{k}}\rangle \end{array}\right)
=\frac{1}{\sqrt{2}}\left(\begin{array}{cccc} 1 & 0 & i & 0\\ 0 & 1 & 0 & -i\end{array}\right)\otimes 1_{6\times 6} \left(\begin{array}{ccc} \mathbf{a}_1 & \cdot & \mathbf{a}_{2m}\end{array}\right)\left(\begin{array}{c} c_1 \\ \vdots \\ c_{2m}\end{array}\right),
\end{equation}
where $c_{1,...,2m}$ are real. 

Following the above procedure, we get
\begin{equation}
\left(\begin{array}{c}\langle b_{(\frac{2\pi}{3},0)}\rangle  \\ \langle b^\dag_{-(\frac{2\pi}{3},0)}\rangle \end{array}\right) = (x_1+ix_2)\mathbf{a}_1+(x_3+ix_4)\mathbf{a}_2+(x_5+ix_6)\mathbf{a}_3+(x_7+i x_8)\mathbf{a}_4,
\end{equation}
where $x_{1,2,...,8}$ are arbitrary real parameters denoting the real and imaginary parts of the condensation, and
\begin{equation}\label{simplestbb2}
\begin{aligned}
\mathbf{a}_1 &= \left(\frac{i}{\sqrt{3}},0,\frac{2 i}{\sqrt{3}},0,\frac{i}{\sqrt{3}},0,0,i,0,0,0,-i\right),\quad
\mathbf{a}_2 = \left(0,-\frac{i}{\sqrt{3}},0,-\frac{2 i}{\sqrt{3}},0,-\frac{i}{\sqrt{3}},i,0,0,0,-i,0\right),\\
\mathbf{a}_3 &=\left(\frac{i}{\sqrt{3}},0,-\frac{i}{\sqrt{3}},0,-\frac{2 i}{\sqrt{3}},0,0,-i,0,-i,0,0\right),\quad
\mathbf{a}_4 =\left(0,-\frac{i}{\sqrt{3}},0,\frac{i}{\sqrt{3}},0,\frac{2 i}{\sqrt{3}},-i,0,-i,0,0,0\right).
\end{aligned}
\end{equation}
Using the real space parton condensation $
\langle b_{\bm{r}}\rangle = e^{i \phi_{\bm{r}} } \langle b_{(\frac{2\pi}{3},0)}\rangle  + e^{-i \phi_{\bm{r}} } \langle b_{(-\frac{2\pi}{3},0)}\rangle$ to write the spin order, we obtain 
\begin{equation}
\bm{S}_{\bm{r}_\tau}
=\bm{n}_\tau + \bm{m}_{1,\tau} \cos 2\phi_{\bm{r}} + \bm{m}_{2,\tau} \sin 2\phi_{\bm{r}},
\end{equation}
where $\bm{n}_\tau$, $\bm{m}_{1,\tau}$ and $\bm{m}_{2,\tau}$ have the expression
\begingroup
\allowdisplaybreaks
\begin{align}
\bm{n}_0&= \left(\frac{1}{3} (x_{1} (x_{3}-2 x_{7})+x_{2} (x_{4}-2 x_{8})-2 x_{3} x_{5}-2 x_{4} x_{6}+x_{5} x_{7}+x_{6} x_{8}),\right.\notag\\
 &\qquad \left.\frac{1}{3} (x_{1} x_{4}-2 x_{1} x_{8}-x_{2} x_{3}+2 x_{2} x_{7}+2 x_{3} x_{6}-2 x_{4} x_{5}+x_{5} x_{8}-x_{6} x_{7}) ,\right. \notag\\
 &\qquad \left.\frac{1}{6} \left(-x_{1}^2+4 x_{1} x_{5}-x_{2}^2+4 x_{2} x_{6}+x_{3}^2-4 x_{3} x_{7}+x_{4}^2-4 x_{4} x_{8}-x_{5}^2-x_{6}^2+x_{7}^2+x_{8}^2\right) \right),\notag\\    
\bm{n}_1&=\left(
 \frac{1}{3} (x_{1} (x_{7}-2 x_{3})+x_{2} (x_{8}-2 x_{4})+x_{5} (x_{3}+x_{7})+x_{6} (x_{4}+x_{8})), \right.\notag\\
 &\qquad \left.\frac{1}{3} (x_{1} (x_{8}-2 x_{4})+x_{2} (2 x_{3}-x_{7})-x_{6} (x_{3}+x_{7})+x_{5} (x_{4}+x_{8})) , \right.\notag\\
 &\qquad \left.\frac{1}{6} \left(2 x_{1}^2-2 x_{1} x_{5}+2 x_{2}^2-2 x_{2} x_{6}-2 x_{3}^2+2 x_{3} x_{7}-2 x_{4}^2+2 x_{4} x_{8}-x_{5}^2-x_{6}^2+x_{7}^2+x_{8}^2\right) \right),\notag\\
\bm{n}_2&=\left(
 \frac{1}{3} (x_{1} (x_{3}+x_{7})+x_{2} (x_{4}+x_{8})+x_{3} x_{5}+x_{4} x_{6}-2 x_{5} x_{7}-2 x_{6} x_{8}),\right.\notag\\ 
 &\qquad \left.\frac{1}{3} (x_{1} (x_{4}+x_{8})-x_{2} (x_{3}+x_{7})-x_{3} x_{6}+x_{4} x_{5}-2 x_{5} x_{8}+2 x_{6} x_{7}) ,\right. \notag\\
 &\qquad \left.\frac{1}{6} \left(-x_{1}^2-2 x_{1} x_{5}-x_{2}^2-2 x_{2} x_{6}+x_{3}^2+2 x_{3} x_{7}+x_{4}^2+2 x_{4} x_{8}+2 x_{5}^2+2 x_{6}^2-2 \left(x_{7}^2+x_{8}^2\right)\right) \right),\notag\\    
\bm{m}_{1,0}&=\left(
 \frac{-x_{1}^2+x_{2}^2+x_{3}^2-x_{4}^2+x_{5}^2-x_{6}^2-x_{7}^2+x_{8}^2}{2 \sqrt{3}} , \frac{x_{1} x_{2}+x_{3} x_{4}-x_{5} x_{6}-x_{7} x_{8}}{\sqrt{3}} , \frac{-x_{1} x_{3}+x_{2} x_{4}+x_{5} x_{7}-x_{6} x_{8}}{\sqrt{3}} \right),\notag\\    
\bm{m}_{1,1}&=\left(
 \frac{2 x_{1} x_{5}-2 x_{2} x_{6}-2 x_{3} x_{7}+2 x_{4} x_{8}-x_{5}^2+x_{6}^2+x_{7}^2-x_{8}^2}{2 \sqrt{3}}, -\frac{x_{1} x_{6}+x_{2} x_{5}+x_{3} x_{8}+x_{4} x_{7}-x_{5} x_{6}-x_{7} x_{8}}{\sqrt{3}} ,\right.\notag\\
 &\qquad \left. \frac{x_{1} x_{7}-x_{2} x_{8}+x_{3} x_{5}-x_{4} x_{6}-x_{5} x_{7}+x_{6} x_{8}}{\sqrt{3}} \right),\notag\\    
\bm{m}_{1,2}&=\left(
 \frac{x_{1}^2-2 x_{1} x_{5}-x_{2}^2+2 x_{2} x_{6}-x_{3}^2+2 x_{3} x_{7}+x_{4}^2-2 x_{4} x_{8}}{2 \sqrt{3}},\frac{x_{1} (x_{6}-x_{2})+x_{2} x_{5}+x_{3} (x_{8}-x_{4})+x_{4} x_{7}}{\sqrt{3}} ,\right. \notag\\
 &\qquad \left. \frac{x_{1} (x_{3}-x_{7})+x_{2} (x_{8}-x_{4})-x_{3} x_{5}+x_{4} x_{6}}{\sqrt{3}} \right),\notag\\    
\bm{m}_{2,0}&=\left(
 \frac{x_{1} x_{2}-x_{3} x_{4}-x_{5} x_{6}+x_{7} x_{8}}{\sqrt{3}} , \frac{x_{1}^2-x_{2}^2+x_{3}^2-x_{4}^2-x_{5}^2+x_{6}^2-x_{7}^2+x_{8}^2}{2 \sqrt{3}} , \frac{x_{1} x_{4}+x_{2} x_{3}-x_{5} x_{8}-x_{6} x_{7}}{\sqrt{3}} \right),\notag\\    
\bm{m}_{2,1}&=\left(
 \frac{-x_{1} x_{6}-x_{2} x_{5}+x_{3} x_{8}+x_{4} x_{7}+x_{5} x_{6}-x_{7} x_{8}}{\sqrt{3}} , \frac{-2 x_{1} x_{5}+2 x_{2} x_{6}-2 x_{3} x_{7}+2 x_{4} x_{8}+x_{5}^2-x_{6}^2+x_{7}^2-x_{8}^2}{2 \sqrt{3}} ,\right.\notag\\
 &\qquad \left. -\frac{x_{1} x_{8}+x_{2} x_{7}+x_{3} x_{6}+x_{4} x_{5}-x_{5} x_{8}-x_{6} x_{7}}{\sqrt{3}} \right),\notag\\    
\bm{m}_{2,2}&=\left(
 \frac{x_{1} (x_{6}-x_{2})+x_{2} x_{5}+x_{3} (x_{4}-x_{8})-x_{4} x_{7}}{\sqrt{3}} ,\frac{-x_{1}^2+2 x_{1} x_{5}+x_{2}^2-2 x_{2} x_{6}-x_{3}^2+2 x_{3} x_{7}+x_{4}^2-2 x_{4} x_{8}}{2 \sqrt{3}} ,\right.\notag\\
  &\qquad \left. \frac{x_{1} (x_{8}-x_{4})+x_{2} (x_{7}-x_{3})+x_{3} x_{6}+x_{4} x_{5}}{\sqrt{3}} \right).
\end{align}
\endgroup
Despite the complicated form, we can verify that 
$|\bm{S}_{\bm{r}_0}|=|\bm{S}_{\bm{r}_1}|=|\bm{S}_{\bm{r}_2}|$, and $\langle \bm{S}_{\bm{r}_0}, \bm{S}_{\bm{r}_1}\rangle = \langle \bm{S}_{\bm{r}_0}, \bm{S}_{\bm{r}_2}\rangle=\langle \bm{S}_{\bm{r}_1}, \bm{S}_{\bm{r}_2}\rangle$ and $\bm{S}_{\bm{r}_0}\cdot (\bm{S}_{\bm{r}_1}\times \bm{S}_{\bm{r}_2})=0$, therefore for each $\bm{r}$, the three sites labeled by the sublattice index $\tau$ form a coplanar $120^\circ$ order. The order has both a $\bm{k}=0$ and a $\bm{k} = \left(\pm \frac{2\pi}{3},0\right)$ (i.e. $\pm \mathrm{K}$ points in the Brillouin zone) wave vector, i.e. it has an enlarged $\sqrt{3}\times \sqrt{3}$ unit cell, and the spins within this enlarged unit cell are of unequal sizes (although within the original sublattice of site $\bm{r}$ the three spins are equal). We have checked the three normal vectors of the coplanar spins and find their orientation are generally different.

\begin{figure}
\centering
\includegraphics[width=0.3\textwidth]{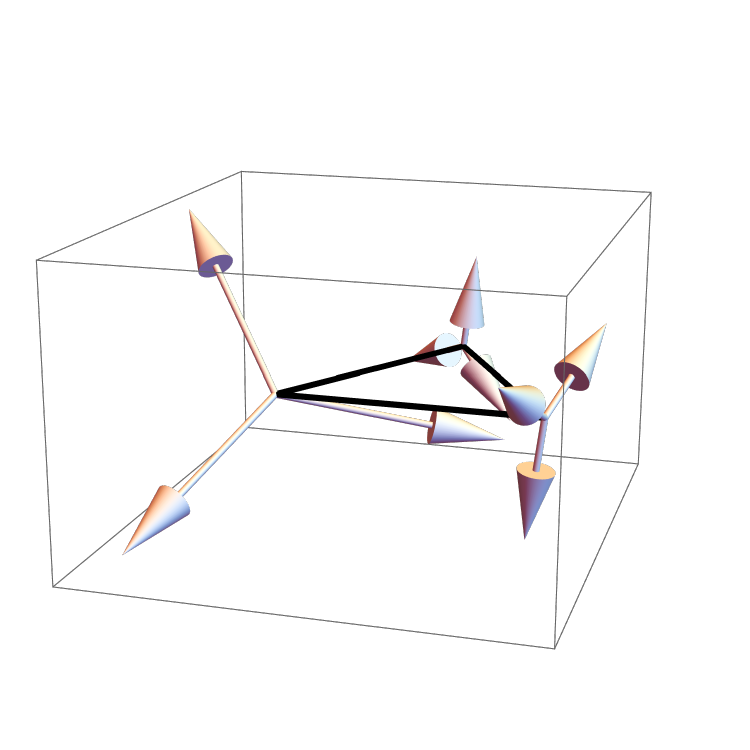}
\includegraphics[width=0.3\textwidth]{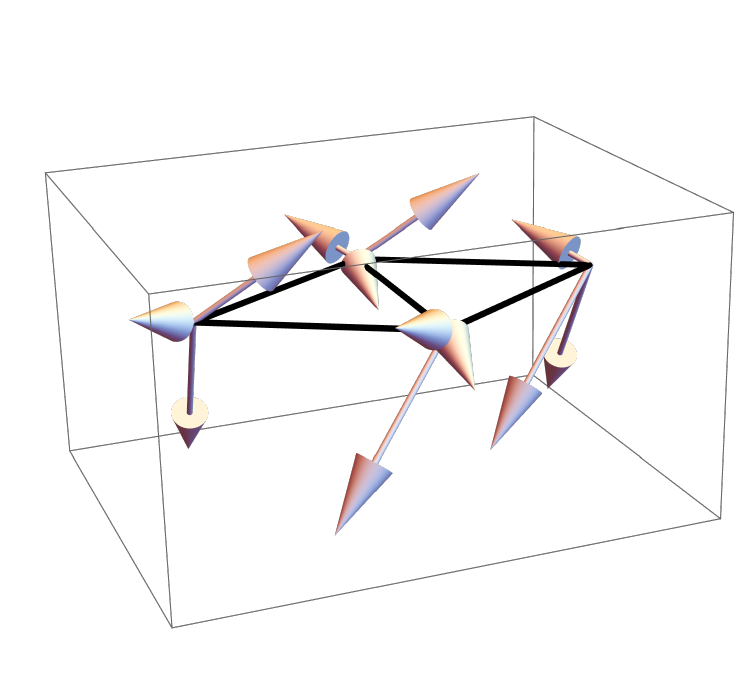}
\caption{Illustration of the sublattice-collinear order in case 4 of $(0,1)$ class and the strip order in case 5 of $(0,1)$ class. The three vectors on each site represents the three spins on the sites $\tau=0,1,2$ within one lattice unit cell. The black bonds denote the (enlarged) magnetic cell.}\label{fig:order_0_1_case3_and_4}
\end{figure}

We call this the $\sqrt{3}\times \sqrt{3}$ umbrella, nonuniform, sublattice-coplanar $120^\circ$ order. This order is illustrated in the left panel of Fig.~\ref{fig:order_0_1_case3_and_4}.

\subsubsection{Case 5, M point}

Define $\mathrm{M}_1 = \left(0,\frac{\pi}{\sqrt{3}a}\right)$, $\mathrm{M}_2 = \left(\frac{\pi}{2a},\frac{\pi}{2\sqrt{3}a}\right)$, $\mathrm{M}_3 = \left(\frac{\pi}{2a},-\frac{\pi}{2\sqrt{3}a}\right)$. This belongs to the ``diagonalizable'' case with quadratic dispersion ($z=2$). The eigenvectors are
\begin{equation}
\text{For }\mathrm{M}_i,i=1,2,3\colon
\quad \mathbf{a}_i,\quad U_{\mathcal{T}}\mathbf{a}^*_i,\quad U_{\mathcal{C}}\mathbf{a}^*_i,\quad U_{\mathcal{C}}U_{\mathcal{T}} \mathbf{a}_i,
\end{equation}
with
\begin{subequations}
\begin{align}
\mathbf{a}_1 = \left(\Delta,0,0,0,A-a,0,0,0,0,0,0,a'+A'\right),\\
\mathbf{a}_2 = 
\left(-a+A,0,\Delta,0,0,0,0,a'+A',0,0,0,0\right),\\
\mathbf{a}_3 = 
\left(0,0,A-a,0,\Delta,0,0,0,0,a'+A',0,0\right).
\end{align}
\end{subequations}
where we defined $\Delta = \sqrt{(a-A)^2+(a'+A')^2}$. Note that $\mathbf{a}_3\propto U^\dag_{{C}_6} \mathbf{a}_1$, $\mathbf{a}_2 \propto U^\dag_{{C}_6}\mathbf{a}_2$, as required by symmetry.

We need to use the above eigenvectors to construct the particle-hole symmetric vectors: define $\mathbf{b}_i = c_1\mathbf{a}_i+c_2U_{\mathcal{T}}\mathbf{a}^*_i+c_3U_{\mathcal{C}}\mathbf{a}^*_i+c_4U_{\mathcal{C}}U_{\mathcal{T}} \mathbf{a}_i$, requiring $\mathbf{b}_i = U_{\mathcal{C}}\mathbf{b}_i^*$ gives
\begin{subequations}
\begin{align}
(\mathbf{b}_1)_{1:6}= \left( \Delta c_1,-\Delta c_2,0,0,(-a+A)c_1+(a'+A')c^*_2,(a-A)c_2+(a'+A')c_1^*\right),\\
(\mathbf{b}_2)_{1:6}= \left(0,0,(-a+A)c_3+(a'+A')c_4^*,(a-A)c_4+(a'+A')c_3^*,\Delta c_3,-\Delta c_4\right),\\
(\mathbf{b}_3)_{1:6}= \left((-a+A)c_5+(a'+A')c_6^*,(a-A)c_6+(a'+A')c_5^*,\Delta c_5,-\Delta c_6,0,0\right),
\end{align}
\end{subequations}
Using the above we obtain the real space parton condensation
\begin{equation}
\langle b_{\bm{r}}\rangle = \sum_{i=1,2,3} e^{i \bm{k}_i\cdot \bm{r}}
(\mathbf{b}_i)_{1:6}
\end{equation}
where $\bm{k}_i$ is the momentum for the $\mathrm{M}_i$ point in the Brillouin zone. We have $\bm{k}_{1,2,3}\cdot \bm{r} = \pi(-r_1+r_2),\pi r_2,\pi r_1$, and
\begin{equation}
\bm{S}_{\bm{r}_\tau}
=\bm{n}_{0,\tau} + \bm{n}_{1,\tau}(-1)^{r_1}+\bm{n}_{2,\tau}(-1)^{r_2}+\bm{n}_{3,\tau}(-1)^{r_1+r_2},
\end{equation}
where $\bm{n}_{0,\tau}$, $\bm{n}_{1,\tau}$, $\bm{n}_{2,\tau}$ and $\bm{n}_{3,\tau}$ are 12 vectors of $c_{1,2,...,6}$, which we omit here. The magnetic order has an enlarged unit cell that is $2\times 2$ times of the original lattice unit cell. We checked that while the order is generically nonplanar and complex, we find that the 12 spins in this $2\times 2$ magnetic unit cell satisfies
\begin{equation}
\begin{aligned}
\bm{S}_{(r_1,r_2)_0} = \bm{S}_{(r_1,r_2+1)_0}&,\quad \bm{S}_{(r_1+1,r_2)_0} = \bm{S}_{(r_1+1,r_2)_0},\quad
\bm{S}_{(r_1,r_2)_1} = \bm{S}_{(r_1+1,r_2)_1},\\ \bm{S}_{(r_1+1,r_2)_1} = \bm{S}_{(r_1+1,r_2+1)_1}&,\quad
\bm{S}_{(r_1,r_2)_2} = \bm{S}_{(r_1+1,r_2+1)_2},\quad \bm{S}_{(r_1,r_2+1)_2} = \bm{S}_{(r_1+1,r_2)_2},
\end{aligned}
\end{equation}
In other words, the three sublattice $\tau=0,1,2$ have different ordering momentum $\mathrm{M}_\tau$, i.e. the sublattice and the magnetic ordering wave vector are \emph{locked}.

We call this the $2\times 2$ sublattice-locked, sublattice-non-uniform strip order.  This order is illustrated in the right panel of Fig.~\ref{fig:order_0_1_case3_and_4}.

\subsection{Class $0-(1,0)$}

We consider both NN and NNN bonds and define 
\begin{equation}\label{class10params}
a=\sin \theta \cos \phi,\quad a' = \sin \theta \sin \phi,\quad A = \cos \theta.
\end{equation}

We checked numerically that the condensation momenta are one of the following: $\Gamma$, $\mathrm{K}$, $\mathrm{M}$, or along the line $\Gamma\mathrm{K}$, or along the line $\mathrm{KM}$. We see that all these points/paths can be viewed as lying on the $k_y=0$ line. The condensation diagram is given in Fig.~\ref{fig:phase_diagram_10_NNN}. As can be seen, when $\theta >\pi/2$ (i.e. $A<0$), we only have condensation at $\Gamma$ or $\mathrm{K}$.

At $\Gamma$, the critical chemical potential is $\mu = \max(a+A,-2(a+A))$; at $\mathrm{K}$, it is $\mu = \max(a-2A+\sqrt{3}|a'|,\frac{1}{2}(-a+2A+\sqrt{3}|a'|))$. As checked numerically, when $A<0$, the critical chemical potential are
\begin{equation}
\mu = \max(\underbrace{-2(a+A)}_{\Gamma},\underbrace{a-2A+\sqrt{3}|a'|}_{\mathrm{K}}).
\end{equation}
The phase boundary for $A<0$ is hence $\phi = \pi/3$ and $\phi = 2\pi/3$. We call them case 1 and case 2. 

\begin{figure}
\centering
\includegraphics[width=0.5\textwidth]{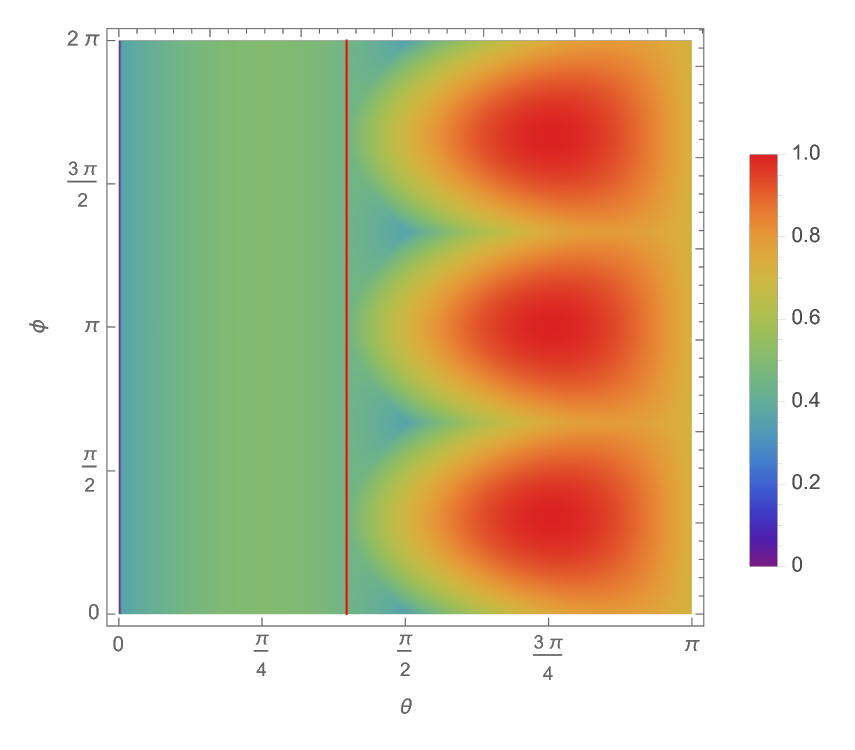}
\caption{The critical chemical potential $\mu_c$ corresponding to Fig.~\ref{fig:phase_diagram_10_NNN}. Color denotes the relative size of $\mu_c$ in arbitrary units. }\label{fig:phase_diagram_10_NNN_critical_mu}
\end{figure}

Keeping the bond parameterization as in Eq.~\eqref{class10params}, the critical chemical potential is given in Fig.~\ref{fig:phase_diagram_10_NNN_critical_mu}. 

By comparing Fig.~\ref{fig:phase_diagram_10_NNN} and Fig.~\ref{fig:phase_diagram_10_NNN_critical_mu}, we see that the largest critical chemical potential (the red regions in Fig.~\ref{fig:phase_diagram_10_NNN_critical_mu}) happen in case 1 (collinear FM order) and case 2 (sublattice-collinear, $\sqrt{3}\times \sqrt{3}$ $120^\circ$ order).

Another tractable case is for $0<\theta < \arctan3\approx 1.24905$, i.e. the part on the left of the red vertical line of the phase diagram in Fig.~\ref{fig:phase_diagram_10_NNN} are all incommensurate. These are due to the energy $E = \frac{\sqrt{\left(a^2+a'^2\right) \cos 2 k_x+ a^2+4 a (A-\mu ) \cos k_x+2 A^2-4 A\mu -a'^2+2 \mu^2}}{\sqrt{2}}$ becoming gapless
at critical momentum $\mu = A+ \sqrt{a^2+a'^2}$ and condensation momentum
\begin{equation}
\mathrm{L} = (\pm \arctan\frac{|a'|}{a},0).
\end{equation}
We call this case 3.

As one can observe from Fig.~\ref{fig:phase_diagram_10_NNN}, for $\arctan3<\theta <\frac{\pi}{2}$ there are also two incommensurate regions, which we call case 4 and 5. The detailed study of these incommensurate phases (cases 3,4,5) will be left to future work.

\subsubsection{Case 1, $\Gamma$ point}

We have four zero energy eigenstates defined by $\mathbf{a}$, $U_{\mathcal{T}}\mathbf{a}^*$, $U_{\mathcal{C}}\mathbf{a}^*$ and $U_{\mathcal{C}}U_{\mathcal{T}}\mathbf{a}$ with $\mathbf{a}$ given in Eq.~\eqref{simplesta}, so everything follows exactly the case 1 of the Class $0-(0,1)$, and we have a collinear ferromagnetic order. 

\subsubsection{Case 2, $\mathrm{K}$ point}

Again first consider $a'>0$. The Hamiltonian $\mathcal{H}(\bm{k})$ at $\mathrm{K} = (\frac{2\pi}{3},0)$ has two zero eigenvalues, whereas the nonhermitian one $J\mathcal{H}(\bm{k})$ has four zero eigenvalues. This belongs to the ``nondiagonalizable'' case, whose low energy gapless dispersion is linear with dynamical exponent $z=1$. Following the procedure outlined in previous sections, we obtain
\begin{equation}
\left(\begin{array}{c}\langle b_{(\frac{2\pi}{3},0)}\rangle  \\ \langle b^\dag_{-(\frac{2\pi}{3},0)}\rangle \end{array}\right) = (x_1+ix_2)\mathbf{a}_1+(x_3+ix_4)\mathbf{a}_2,
\end{equation}
where
\begin{equation}
\mathbf{a}_1 = \left(\frac{1}{\sqrt{2}},0,-\frac{1}{\sqrt{2}},0,\frac{1}{\sqrt{2}},0,0,-\frac{i}{\sqrt{2}},0,\frac{i}{\sqrt{2}},0,-\frac{i}{\sqrt{2}}\right),\quad
\mathbf{a}_2 = \left(0,-\frac{1}{\sqrt{2}},0,\frac{1}{\sqrt{2}},0,-\frac{1}{\sqrt{2}},-\frac{i}{\sqrt{2}},0,\frac{i}{\sqrt{2}},0,-\frac{i}{\sqrt{2}},0\right).
\end{equation}
From this we obtain the real space spin
\begin{equation}
\begin{aligned}
\bm{S}_{\bm{r}_\tau} = \Big(\sin (2\phi_{\bm{r}})\frac{x_1^2-x_2^2-x_3^2+x_4^2}{2}+\cos (2 \phi_{\bm{r}}) (x_1 x_2-x_3 x_4),& \cos (2 \phi_{\bm{r}}) \frac{x_1^2-x_2^2+x_3^2-x_4^2}{2}-\sin (2 \phi_{\bm{r}}) (x_1 x_2+x_3 x_4),\\
&\sin (2 \phi_{\bm{r}}) (x_1 x_3-x_2 x_4)+\cos (2 \phi_{\bm{r}}) (x_1 x_4+x_2 x_3)\Big),
\end{aligned}
\end{equation}
where $\phi_{\bm{r}} = \left(\frac{2\pi}{3},0\right)\cdot \bm{r}$. We check that this is a coplanar $120^\circ$ order. we call this the sublattice-uniform coplanar $120^\circ$ order. This order is illustrated in Fig.~\ref{fig:pd}(d) in the main text.

The magnetic order for the $a'<0$ can be obtained in the same way, which has the opposite vector spin chirality from the $a'>0$ case.

\end{document}